\documentclass{osa-article}

\journal{osajournal}
\articletype{Research Article}

\begin{document}

\title{Deep learning-augmented Computational Miniature Mesoscope}

\author{Yujia Xue\authormark{1.†}, Qianwan Yang\authormark{1,†}, Guorong Hu\authormark{1}, Kehan Guo\authormark{1} and Lei Tian\authormark{1,2,3,*}}

\address{\authormark{1}Department of Electrical and Computer Engineering, Boston University, Boston, MA 02215, USA.\\
\authormark{2}Department of Biomedical Engineering, Boston University, Boston, MA 02215, USA.\\
\authormark{3}Neurophotonics Center, Boston University, Boston, MA 02215, USA.\\
\authormark{†}These authors contributed equally to this paper.}
\email{\authormark{*}leitian@bu.edu} 

\begin{abstract}
 Fluorescence microscopy is essential to study biological structures and dynamics. However, existing systems suffer from a tradeoff between field-of-view (FOV), resolution, and complexity, and thus cannot fulfill the emerging need of miniaturized platforms providing micron-scale resolution across centimeter-scale FOVs. 
 To overcome this challenge, we developed Computational Miniature Mesoscope (CM$^2$) that exploits a computational imaging strategy to enable single-shot 3D high-resolution imaging across a wide FOV in a miniaturized platform. 
 Here, we present CM$^2$ V2 that significantly advances both the hardware and computation. 
 We complement the 3$\times$3 microlens array with a hybrid emission filter that improves the imaging contrast by 5$\times$, and design a 3D-printed freeform collimator for the LED illuminator that improves the excitation efficiency by 3$\times$.
 To enable high-resolution reconstruction across a large volume, we develop an accurate and efficient 3D linear shift-variant (LSV) model for characterizing spatially varying aberrations. 
 We then train a multi-module deep learning model, CM$^2$Net, using only the 3D-LSV simulator. 
 We quantify the detection performance and localization accuracy of CM$^2$Net for reconstructing fluorescent emitters under different conditions in simulation.
 We then show that CM$^2$Net generalizes well to experiments and achieves accurate 3D reconstruction across a $\sim$7-mm FOV and 800-$\mu$m depth, and provides $\sim$6-$\mu$m lateral and $\sim$25-$\mu$m axial resolution. This provides $\sim$8$\times$ better axial resolution and $\sim$1400$\times$ faster speed as compared to the previous model-based algorithm. We anticipate this simple and low-cost computational miniature imaging system will be impactful to many large-scale 3D fluorescence imaging applications.
\end{abstract}

\section{Introduction}
Fluorescence microscopy is indispensable to study biological structures and dynamics\cite{peronComprehensiveImagingCortical2015a}. 
However, the emerging need of compact, lightweight platforms achieving micron-scale resolution across centimeter-scale fields-of-view (FOVs) has created two new challenges. 
The first challenge is to overcome the barrier of imaging at large scale while preserving resolution\cite{weisenburgerGuideEmergingTechnologies2018a}.
Recently developed table-top systems\cite{fanVideorateImagingBiological2019a,kauvarCorticalObservationSynchronous2020a} have enabled multiscale measurements with sufficient resolution, however, they are complex and bulky. 
The second challenge is to perform large-scale imaging in a compact, lightweight platform.
Miniaturized fluorescence microscopes, i.e. miniscopes\cite{aharoniAllLightThat2019a}, have enabled neural imaging in freely moving animals. 
However, most of the miniscopes rely on a gradient index (GRIN) objective lens\cite{aharoniAllLightThat2019a} that limits the FOVs to <1 mm$^2$.  
Wide-FOV miniscopes have recently been developed by replacing the GRIN with a compound lens\cite{scottImagingCorticalDynamics2018a,guoMiniscopeLFOVLargeField2021a}, but at the cost of degraded resolution and increased size, weight, and system complexity.  
In general, fundamental physical limits preclude meeting the joint requirements of FOV, resolution, and miniaturization using conventional optics.

\begin{figure}[tbp]
\centering\includegraphics[width=0.9\linewidth]{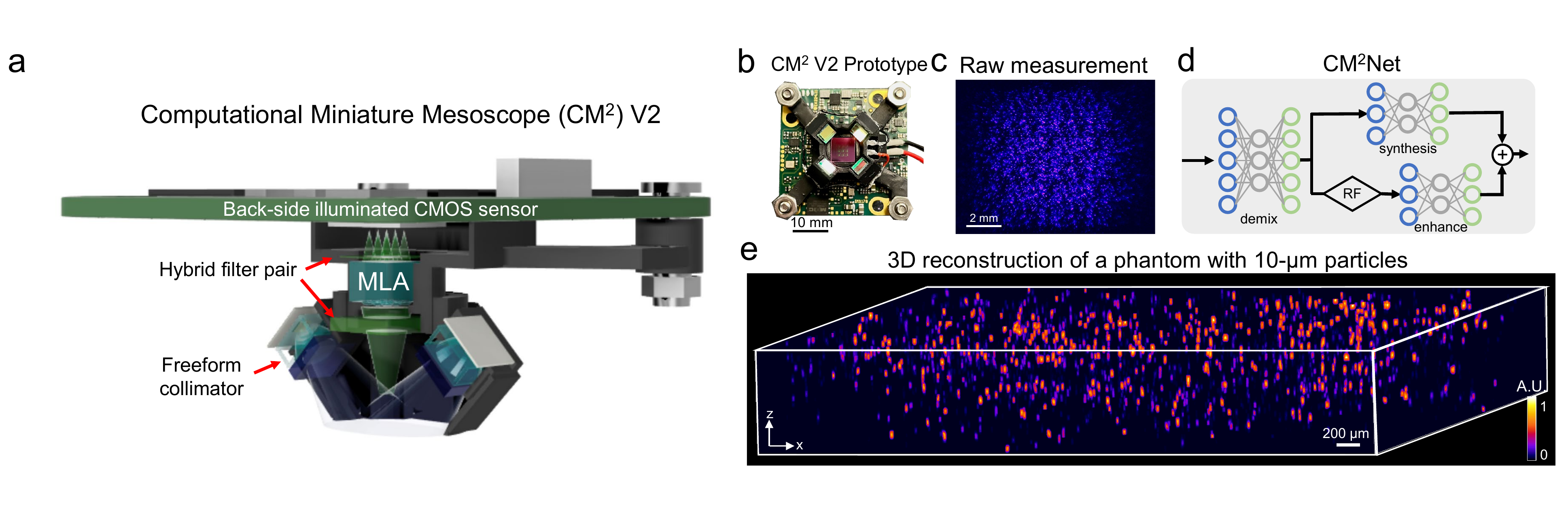}
\caption{Overview of the Computational Miniature Mesoscope (CM$^2$) V2. (a) The CM$^2$ V2 hardware platform features miniature LED illuminators for high-efficiency excitation, hybrid filters for spectral leakage rejection, and a BSI CMOS sensor for high-SNR measurement. (b) A photo of the assembled CM$^2$ V2 prototype. (c) An example CM$^2$ measurement from a volume consisting of 10-$\mu$m fluorescent beads. (d) The CM$^2$Net combines view demixing (demix), view synthesis (synthesis) and lightfield refocusing (RF) enhancement (enhance) modules to achieve high-resolution, fast, and artifact-free 3D reconstruction. (e) The CM$^2$Net reconstruction from the measurement in (c), spanning a 7-mm FOV and 800-$\mu$m depth range.}
\label{fig1}
\end{figure}

Computational imaging techniques have shown unique capabilities of overcoming the limitations of conventional optics by jointly designing optics and algorithms. 
Lightfield microscopy (LFM)\cite{levoyLightFieldMicroscopy2006a} and related technologies\cite{llavadorViewImagesUnprecedented2018a,guoFourierLightfieldMicroscopy2019a}, achieve single-shot high-resolution 3D fluorescence imaging\cite{prevedelSimultaneousWholeanimal3D2014a,pegardCompressiveLightfieldMicroscopy2016a}. 
LFM works by attaching a microlens array (MLA) to an existing microscope to collect both spatial and angular information, which enables reconstructing the 3D fluorescence from a single shot. 
While miniaturized LFMs\cite{skocekHighspeedVolumetricImaging2018a,yannyMiniscope3DOptimizedSingleshot2020a}, enable single-shot 3D imaging on modified miniscope platforms, the FOV is limited by the GRIN lens. 
Lensless imaging is another computational imaging technique for single-shot 3D imaging, where a mask\cite{adamsVivoLenslessMicroscopy2022a} or a diffuser (random microlens array)\cite{kuoOnchipFluorescenceMicroscopy2020a,tianGEOMScopeLargeFieldofView2021a} is placed directly in front of a CMOS. However, the removal of focusing optics imposes penalties to the measurement’s contrast and signal-to-noise ratio (SNR)\cite{kauvarCorticalObservationSynchronous2020a}, severely limiting the sensitivity for imaging weak fluorescent signals\cite{kauvarCorticalObservationSynchronous2020a}.

We recently developed Computational Miniature Mesoscope (CM$^2$)\cite{xueSingleshot3DWidefield2020} that aims to overcome all the key limitations of FOV, resolution, contrast, SNR, and size and weight in existing miniature fluorescence imaging systems. 
The CM$^2$ combines the merits of both LFM and lensless designs. 
It places a 3 $\times$ 3 MLA directly in front of a CMOS sensor for imaging, like the lensless design. This ensures compactness and light weight while further exploiting the microlens’s focusing power for providing high image contrast. 
The CM$^2$ captures an image with multiple views, which enables robust recovery of 3D fluorescence in a single shot, like the LFM. 
Previously, we demonstrated CM$^2$ V1 that achieved 3D fluorescence imaging in a 7 $\times$ 8 mm$^2$ FOV with 7-$\mu$m lateral and 200-$\mu$m axial resolution\cite{xueSingleshot3DWidefield2020}. 
CM$^2$ V1 is the first standalone computational miniature fluorescence microscope with an integrated illumination module. 
Using a four-LED array in an oblique epi-illumination geometry, CM$^2$ V1 can uniformly illuminate a 1-cm$^2$ FOV and achieves a $\sim$24\% light efficiency. 
In this work, we significantly advance the CM$^2$ technology and report CM$^2$ V2 that integrates innovations in both hardware and computation to address limitations in light efficiency, image contrast, reconstruction quality and speed, as summarized in Fig. \ref{fig1}.

On the hardware, we present updates that significantly improve the image contrast and light efficiency. 
First, we complement the 3$\times$3 MLA with a hybrid emission filter \cite{sasagawaHighlySensitiveLensfree2018a} (Fig.~\ref{fig1}a) that suppresses the spectral leakage suffered by the V1 system.
Second, we design and 3D-print a miniature freeform LED collimator (Fig. \ref{fig1}a) that improves the excitation efficiency while preserving compactness and light weight.
This new illuminator achieves $\sim$80\% efficiency, a $\sim$3$\times$ improvement over the V1 design, and provides a confined uniform illumination with up-to 75~mW excitation power across an 8-mm diameter FOV. 
Built around a back-side illuminated (BSI) CMOS (Fig.~\ref{fig1}a, b), CM$^2$ V2 achieves a 5$\times$ improvement in image contrast and captures high-SNR measurements in various experimental conditions (an example image shown in Fig.~\ref{fig1}c).

On the computation, we present a deep learning model, termed CM$^2$Net, to achieve high-quality 3D reconstruction across a wide FOV with significantly improved axial resolution and reconstruction speed. 
Deep learning has recently emerged as the state-of-the-art for solving many inverse problems in imaging\cite{barbastathisUseDeepLearning2019a}. 
For example, deep learning techniques have been developed for LFM to achieve high-resolution 3D reconstruction\cite{yannyDeepLearningFast2022a,wangRealtimeVolumetricReconstruction2021a}. 
To devise a robust and accurate model, we consider several key features in the image formation of CM$^2$. The measurement contains lightfield information in 3 $\times$ 3 overlapped views, and the system is 3D linear shift variant (LSV)~\cite{xueSingleshot3DWidefield2020}. CM$^2$Net solves the single-shot 3D reconstruction problem using three functional modules (Fig. \ref{fig1}d). The “view demixing” module separates the single measurement into 3 $\times$ 3 non-overlapping views by exploiting the distinct aberration features from the array point spread functions (PSF). The “view-synthesis” module and the “lightfield refocusing enhancement” module jointly perform high-resolution 3D reconstruction across a wide FOV by utilizing complementary information.

To incorporate the 3D-LSV information into the trained CM$^2$Net, we develop an accurate and efficient 3D-LSV forward model for synthesizing CM$^2$ measurements. 
Our 3D-LSV model is based on a low-rank approximation using a small number of experimentally calibrated PSFs taken on a sparse 3D grid. 
A key difference between our 3D-LSV model and the depth-wise LSV model~\cite{yannyMiniscope3DOptimizedSingleshot2020a} is the reduced model complexity by a global decomposition and the resulting shared basis PSFs. 
We show that the added axial interpolation in our 3D model achieves “axial super-resolution” beyond the large axial step used in the PSF calibration. 
We generate all the training data using this 3D-LSV simulator to train CM$^2$Net, which bypasses the need for physically acquiring a large-scale training data set in experiments.

We first quantitatively evaluate CM$^2$Net’s performance for reconstructing 3D fluorescent emitters in simulation. Our results demonstrate that CM$^2$Net is robust to variations in the imaging FOV and fluorescent emitter’s size, intensity, 3D location, and seeding density. 
Our ablation studies highlight that the view-demixing module significantly reduces the false positive rates in the reconstruction and that the reconstruction module, consisting of the view-synthesis-net and lightfield refocusing enhancement-net, learns complementary information and enables accurate 3D reconstructions across a wide FOV. 
We quantify the CM$^2$Net’s detection performance and localization accuracy on fluorescent emitters with seeding densities and SNRs approximately matching in-vivo cortex-wide one-photon Calcium imaging across $\sim$7 mm FOVs. 
CM$^2$Net achieves averaged recall and precision of 0.7 and 0.94 respectively, comparable to the state-of-the-art deep learning-based neuron detection pipeline~\cite{bao2021segmentation}. 
It achieves averaged lateral and axial root mean squared localization error (RMSE) of 4.17 $\mu$m and 11.2 $\mu$m respectively, indicating close to single-voxel localization accuracy.
We further perform numerical studies to show that the trained CM$^2$Net generalizes well to complex neural structures, including sparsely and densely labeled neurons across the entire mouse cortex and brain vessel networks.

We show that the 3D-LSV simulator-trained CM$^2$Net generalizes well to experiments (an example reconstruction on 10-$\mu$m beads shown in Fig.~\ref{fig1}e). We  demonstrate CM$^2$Net’s robustness to variations in emitter’s local contrast and SNR on mixed 10-$\mu$m and 15-$\mu$m beads. 
Notably, CM$^2$Net enhances the axial resolution to $\sim$25-$\mu$m -- $\sim$8$\times$ better than the model-based reconstruction. The 3D reconstructions are validated against table-top widefield measurements. The reconstruction quality is quantitatively evaluated and shown nearly uniform detection performance across the whole FOV with few mis-detections. 
In addition, CM$^2$Net reduces the reconstruction time to <4 seconds for a volume spanning a 7-mm FOV and 0.8-mm depth on a standard 8 GB GPU, which is $\sim$1400$\times$ faster speed and $\sim$19$\times$ less memory cost than the model-based algorithm.

Overall, our contribution is a novel deep learning-augmented computational miniaturized microscope that achieves single-shot high-resolution ($\sim$6-$\mu$m lateral and $\sim$25-$\mu$m axial resolution) 3D fluorescence imaging across a mesoscale FOV. 
Built on off-the-shelf and 3D-printed components, we expect this simple and low-cost miniature system can find utility in a wide range of large-scale 3D fluorescence imaging and neural recording applications.

\section{Methods}
\subsection{The CM\textsuperscript{2} V2 hardware platform}
\begin{figure}[htbp]
\centering\includegraphics[width=\linewidth]{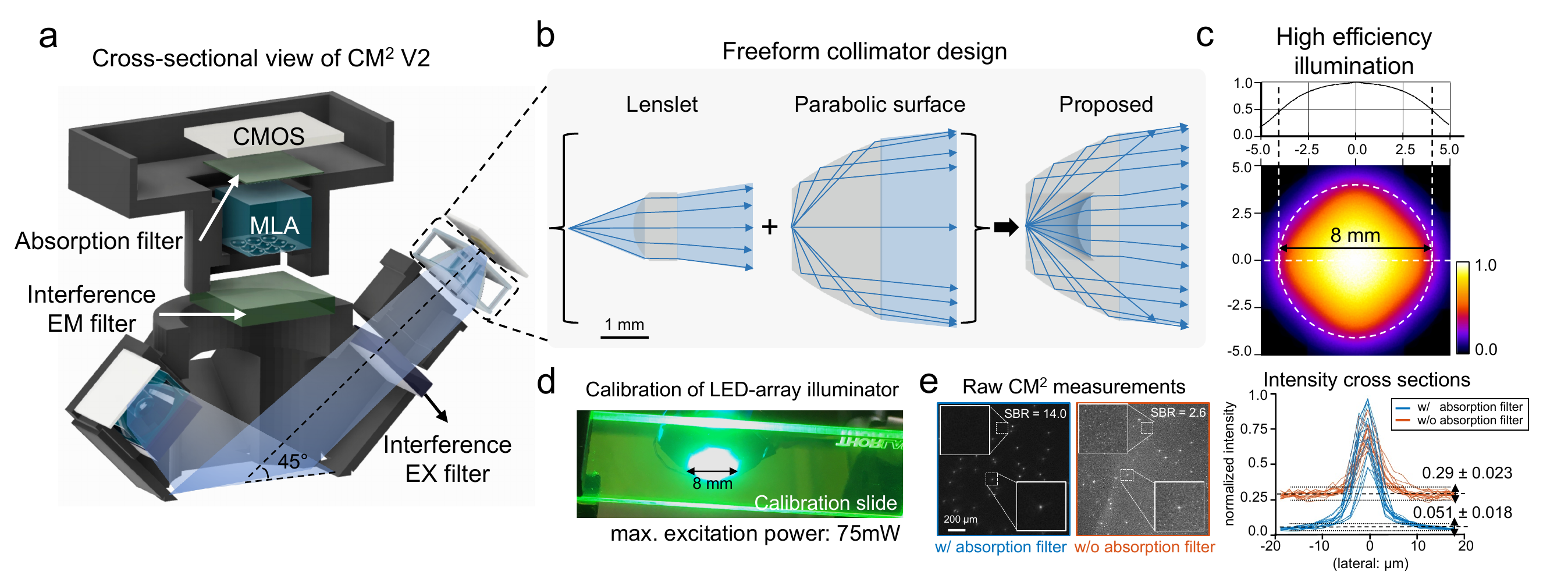}
\caption{CM$^2$ V2 hardware platform. (a) A cross-sectional view of the CM$^2$ V2 platform. The platform incorporates an MLA, freeform illuminators, a hybrid interference-absorption emission filter pair, and a BSI CMOS sensor. (b) The freeform LED collimator combines a singlet and a TIR parabolic surface. (c) Zemax simulation of the four-LED array demonstrates high-efficiency, uniform excitation onto a confined 8-mm circular region. (d) Experimental validation of the illumination module. (e) The hybrid emission filter pair improves the raw measurement’s SBR by > 5$\times$ (sample: 10-$\mu$m fluorescent beads in clear resin). Intensity profiles taken from several fluorescent beads show the SBR improvement by the hybrid filter design.}
\label{fig2}
\end{figure}

CM$^2$ V2 is a standalone miniature fluorescence microscope that is built with off-the-shelf and 3D-printed components, as illustrated in Figs. \ref{fig1}a and \ref{fig2}a. It mainly consists of two parts, including a newly designed illumination module and an upgraded imaging module. As compared to the V1 platform, the V2 platform features freeform LED-collimators that improve the illumination efficiency by $\sim$3$\times$, a hybrid emission filter design that improves the image contrast by $\sim$5$\times$. 

To design the illumination module, our goal is to achieve $\sim$50 mW total excitation power across a centimeter-scale FOV, which is sufficient for one-photon widefield Calcium imaging in mouse brains\cite{kauvarCorticalObservationSynchronous2020a}. 
In addition, the illumination module needs to be highly efficient without incurring excessive heat burden. 
Our solution is to incorporate a compact, lightweight freeform collimator in-between the surface-mounted LED (Lumileds, LXML-PB01-0040) and the excitation filter (Chroma Technology, no. 470). 
The collimator is based on a refraction-reflection freeform design\cite{maFreeformIlluminationLens2015a}. It consists of an inner refractive lenslet and an outer parabolic reflective surface (Fig. \ref{fig2}b). 
The lenslet collimates the light within a $\sim$52-degree conical angle. The parabolic surface satisfies the total internal reflection (TIR) condition and collimates the light emitted at high angles. The LED is placed around the shared focal point of the lenslet and the parabolic refractor. Each collimator is $\sim$4 $\times$ 4 $\times$ 1 mm$^3$ in size, weighs $\sim$0.03 grams, and is 3D-printed with clear resin (printed on Formlabs Form 2, no. RS-F2-GPCL-04). The design achieves an efficiency of $\sim$80\% in Zemax simulation, which considers the finite-sized LED emitter, broadband LED emission spectrum, and angle-dependent transmission spectrum of the excitation filter.

The entire illumination module consists of four LED illuminators placed symmetrically around the imaging module. After performing optimization in Zemax, the LED illuminator is placed $\sim$6.7 mm away from the imging optical axis and tilted by $\sim$45 degrees to direct the light towards the central FOV. The Zemax simulation shows that this design provides nearly uniform illumination confined in an 8-mm circle (Fig. \ref{fig2}c). The experimental validation on a green fluorescence calibration slide (Thorlabs, no. FSK2) closely matches with the simulation (Fig. \ref{fig2}d). The total excitation power is measured to be up-to 75mW (at maximum driving current of 350 mA) at $\sim$470 nm excitation wavelength. 

The imaging module is built around an off-the-shelf 3 $\times$ 3 MLA (Fresnel Technologies Inc., no. 630) to form a finite-conjugate imaging geometry with $\sim$0.57 magnification. The lateral resolution is primarily limited by the NA of a single microlens, which is $\sim$6 µm measured experimentally (see Supplement 1). 
We incorporate an interference-absorption emission filter pair to improve the signal-to-background ratio (SBR) in the raw measurement. 
An interference filter (Chroma Technology, no. 535/50) is placed in front of the MLA. 
An additional long-pass absorption filter (Edmund Optics, Wratten color filter no. 12) is placed after the MLA to  suppress the leakage light. 
The emission spectra of the emission and absorption filters are optimized for the green fluorescence, as detailed in Supplement 1. 
As compared to the interference-filter only measurement, this hybrid filter design improves the SBR by >5$\times$ on a phantom consisting of 10-$\mu$m fluorescence beads (Fig. \ref{fig2}e). This improvement makes the new CM$^2$ V2 platform more robust in low-light fluorescence imaging conditions.

The CM$^2$ V2 is built around a backside-illuminated (BSI) CMOS sensor (IDS Imaging, IMX178LLJ), which gives a 4.15-$\mu$m effective pixel size. The dome-shaped 3D-printed housing (printed on Formlabs Form 2, black resin, no. RS-F2-GPBK-04) provides mechanical support and light shielding. The size and weight of CM$^2$ V2 is only limited by the CMOS sensor. The CM$^2$ V2 prototype is $\sim$36 $\times$ 36 $\times$ 15 mm$^3$ in size, including the commercial CMOS PCB board. The custom parts excluding the PCB is $\sim$20 $\times$ 20 $\times$ 13 mm$^3$ in size and weighs only $\sim$2.5 grams. 

\subsection{3D Linear Shift Variant model of the CM\textsuperscript{2}}
\label{sec2.2}
Our goal is to build an accurate and efficient 3D linear shift variant (LSV) model to describe the CM$^2$ image formation. Using the synthetic data simulated from this model, we will later train the proposed CM$^2$Net to perform 3D reconstruction. In this section, we describe a sparse PSF calibration procedure and a low-rank approximation-based 3D-LSV model. 

\begin{figure}[t]
\centering\includegraphics[width=0.85\linewidth]{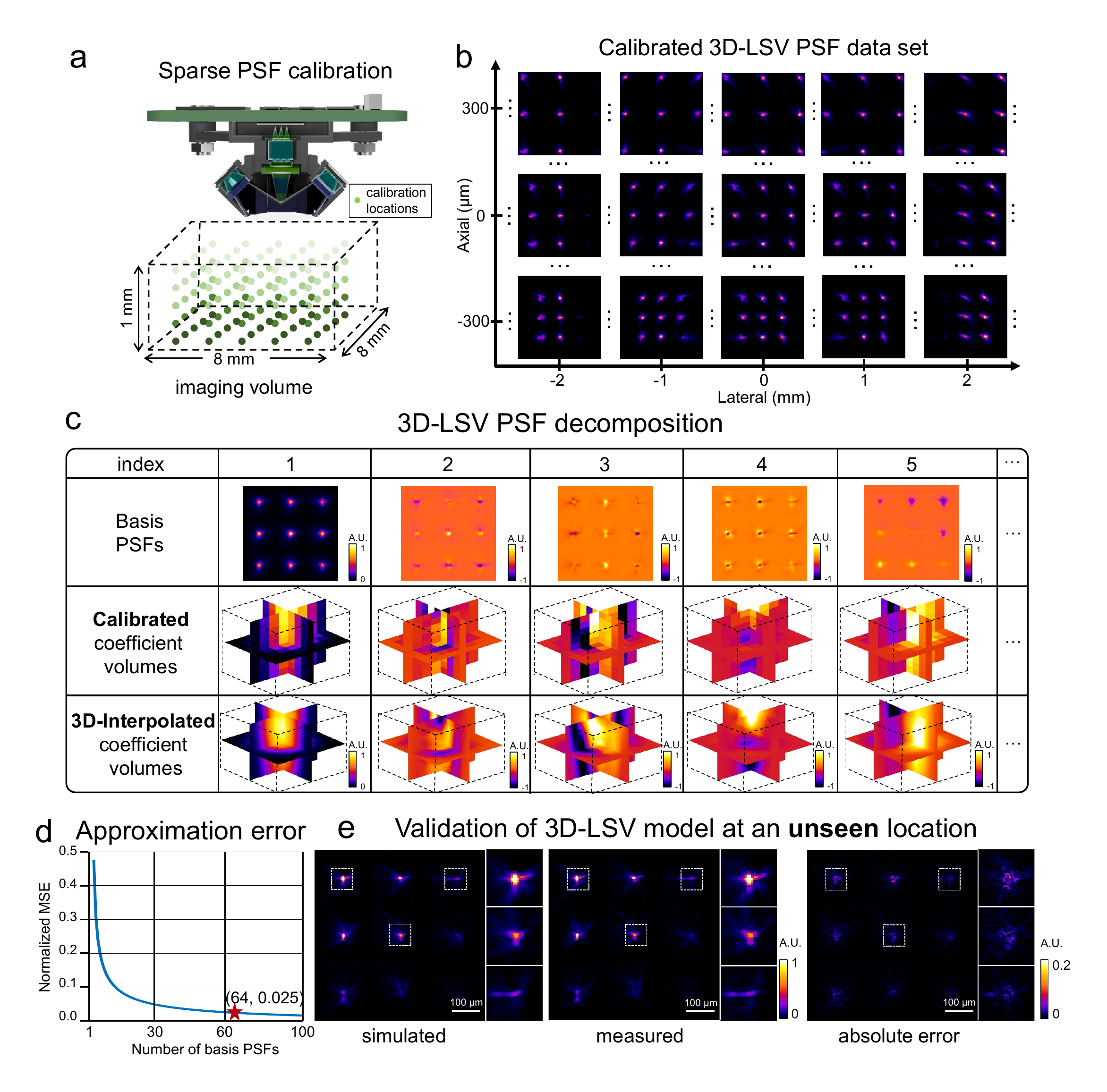}
\caption{3D Linear Shift Variant (LSV) model of the CM$^2$. (a) Illustration of the sparse PSF calibration process. A point source is scanned through the 8 $\times$ 8 $\times$ 1 mm$^3$ imaging volume with a 1-mm lateral and 100-$\mu$m axial steps, generating in-total 891 calibrated PSFs. (b) Example preprocessed calibrated PSFs. The shift variance in 3D is clearly visible. (c) Results of the low-rank decomposition. Row 1-2: the computed basis PSFs and coefficient volumes respectively from the decomposition on the calibrated PSFs. Row 3: the 3D-interpolated coefficient volumes. (d) 64 basis PSFs are chosen for our 3D-LSV model that yields a small 0.025 normalized mean squared error (MSE). (e) Validation of the simulated PSF using our 3D-LSV model at an unseen location. The error between the numerically simulated and experimentally measured PSFs is small, as quantified by the pixel-wise absolute error.}
\label{fig3}
\end{figure}

First, to calibrate the spatially varying PSFs, we scan a 5-$\mu$m point source on a 3-axis translation stage. The point source is scanned across an 8 mm $\times$ 8 mm $\times$ 1 mm volume with steps of 1 mm laterally and 100 $\mu$m axially, which yields a stack of 9 $\times$ 9 $\times$ 11 calibrated PSFs, as illustrated in Fig. \ref{fig3}a. Several example calibrated PSFs are shown in Fig. \ref{fig3}b, which highlight the following key features of the CM$^2$ image formation. At a given lateral position, the off-axis foci shift laterally with the depth, akin to the lightfield. At a given depth, the PSFs are still shift variant because of the spatially varying aberrations from the microlenses and the missing side foci at large off-axis locations (when lateral location > 1.7 mm)\cite{xueSingleshot3DWidefield2020}. As a result, to fully characterize the CM$^2$ 3D PSF, it necessitates a 3D-LSV forward model. Unfortunately, scanning the point source on the entire dense grid at our desired 3D resolution (4.15 µm $\times$ 4.15 µm $\times$ 10 µm) across the targeted imaging volume ($\sim$8 mm $\times$ 8 mm $\times$ 1 mm) would require $\sim$370 million PSF measurements, which is impractical. Next, we describe a computational procedure to address this challenge.

We develop a low-rank approximation based 3D-LSV model for simulating the CM$^2$ measurement in the following steps (more details in Supplement 1):
\begin{enumerate}
\item We denote the sparsely calibrated PSFs as $H(u,v;x,y,z)$, where $(u,v)$ are the pixel coordinates of the PSF image, and $(x,y,z)$ is the 3D location of the point source. In total, the calibrated PSF set contains $N$ = 891 images (the effect of the PSF calibration grid is studied in Supplement 1). Each raw PSF image contains $\sim$6.4M pixels, which is too large to be directly operated on for the low-rank decomposition. To address this issue, we develop a memory efficient scheme by exploiting the highly confined foci in the PSF image. We remove most of the dark regions in the images and then align the cropped foci. The alignment step essentially compensates for the depth-dependent lateral shift in the off-axis foci. We denote this “compressed” and aligned PSF calibration set as $H_c (u',v';x,y,z)$, where $(u',v')$ are the new pixel coordinates after cropping and alignment.
\item We approximate the $N$ calibrated PSFs by a rank-$K$ singular value decomposition (SVD):
\begin{equation}
H(u',v';x,y,z) \approx
 \sum_{i=1}^{K}M^i(x,y,z)H^i_b(u',v'),
\label{eq1}
\end{equation}
where $H_b^i (u',v'),\{i=1, \ldots ,K\}$ denotes the $i^\mathrm{th}$ basis PSF and $M^i (x,y,z)$ is the corresponding coefficient volume. Equation \eqref{eq1}  approximates the set of calibrated PSFs as a linear combination of $K$ basis PSFs. The first five basis PSFs and coefficient volumes are shown in the first two rows in Fig. \ref{fig3}c. We choose $K$ = 64 that has a small $\sim$2.5\% approximation error on the calibration set (Fig. \ref{fig3}d). The choice of $K$ incurs a tradeoff between the model accuracy and computational cost. In addition, this low-rank approximation also helps suppress noise in the raw PSF measurements (more details in Supplement 1).

\item To obtain the coefficient volumes at any uncalibrated 3D location, we perform 3D bilinear interpolation from the sparse calibration grid to the dense reconstruction grid.
This procedure relies on the assumption that the PSFs are slowly varying in 3D\cite{debarnotLearningLowdimensionalModels2020}, which means that 1) the basis PSFs can be accurately estimated from a sparse set of PSF measurements, and 2) the decomposition coefficients are smooth in 3D. The interpolated coefficient volumes for the first five basis PSFs are shown in the third row in Fig. \ref{fig3}c.

\item The final 3D-LSV model is computed by $K$ weighted 2D depth-wise convolutions in the lateral dimension $\circledast_{u,v}$, followed by a summation along the axial dimension $z$:
\begin{equation}
g(u,v) =
 \sum_{z}\sum_{i=1}^{K}[M^i(u,v,z)O(u,v,z)]\circledast_{u,v}H^i_b(u,v,z).
\label{eq2}
\end{equation}
Here, $O(u,v,z)$ is the 3D fluorescence distribution of the object. 
Both the basis PSF $H_b^i (u,v,z)$ and coefficient volume $M^i (u,v,z)$ have been placed back to the original sensor pixel coordinates by accounting for the expected lateral shift at each depth $z$. The pixel coordinates $(u,v)$ in the image and the object space coordinates $(x,y)$ are related by the magnification $M$ by $u=Mx$, $v=My$. 
\end{enumerate}

\subsection{CM\textsuperscript{2}Net design}

\begin{figure}[h]
\centering\includegraphics[width=\linewidth]{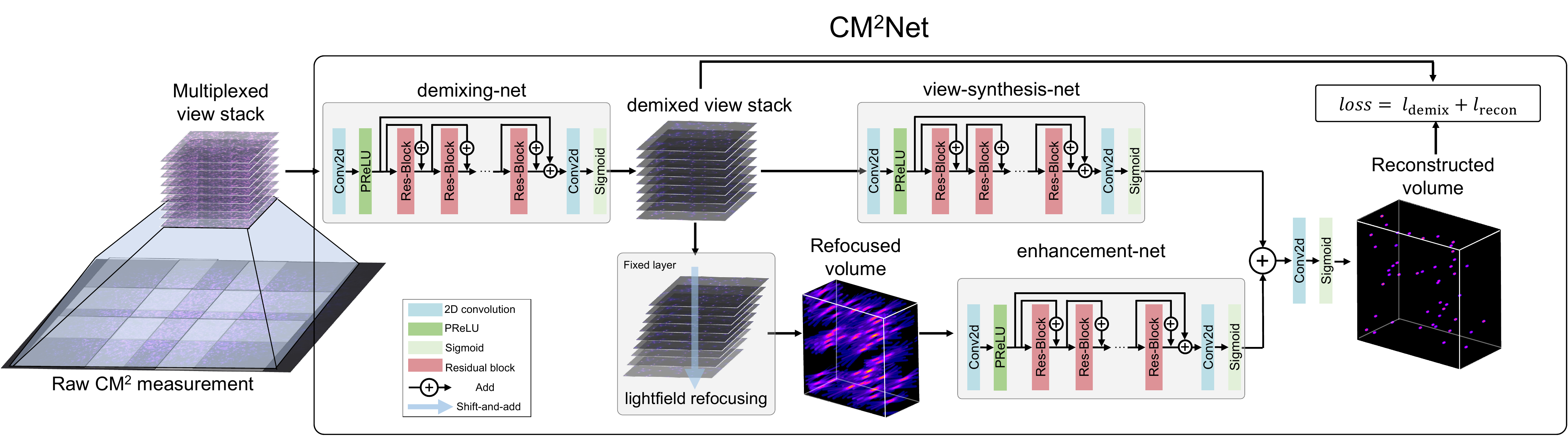}
\caption{CM$^2$Net structure. The raw CM$^2$ measurement is first preprocessed to form a multiplexed view stack. The demixing-net removes the crosstalk artifact and outputs the demixed view stack by learning view-dependent aberrations. The demixed view stack is processed by the “shift-and-add” lightfield refocusing algorithm to form a geometrically refocused volume. The enhancement-net branch removes the refocusing artifacts and enhances the reconstructed 3D resolution. The view-synthesis-net branch directly processes the demixed views to perform the 3D reconstruction. The sum of the output from the two branches is further processed to form the final reconstruction. CM$^2$Net is trained with a mixed loss function combining the demixing and reconstruction losses.}
\label{fig4}
\end{figure}

To enable fast and accurate 3D reconstruction from a CM$^2$ measurement, we implement a modular deep learning model, CM$^2$Net, to incorporate the key feature of the CM$^2$ physical model. 
Each CM$^2$ image contains 3 $\times$ 3 multiplexed views to capture  projection information about the 3D object\cite{xueSingleshot3DWidefield2020}. 
This multi-view geometry introduces two challenges to the network design. First, the image features needed for 3D reconstruction are non-local, instead they are separated by a few thousands of pixels. To fully capture the non-local information requires a sufficiently large receptive field, which is not easily achieved by a standard convolutional neural network. 
Second, the view-multiplexing requires the network not only to reconstruct 3D information, but also to remove crosstalk artifacts. 
To address these challenges, CM$^2$Net combines three modules to break the highly ill-posed inverse problem into three simpler tasks, including view-demixing, view-synthesis, and lightfield-refocusing enhancement, as illustrated in Fig. \ref{fig4}. 

The first module, view ``demixing-net'', demultiplexes a CM$^2$ image into nine “demixed” views, each corresponding to the image captured by a single microlens without crosstalks from other microlenses. 
To perform this task, demixing-net synthesizes the information contained in the entire CM$^2$ measurement. 
To facilitate this process, we first construct a “view stack” by cropping and view-aligning nine patches from the raw measurement based on the chief ray of each microlens (Fig. \ref{fig4}). 
This input view stack contains multiplexed information, which demixing-net seeks to demultiplex. The ground-truth output is the demixed view stack containing nine crosstalk-free images, which is made possible on simulated training data using our 3D-LSV model. 
Our results show that this task can be accurately performed by utilizing the distinctive aberration features from different microlenses. In Supplement 1, we further perform ablation study on demixing-net and highlight that it significantly reduces the false positives in the reconstructions.

The demixed view stack is akin to a 3 $\times$ 3 view lightfield measurement, which is processed by two reconstruction branches. 
The first branch is “view-synthesis-net”, which directly performs the 3D reconstruction based on disparity information in the views, as inspired by deep-learning enhanced LFM\cite{wangRealtimeVolumetricReconstruction2021a}. 
The second branch explicitly incorporates the geometrical optics model of lightfield. 
The demixed views are first processed by the lightfield refocusing algorithm\cite{ng2005light} to generate a refocused volume, and then is fed into “enhancement-net” to remove artifacts and enhance the reconstructed resolution. 
The refocused volume already provided most of the 3D object information but suffers from three artifacts, including severe axial elongation due to limited angular coverage, boundary artifacts from the “shift-and-add” operation, and missing object features at the peripheral FOV regions due to inexact view matching between the 3 $\times$ 3 MLA. 
To achieve the best performance, the outputs from the two branches are summed and further processed to yield the final 3D reconstruction. 
To highlight the effectiveness of this design, we conduct ablation studies and visualize the respective activation maps of the two branches in Supplement 1. Our results show that the lightfield refocusing enhancement-net achieves high-quality reconstruction at the central FOV region, and the view-synthesis-net improves the performance at the peripheral FOV regions. Together, the two reconstruction branches utilize complementary information to achieve high-resolution reconstruction across a wide FOV.

Overall, CM$^2$Net is trained entirely on simulated data from our 3D-LSV model. 
The loss function combines a demixing loss and a reconstruction loss: $\mathrm{loss}= \alpha_1 l_\mathrm{demix}+\alpha_2 l_\mathrm{rec}$, which promotes the fidelity of the demixed views and the 3D reconstruction results, respectively.
For both loss components, we use the binary cross entropy (BCE) since it promotes sparse reconstructions\cite{li2018deep}, which is defined by
$\mathrm{BCE}(y,\hat{y})=\sum_{i}y_{i}\log(\hat{y}_i)+(1-y_{i})\log(1-\hat{y}_i)$, and the summation is over all the voxels indexed by $i$,  and $y$ and $\hat{y}$ denote the ground-truth and reconstructed intensity, respectively. The weights of the two loss functions $(\alpha_1,\alpha_2)$ are set to be $(1, 1)$ after performing hyperparameter tuning, which concluded that the demixing and reconstruction losses have equal importance.

CM$^2$Net is implemented in Python 3.7 with TensorFlow 2.3. The multiple sub-networks are trained together in an “end-to-end” fashion on an Nvidia P100 GPU (16 GB) with a batch size of 2. We use Adam optimizer with an adaptive learning rate schedule. The initial learning rate is 10$^{-4}$ and automatically decreases by a factor of 0.9 after the loss on a small validation set ($\sim$400 patches) plateaus for 2 consecutive epochs. The training takes $\sim$48 hours to complete. Additional implementation details are provided in Supplement 1.

\subsection{Synthetic training data generation}
\label{sec2.4}
We generate a large-scale training dataset for CM$^2$Net based on the 3D-LSV model (Eq. \eqref{eq2}) from a set of synthetic volumes. 
The FOV of each synthetic volume follows a uniform random distribution between 6.5 mm and 7.5 mm, $U$[6.5 mm, 7.5 mm]. 
The degree of view-multiplexing is determined by the FOV\cite{xueSingleshot3DWidefield2020}.
For FOV < 2.7 mm, no view-multiplexing is present. 
As the FOV increases, the overlap between neighboring views increases approximately quadratically. 
At the largest FOV = 7.5 mm,  $\sim$64\% of overlap is present. 
The depth range is fixed at 800 $\mu$m. 
The volumes are sampled at 4.15 $\mu$m laterally (matching the effective pixel size of CM$^2$ V2) and 10 $\mu$m axially (10$\times$ higher sampling than the physical scanning step size).

We randomly place spherical emitters into the volumes by the following steps. Due to the large sampling grid size, we first generate each ground-truth emitter on a 5$\times$ finer grid (0.83 $\mu$m $\times$ 0.83 $\mu$m $\times$ 2 $\mu$m). Next, we perform 5 $\times$ 5 $\times$ 5 average binning to make the ground-truth volume having the same grid size as the final reconstruction. 
The emitter's diameter follows a uniform random distribution $U$[8 $\mu$m, 20 $\mu$m], which  approximately matches the typical size of neuronal cell bodies. 
The emitter’s intensity is set by the surface area, i.e. proportional to the diameter squared, which matches with our experimental measurements.
The size range used in our data leads to a 6.25$\times$ intensity variation range. 
To further vary the emitter's intensity at a given size, a random scaling factor following a uniform distribution $U$[0.8, 1.2] is added, which approximately matches with the contrast from one-photon fluorescence microscopes on Calcium indicators\cite{songNeuralAnatomyOptical2021a}. 
The emitter density in each volume follows a uniform random distribution $U$[10, 100] (number of emitters / mm$^2$), which simulates different fluorescence labeling densities used in cortex-wide neuronal imaging applications~\cite{kauvarCorticalObservationSynchronous2020a,fanVideorateImagingBiological2019a}. 

We first generate noise-free measurements using the 3D-LSV model. We then add realistic levels of mixed Gaussian and Poisson noise. The parameters for the additive Gaussian noise (normalized mean = 0.048, standard deviation = 0.017) are estimated by multiple dark measurements taken with the same acquisition parameters as the real experiments (30 ms exposure time, 40 dB gain). The Poisson noise is added by estimating the expected photon budget ($\sim$500 peak number of photons, and a unit effective gain) in typical widefield one-photon imaging\cite{kauvarCorticalObservationSynchronous2020a}. To train the view demixing-net, we generate the ground-truth non-overlapping views using the same 3D-LSV model with the single-microlens PSF. 

After synthesizing the measurements, we crop the overlapped views (1920 $\times$ 1920 pixels) based on the chief ray location of each microlens at the in-focus image plane. Next, we stack the 9 cropped views to form a 1920 $\times$ 1920 $\times$ 9 multi-channel input to CM$^2$Net.
Finally, CM$^2$Net  is trained on 9700 uniformly cropped patches (320 $\times$ 320 pixels)  from 270 synthetic objects. 

\section{Results}
\subsection{3D-LSV simulator enables accurate 3D reconstruction across a wide FOV}

\begin{figure}[!t]
\centering\includegraphics[width=\linewidth]{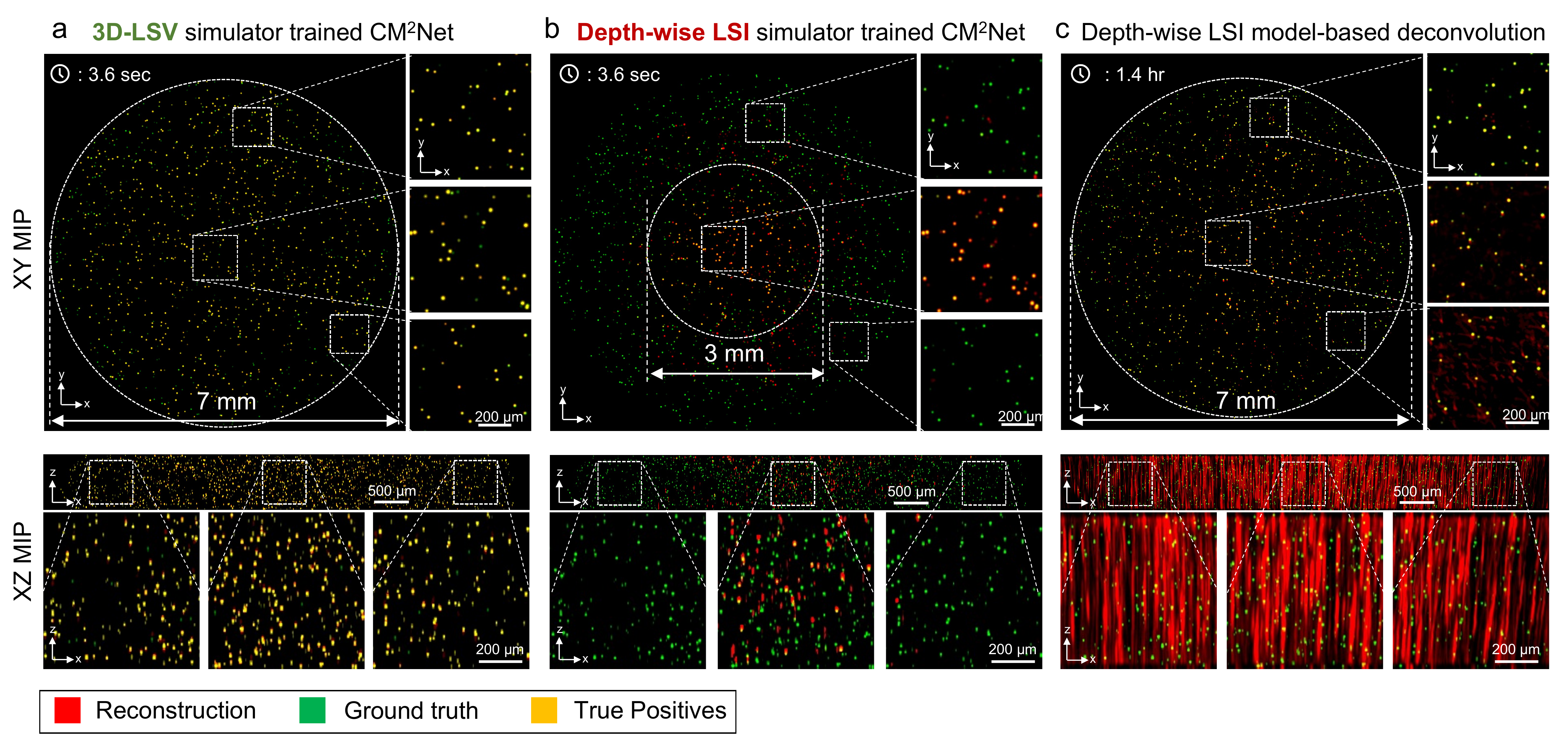}
\caption{The 3D-LSV simulator enables wide-FOV, high-resolution 3D reconstruction. Reconstructions from (a) LSV-CM$^2$Net trained with the 3D-LSV simulator, (b) LSI-CM$^2$Net trained with the depth-wise LSI simulator, and (c) depth-wise LSI model-based deconvolution. LSV-CM$^2$Net provides accurate and high 3D resolution reconstruction across the entire volume. LSI-CM$^2$Net suffers from many false negatives. The model-based deconvolution suffers from severe axial elongations. In all figures, the positive $z$-axis points at the direction towards the CM$^2$ system.}
\label{fig5}
\end{figure}

To demonstrate our 3D-LSV model is essential to achieve accurate 3D reconstruction across a wide FOV, we compare two CM$^2$Net trained with two different forward models. 
The first network, termed LSV-CM$^2$Net, is trained by our 3D-LSV model. 
The second network, termed LSI-CM$^2$Net, is trained by our previous depth-wise LSI model\cite{xueSingleshot3DWidefield2020}, which assumes the on-axis PSF is invariant at each depth.
We also benchmark the network reconstructions against the depth-wise LSI model-based deconvolution algorithm\cite{xueSingleshot3DWidefield2020}.

The 3D reconstructions on a cylindrical volume ($\sim$7-mm diameter, 0.8-mm depth) from LSV-CM$^2$Net, LSI-CM$^2$Net, and model-based deconvolution are shown in Fig. \ref{fig5}a, \ref{fig5}b, and \ref{fig5}c, respectively. 
In each figure, we overlay the reconstruction (in red) onto the ground truth (in green) and visualize the XY and XZ maximum intensity projections (MIP). 
When the reconstruction matches with the ground truth (i.e. True Positives), the overlayed region appears in yellow. When the reconstruction misses certain particles (i.e. False Negatives), the region appears in green. When the reconstruction creates false particles (i.e. False Positives) or suffers from axial elongations, the region appears in red. 
By visual inspection, LSV-CM$^2$Net can accurately reconstruct the entire 7-mm FOV throughout the 0.8-mm depth range. as highlighted by the three zoom-in regions of the XZ MIPs taken from the central and two peripheral regions. 
In contrast, the LSI-CM$^2$Net suffers from severe artifacts especially beyond the central 3-mm diameter region.
The model-based reconstruction matches well with the ground truth across the entire volume, but suffers from severe axial elongations\cite{xueSingleshot3DWidefield2020}.

A major improvement of CM$^2$Net over the model-based deconvolution is the significantly reduced axial elongation, as also shown in our experiments in Section \ref{sec3.3}. In addition, CM$^2$Net  dramatically reduces the reconstruction time and memory burden. To perform the large-scale reconstruction in Fig. \ref{fig5} ($\sim$230 million voxels), the model-based method requires $\sim$1.4 hours and $\sim$150-GB RAM. In contrast, CM$^2$Net takes only $\sim$3.6 seconds on an entry-level GPU (Nvidia RTX 2070, 8GB RAM), which is $\sim$1400$\times$ speed-up and $\sim$19$\times$ memory cost reduction.

To demonstrate the potential applications of CM$^2$Net to reconstruct complex brain structures, we perform simulation studies on imaging 3D neuronal populations and mouse brain vessels in Supplement 1.
Our results on neuronal imaging show that CM$^2$Net can achieve high reconstruction performance on both sparsely (20 neurons / mm$^2$) labeled and densely (100 neurons / mm$^2$) labeled neuronal populations across a cortex-wide (7.5 $\times$ 6.6 mm) FOV and is robust to the complex brain geometry.
Our result on a mouse blood vessel network highlights a few key properties of the particle-dataset trained CM$^2$Net. First, the view-demixing module can perform  reliable demultiplexing on axially overlapping small vessels. The demixing result highly matches with the ground truth, demonstrating that the demixing network is robust to overlapping views from continuous objects, even though it is trained entirely on sparse fluorescent beads. Second, the lightfield refocused volume on demixed views can correctly resolve complex 3D geometry, which lays the foundation for the final 3D reconstruction. Third, the reconstruction module can correctly reconstruct the vessel network, albeit with discontinuity artifacts. We attribute the artifacts to the sparsity constraint implicitly enforced by the particle-dataset trained CM$^2$Net on the 3D reconstruction. Overall, CM$^2$Net can provide high-quality reconstruction on brain vasculature across a wide (6 $\times$ 4 mm) FOV and can resolve the complex 3D geometry. 

\subsection{Quantitative analysis of CM\textsuperscript{2}Net performance}
\label{sec3.2}

We quantitatively show that the trained CM$^2$Net can provide high-quality reconstruction and is robust to variations in the emitter's lateral location (FOV), seeding density, depth, size and intensity in simulation.
To perform the evaluation, we simulate a testing set consisting of 180 volumes that uniformly falls in nine density ranges [10:10:100] (number of emitters/mm$^2$). The data synthesis procedure follows the procedure in Section \ref{sec2.4}.

We quantify the detection capability of CM$^2$Net using recall, precision, F1 score, and Jaccard index.
Recall measures the sensitivity / detection rate by the ratio between the correctly reconstructed and the actual total number of emitters. Precision measures the specificity by the ratio between the correctly reconstructed and the total reconstructed number of emitters. 
F1-score and Jaccard index combine these two complementary metrics. 
In addition, we quantify the 3D localization accuracy by lateral and axial root mean squared localization error (RMSE)~\cite{sage2019super}.
A global threshold needed to binarize the reconstructed volume when computing the metrics is set by maximizing the F1-score on the testing set\cite{bao2021segmentation}. More details on the quantitative metrics are provided in Supplement 1.
We compute the statistics of each metric at a given condition (e.g. a lateral location), when all other parameters (e.g. emitter’s depth, density, size) are randomized.

\begin{figure}[!tbh]
\centering\includegraphics[width=\linewidth]{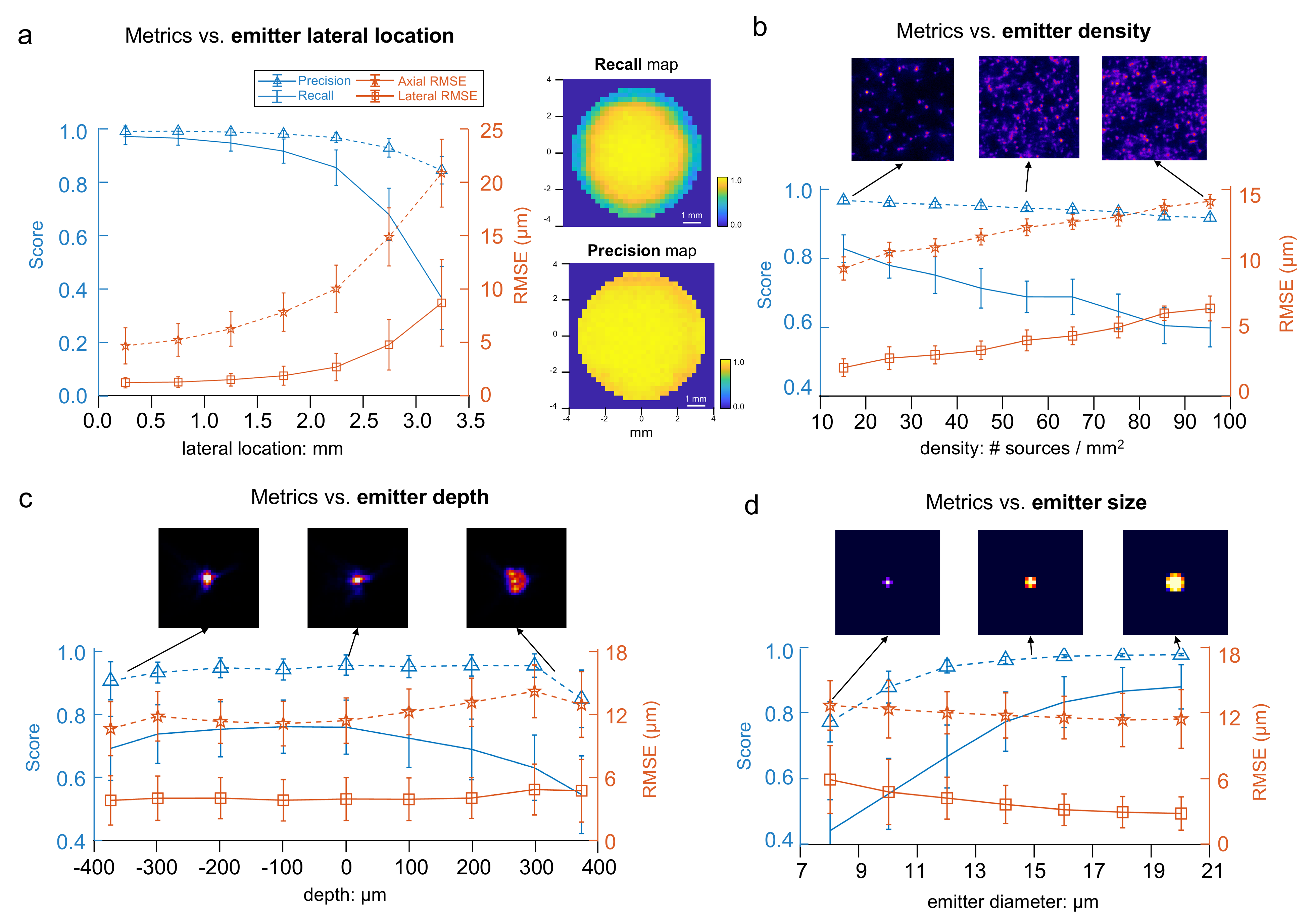}
\caption{Quantitative evaluation of CM$^2$Net in simulation. Detection performance quantified by recall and precision (blue) and localization accuracy by lateral and axial RMSE (orange) when varying the emitter's (a) lateral location, (b) seeding density, (c) depth, and (d) size. 
(a) CM$^2$Net achieves precision >$\sim$0.85 for lateral location <3.5 mm (FOVs < 7 mm). Recall drops from 0.97 at the central FOV to 0.35 near the edge. Lateral / axial RMSE increases from 1.24 $\mu$m / 4.7 $\mu$m at the central FOV to 8.7 $\mu$m / 21 $\mu$m near the edge. The recall and precision maps (250 $\mu$m patch size) show nearly isotropic, high detection performance across the central 6-mm FOV. 
(b) Precision is >0.92 for all emitter densities. Recall decreases from $\sim$0.83 at the lowest density to $\sim$0.61 at the highest density. Lateral / axial RMSE  degrades linearly from 2 $\mu$m / 9.3 $\mu$m at 10 emitters/mm$^2$ to 6.4 $\mu$m / 14 $\mu$m at 100 emitters/mm$^2$. Three example image patches are shown to visualize the density variations. 
(c) Precision is >0.85 throughout the depth, and recall is >0.7 within [-400 $\mu$m, 200 $\mu$m] depth range and drops to $\sim$0.54 at 400 $\mu$m.  Lateral / axial RMSE is <5 $\mu$m / 14 $\mu$m for all depths. The foci from the central microlens at -400 $\mu$m, 0 $\mu$m, and 400 $\mu$m are shown in the top panel to visualize the depth-dependent aberrations. 
(d) Precision is >0.9 and recall  is >0.71 for emitter’s diameter >11 $\mu$m. Lateral / axial RMSE   decreases almost linearly from 5.9 $\mu$m / 12.7 $\mu$m for 8-$\mu$m emitters to 2.8 $\mu$m / 11.5 $\mu$m for 20-$\mu$m emitters. Both the detection rate and localization accuracy degrade for smaller emitters since the local SNR and emitter’s intensity scales with the diameter squared. The top panel shows example reconstructed emitters of diameter [8, 14, 20] $\mu$m.}
\label{fig6}
\end{figure}

First, the performance at different lateral locations are evaluated in Fig.~\ref{fig6}a.
We aggregate the emitters into 7 bins [0 mm: 0.5 mm: 3.5 mm] (distance from the center) and compute the statistics.
The averaged precision and recall (blue) remain >0.93 and >0.68 when the distance is <3 mm (i.e. FOV < 6 mm). 
Precision and recall reduce to $\sim$0.85 and $\sim$0.37, respectively, when the  distance is $\sim$3.5 mm (FOV = 7 mm). 
Lateral / axial RMSE (orange) is less than 5 $\mu$m / 15 $\mu$m within the 6-mm FOV and degrade to 8.7 $\mu$m / 21 $\mu$m at the edge. 
The standard deviation (the error bar) increases with the distance, indicating that the reconstruction is more consistent at the central FOV. 
To better visualize the detection performance, we calculate recall and precision maps in Fig. \ref{fig6}a (details in Supplement 1). The precision map shows that CM$^2$Net provides nearly isotropic, high specificity within the 7-mm FOV. The recall map shows that CM$^2$Net provides a high detection rate in the central 6-mm FOV, and degrades at the outer regions.

To understand the origin of the degradation in the peripheral FOV, we perform ablation studies by feeding the CM$^2$Net’s reconstruction module with the ground-truth demixed views (see Supplement 1). 
The result shows consistently high recall (>0.89) for the entire 7-mm FOV, showing the robustness of the reconstruction module. This implies that the degraded recall is due to imperfect view-demixing at outer FOV regions. 
To further diagnose the system, we compare the intensity distribution of the point source for PSF calibration and the recall map, and find qualitative correspondence. 
We hypothesize that the training of view-demixing net is impacted by the rapid intensity fall-off ($\sim$85\% drop at the 7-mm FOV edge) of the imperfect point source.

We evaluate the metrics for different emitter densities at [10: 10: 100] (emitters / mm$^2$) in Fig. \ref{fig6}b.
As expected, both precision and recall decrease, whereas both lateral and axial RMSEs increase with the density. Precision remains >0.92 for all emitter densities, indicating very few false positives in the reconstruction despite the large (10$\times$ span) density variations.  Recall decreases approximately linearly from $\sim$0.83 at 10 emitters/mm$^2$ to $\sim$0.61 at 100 emitters/mm$^2$.  Lateral / axial RMSE increases approximately linearly from 2 $\mu$m / 9.3 $\mu$m at the lowest density to 6.4 $\mu$m / 14 $\mu$m at the highest density. This means that as the density increases, CM$^2$Net suffers from more false negatives and lower localization accuracy.

We evaluate the metrics for different emitter’s depths in 9 bins: [-400 $\mu$m, -350 $\mu$m), [-350 $\mu$m: 100 $\mu$m: 350 $\mu$m),  [350 $\mu$m, 400 $\mu$m] in Fig. \ref{fig6}c. The smaller bin size in the first and last bins are due to the limited depth range used in the study.
Precision is consistently >0.85 for the entire range. 
Recall is >0.7 within [-400 $\mu$m, 200 $\mu$m], and gradually decreases to $\sim$0.54 at 400 $\mu$m.  
To explain the decrease of recall at these large defocus depths, we visualize the on-axis PSF and show that it degrades more severely as the source moves closer to the MLA. This results in lower SNRs in the measurement, which leads to more false negatives. 
Lateral / axial RMSE is <~5$\mu$m / 14 $\mu$m for all depths, and degrades only slightly at large defocus. The slight drops in the first and last bins are attributed to the smaller sample size (in combination of the smaller bin size and fewer true positives) that introduces errors in the statistics. We observe that the minimum RMSE is centered around the 100-µm bin, which suggests that there may be a slight defocus between our nominal and the actual focal plane. 
Overall, the RMSE analysis shows that 3D localization is generally robust to defocus within the 800-µm depth range.

Finally, we quantify the metrics for different emitter diameters at  [7 $\mu$m: 2 $\mu$m: 21 $\mu$m]. For diameters ranging 11-20 $\mu$m, precision is >0.9 and recall is >0.71. As the diameter decreases, both precision and recall drop approximately linearly to 0.55 and 0.48 for 8-$\mu$m emitters, respectively. Lateral / axial RMSE decreases from 5.9 $\mu$m / 12.7 $\mu$m for 8-$\mu$m emitters to 2.8 $\mu$m / 11.5 $\mu$m for 20-$\mu$m emitters.
We attribute the worse performance for smaller emitters to two factors. First, since the emitter’s intensity is proportional to the size squared, the SNR rapidly decreases as the size reduces. 
Second, due to the coarse sampling in the reconstruction, the number of voxels for each emitter is <5 when diameter is <11~$\mu$m (see the top panel of Fig. \ref{fig6}d).

The averaged precision and recall for the entire testing set is $\sim$0.7, $\sim$0.94 respectively, comparable to the state-of-the-art deep learning neuron detection algorithm\cite{bao2021segmentation}.
The averaged lateral and axial RMSEs are 4.17 $\mu$m and 11.2 $\mu$m respectively, close to the reconstruction grid size (lateral 4.15 $\mu$m and axial 10 $\mu$m), which indicates that the localization accuracy is close to one voxel.
This study establishes that CM$^2$Net can detect emitters with few “hallucinated” sources ($\sim$4\% false positive rate on average) and high detection rates ($\sim$30\% false-negative rate on average) with good localization accuracy in a broad range of conditions.

\subsection{CM\textsuperscript{2}Net achieves high 3D resolution, wide-FOV reconstruction in experiments}
\label{sec3.3}

We demonstrate that the generalization capability of simulator-trained CM$^2$Net enables high 3D resolution reconstruction in experiments with high detection performance.

We first image a cylindrical volume embedded with 10-$\mu$m green-fluorescent beads (details about the sample preparation and experimental setup are provided in Supplement 1). 
The phantom is estimated to have 10-20 emitters / mm$^2$.
First, to remove the non-uniform background and match the intensity statistics with the simulation data, we preprocess the raw experimental measurements with histogram matching. Next, we manually cropped 9 views to input to CM$^2$Net for 3D reconstruction (more details in Supplement 1).

\begin{figure}[!t]
\centering\includegraphics[width=.78\linewidth]{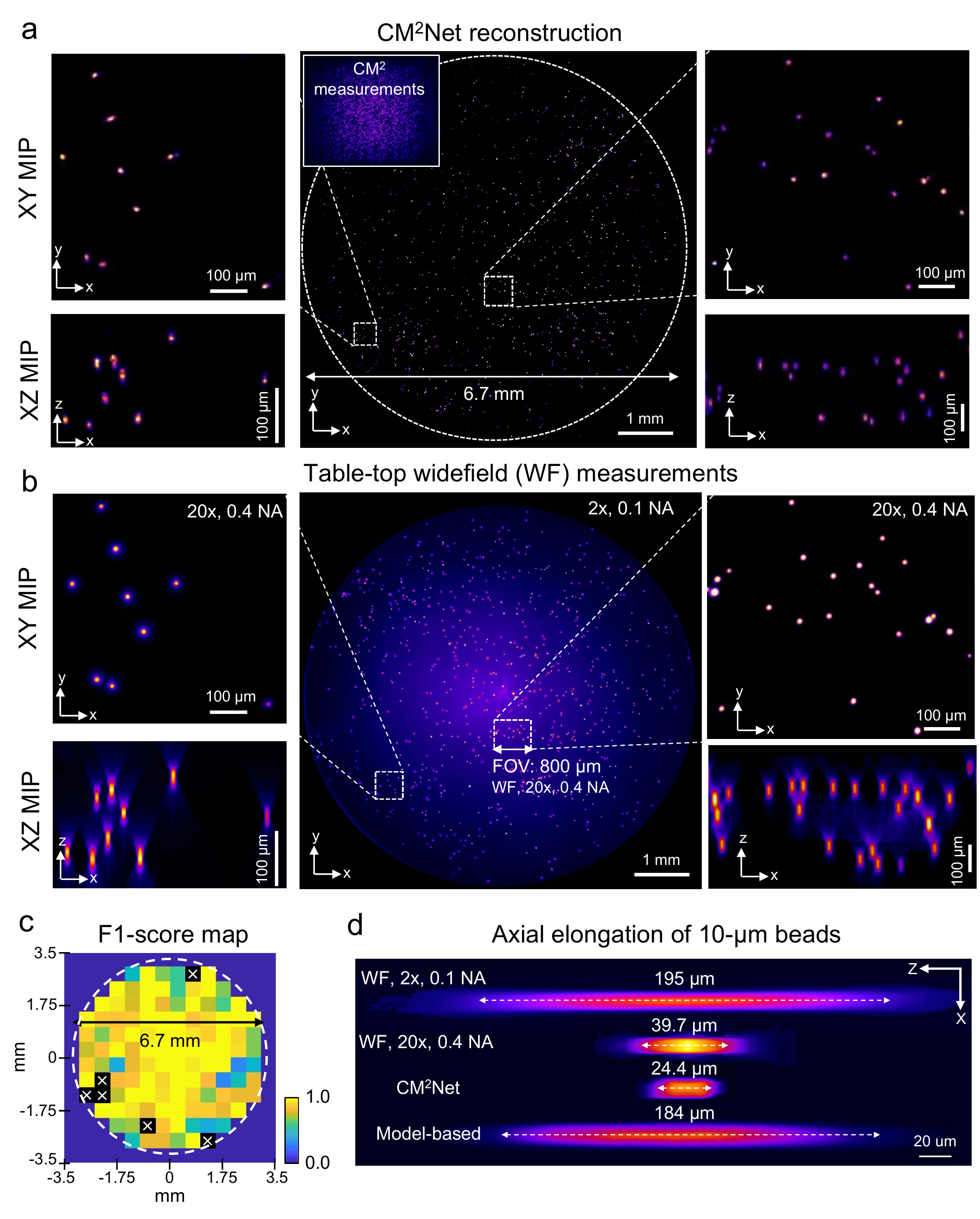}
\caption{CM$^2$Net achieves high 3D resolution reconstruction across a wide FOV in experiment. (a) Visualization of CM$^2$Net reconstruction. The measurement shown as the inset (sample: 10-$\mu$m fluorescent beads in a cylindrical volume with $\sim$6.7-mm diameter and $\sim$0.5-mm depth). 
(b) Validation using 2$\times$, 0.1 NA and 20$\times$, 0.4 NA objective lenses on a widefield (WF) microscope. CM$^2$Net provide high-quality reconstruction across the 6.7-mm FOV as validated by the 2$\times$ measurement. 
Both lateral and axial reconstructions are in good agreement with the high-resolution 20$\times$ measurement. 
(c) F1-score map computed by comparing the XY MIPs of CM$^2$Net reconstruction and WF 2$\times$ measurement (500~$\mu$m patch size). 'x' marks F1-score~=~0, resulting from either the WF measurement or CM$^2$Net reconstruction is empty. (d) 
The axial elongations are 195 $\mu$m, 184 $\mu$m, 39.7 $\mu$m, and 24.4~$\mu$m for WF 2$\times$, model-based deconvolution, WF 20$\times$, and CM$^2$Net reconstruction, respectively.}
\label{fig7}
\end{figure}

The CM$^2$Net reconstruction is shown in Fig.~\ref{fig7}a and is validated against widefield measurements from a standard table-top epi-fluorescence microscope in Fig.~\ref{fig7}b. 
First, we validate the full-FOV reconstruction by comparing the XY MIPs of the reconstruction and the widefield $z$-stack from a 2$\times$, 0.1 NA objective. 
To further assess the reconstruction at greater resolution, two zoom-in XY MIPs of the reconstruction from the central and edge of the FOV are compared with the high-resolution $z$-stack from a 20$\times$, 0.4 NA objective. 
By visual inspection, the reconstruction matches well to the widefield measurements. 
The reconstruction quality maintains at peripheral FOV regions, a marked improvement over our previous model-based reconstruction\cite{xueSingleshot3DWidefield2020}. 

A major goal we aim to achieve using CM$^2$ is 3D high-resolution imaging across a wide FOV. 
To highlight this capability, we compare the FOV achieved by CM$^2$ V2 ($\sim$7 mm) with the 2$\times$ objective ($\sim$8 mm) and 20$\times$ objective ($\sim$800 $\mu$m) (marked in Figs. \ref{fig7}a and \ref{fig7}b.)
Representative axial profiles of the 10-$\mu$m beads reconstructed by CM$^2$Net, model-based deconvolution, and 2$\times$ and 20$\times$ widefield measurements are compared in Fig. \ref{fig7}d. 
CM$^2$Net achieves an axial elongation of $\sim$24 $\mu$m, which is $\sim$8$\times$ better than the model-based deconvolution ($\sim$184 $\mu$m) and outperforms 20$\times$, 0.4 NA measurement ($\sim$39.7 $\mu$m).

To quantify the detection performance, we compute recall, precision, and F1-score by comparing the XY MIPs of the CM$^2$Net reconstruction and widefield 2$\times$ measurement. CM$^2$Net achieves recall $\sim$0.78 and precision $\sim$0.80. In comparison, the recall and precision in simulation at the corresponding density is $\sim$0.83 and $\sim$0.97, respectively. 
The simulator-trained CM$^2$Net degrades slightly on experiment with $\sim$5\% higher false-negative rate and $\sim$17\% higher false-positive rate. 
We attribute the reduced performance to the undesired extra views in the experimental measurements (see the analysis in Supplement 1).

To quantify spatially variations in performance, we construct the recall and precision maps in Supplement 1. Recall in most regions are >0.75, indicating <25\% false-negative rates. Precision is >0.8 except for a few patches with <2 beads, indicating only a few false positives in the reconstruction. 
In Fig. \ref{fig7}c, we show the F1-score map generally achieves a high value of >0.75. 
An overlay between the full-FOV reconstruction and the widefield 2$\times$ measurement is shown in Supplement 1 to provide further visual inspections.

This experiment shows that CM$^2$Net provides high 3D resolution reconstruction across a wide FOV with high sensitivity and precision. 
The 24-$\mu$m axial elongation achieved by CM$^2$Net is $\sim$4$\times$ better than the 100-$\mu$m axial spacing in the PSF calibration. 
This shows that the axial interpolation in our 3D-LSV model is  effective to achieve axial “super resolution” in real experiments. Both the recall and precision agree with the simulation, validating that the 3D-LSV simulator-trained CM$^2$Net can generalize well to experimental measurement.

\subsection{Experimental demonstration on mixed size fluorescent beads}

\begin{figure}[!t]
\centering\includegraphics[width=.8\linewidth]{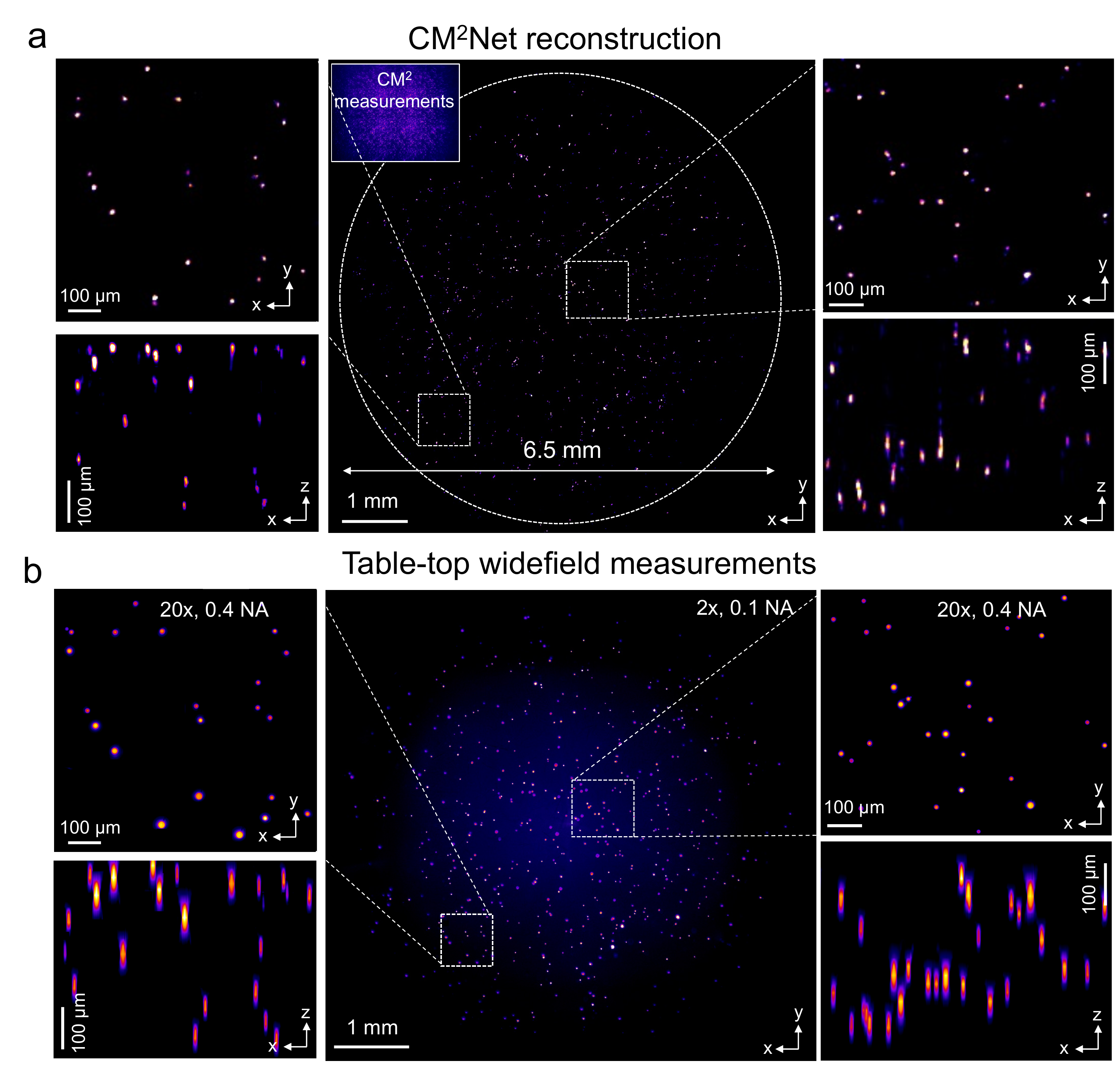}
\caption{Experiment on mixed fluorescent beads. (a) Visualization of the CM$^2$Net reconstruction. The CM$^2$ measurement shown as the inset. (b) Validation measurements from widefield 2$\times$, 0.1 NA and 20$\times$, 0.4 NA objectives (sample: mixed 10-$\mu$m and 15-$\mu$m fluorescent beads in a cylindrical volume with $\sim$6.5-mm diameter and $\sim$0.8-mm depth). 
The CM$^2$Net full-FOV reconstruction is in good agreement with the 2$\times$ measurement. The lateral and axial reconstructions are validated by the high-resolution 20$\times$ measurement in both central and peripheral FOV regions.}
\label{fig8}
\end{figure}

Fluorescent emitters with different sizes and brightness result in different local contrast and SNRs in the CM$^2$ measurement\cite{xueSingleshot3DWidefield2020}. This is an important consideration as we develop CM$^2$ towards realistic biological applications. To demonstrate this capability, we conduct proof-of-concept experiments on mixed size fluorescent beads. 
Our result shows that CM$^2$Net can robustly handle such sample variations in real experiments.

We image a cylindrical volume (diameter $\sim$6.5 mm, depth $\sim$0.8 mm) embedded with mixed 10-$\mu$m and 15-$\mu$m  beads (details in Supplement 1). The phantom is estimated to have 10-20 emitters / mm$^2$. 
In the CM$^2$  measurements, the 15-$\mu$m beads are $\sim$2.2$\times$ brighter than 10-$\mu$m beads, matching with their surface area ratio and our synthetic data model in Sec.~\ref{sec2.4}. 
The CM$^2$Net reconstruction is shown in Fig. \ref{fig8}a. 
The 3D reconstruction is validated against widefield measurements in Fig. \ref{fig8}b. 
First, we assess the full-FOV reconstruction by comparing the XY MIPs of the CM$^2$Net reconstruction and the 2$\times$ $z$-stack measurement. 
We further compare two zoom-in regions from the center and corner FOVs with the high-resolution 20$\times$ $z$-stack measurement. 
By visual inspection, CM$^2$Net reliably reconstructs both 10-$\mu$m and 15-$\mu$m beads. 
The XZ MIPs of the CM$^2$Net 3D reconstruction are in good agreement with the 20$\times$ $z$-stack measurements. 
The axial confinement on both 10-$\mu$m and 15-$\mu$m beads by CM$^2$Net are better than the 20$\times$ measurements.

To quantitatively assess the CM$^2$Net reconstruction, we compute recall, precision, and F1-score maps in Supplement 1 by comparing the XY MIPs from the CM$^2$Net reconstruction and widefield 2$\times$ measurement. 
CM$^2$Net achieves averaged recall $\sim$0.73 and precision $\sim$0.84 across the 6.5-mm FOV. 
As compared to the mono-10-$\mu$m bead experiment, we attribute the slightly decreased recall to the greater intensity and SNR variations. We attribute the increased precision to the reduced FOV and less contamination from the extra views (see analysis in Supplement 1). 
An overlay between the full-FOV reconstruction and the 2$\times$ measurement is shown in Supplement 1 to provide further visual inspections. The results show that CM$^2$Net is robust to the emitter size and intensity variations in the experiment.

This experiment again highlights the wide-FOV and high-resolution 3D imaging capability of CM$^2$ V2. Our training data containing randomized emitter sizes and intensities are effective to make CM$^2$Net robust to experimental variations. As a result, CM$^2$Net can provide high-quality 3D reconstruction with good sensitivity and precision on mixed size emitters that have large differences in the feature size and local SNR.

\section{Conclusion}
In summary, we have presented a new Computational Miniature Mesoscope (CM$^2$) system, which is a deep learning-augmented miniaturized microscope for single-shot 3D high-resolution fluorescence imaging. The system reconstructs emitters across a $\sim$7-mm FOV and 800-$\mu$m depth with high sensitivity and precision, and achieves $\sim$6-$\mu$m lateral and $\sim$25-$\mu$m axial resolution.

The main hardware advancement in CM$^2$ V2 includes a novel 3D-printed freeform illuminator that increases the excitation efficiency by $\sim$3$\times$. Each 3D-printed LED collimator can provide up-to 80\% light efficiency yet weighs only 0.03 grams. It is low-cost and rapidly fabricated on a table-top 3D printer. In addition, we adapted a hybrid emission filter design that suppresses the excitation leakage and improves the measurement SBR by more than 5$\times$.

The computational advancement includes three main parts. First, we developed an accurate and computationally efficient 3D-LSV forward model that characterizes the spatially varying PSFs across the large (cm$^2$ $\times$ mm-scale) imaging volume supported by the CM$^2$.  Second, we developed a multi-module CM$^2$Net that achieves robust, high-resolution 3D reconstruction from a single-shot CM$^2$ measurement. Third, using the 3D-LSV simulator to generate the entire training dataset, CM$^2$Net provides high detection sensitivity and precision and good localization accuracy on fluorescent emitters across a wide FOV and generalize well to experiments.
In addition, our numerical studies show that  CM$^2$Net can achieve high reconstruction performance on both neuronal populations and vascular structures across a cortex-wide FOV and is robust to the complex mouse brain geometry.

Our demonstration on the utility of freeform optics fabricated by 3D printing may be a fruitful area for future research, especially for miniature microscopes and other miniature optical devices. In recent years, freeform optics has emerged as the ideal solution to bypass many limitations in conventional optics, such as compactness and imaging performance\cite{rollandFreeformOpticsImaging2021a}. At the same time, non-conventional optics has been enabled by novel 3D printing process, such as  micro-\cite{yannyMiniscope3DOptimizedSingleshot2020a,hongThreedimensionalPrintingGlass2021a}, diffractive\cite{orange-kedem3DPrintableDiffractive2021a}, and volume optics\cite{orange-kedem3DPrintableDiffractive2021a}. We envision that novel 3D-printed freeform optics can be incorporated in future CM$^2$ platforms to enhance its imaging capability.

The CM$^2$ V2 platform is built on a back-side illuminated (BSI) CMOS sensor, which significantly improves the measurement’s SNR and dynamic range over the conventional CMOS sensor in the V1 platform. The size and weight of the CM$^2$ V2 prototype is limited by the availability of miniature BSI CMOS sensor. However, we do not anticipate this to be a major roadblock for future development as encouraged by the recent development of MiniFAST\cite{juneauMiniFASTSensitiveFast2020a} BSI CMOS-based miniscope. With further advancement on high-speed data transmission and high pixel-count BSI CMOS sensor platform, we expect CM$^2$ can be further miniaturized to be suitable for “wearable” in-vivo neural recordings on mice and other small animals.

Our 3D-LSV model is essential to achieve high 3D resolution reconstruction across a large imaging volume. A notable result we have demonstrated is that the axial resolution is not limited by the axial step used for the 3D PSF calibration. This allowed us to bypass the large data requirement in the alternative depth-wise LSV framework\cite{yannyMiniscope3DOptimizedSingleshot2020a,kuoOnchipFluorescenceMicroscopy2020a,yannyDeepLearningFast2022a} and to perform data-efficient PSF calibrations across a centimeter-scale FOV and millimeter-scale depth range. We expect the same sparse 3D PSF calibration, low-rank decomposition, and 3D interpolation procedure are applicable to other computational 3D microscopy techniques, such as LFM and lensless imaging.
In addition, it may be possible to develop hybrid 3D PSF calibration procedures by combining physical measurements and numerical modeling to further improve the model accuracy, as recently shown in high-resolution LFM imaging\cite{hua2021high,wu2021iterative}.

Our CM$^2$Net incorporates both the view-multiplexing and lightfield information in the CM$^2$ image formation. We have shown that the view-demixing module significantly suppresses the false positives in the 3D reconstruction. The simulator-training scheme was essential to enable the training of the view-demixing sub-network. This highlights several key advantages of simulator-based training over experiment-based training schemes. It not only forgoes the laborious physical data collection process, but also enables access to novel data pairs that are impractical to collect experimentally. The reconstruction module combining the lightfield-refocusing enhancement and view-synthesis branches is shown to learn complementary information from the demixed views and enable highly accurate 3D reconstructions, and is readily applicable to other LFM modalities.

Our numerical study on reconstructing brain vasculature indicates that the emitter-dataset trained CM$^2$Net implicitly enforces a sparsity constraint to the 3D reconstruction that produces discontinuity artifacts. We trained  CM$^2$Net  on individual emitters since our targeted application is to image  neurons labeled with genetically encoded calcium indicators in mouse brains\cite{zhang2021fast}. To better adapt CM$^2$Net to other complex structures such as blood vessels, one can perform transfer learning on a dataset tuned to the specific application. Conveniently, Our 3D-LSV simulator is directly applicable to generate non-emitter data, as shown in our study.

An outstanding challenge to expand the utility of the CM$^2$ is tissue scattering\cite{xueSingleshot3DWidefield2020}. There are several promising solutions we envision that are applicable to CM$^2$, such as miniature structured illumination technique\cite{supekarMiniatureStructuredIllumination2021a} and scattering-incorporated 3D reconstruction frameworks\cite{zhangComputationalOpticalSectioning2021a,tahirAdaptive3DDescattering2022a}, which will be investigated in our future work.

\bigskip

\begin{backmatter}

\bmsection{Funding}
National Institutes of Health (NIH) (R21EY030016, R01NS126596).

\bmsection{Acknowledgments}
The authors thank Boston University Shared Computing Cluster for proving the computational resources, and Boston University Photonics Center for providing the access to Zemax.

\bmsection{Disclosures}
The authors declare no conflicts of interest.

\bmsection{Data Availability Statement}
The CM$^2$ V2 hardware, the CM$^2$Net implementation, and the pretrained model are available at: \href{https://github.com/bu-cisl/Computational-Miniature-Mesoscope-CM2}{https://github.com/bu-cisl/Computational-Miniature-Mesoscope-CM2}.

\bmsection{Supplemental document}
See Supplement 1 for supporting content.

\end{backmatter}

\bibliography{reference}
\end{document}


\maketitle

\section{CM$^2$ V2 Hardware Design and Prototype}
\subsection{Additional Details on CM$^2$ V2 Hardware Design}

The CM$^2$ V2 platform consists of two main parts, including the imaging module and the illumination module. The overall design is visualized in Figs 1a and 2a in the main text. Built from 3D-printed components and off-the-shelf optics and electronics, the CM$^2$  V2 platform has an overall dimension of 36 mm $\times$ 36 mm $\times$ 15 mm including the backside-illuminated (BSI) CMOS PCB board (IDS Imaging, monochrome BSI CMOS IMX178LLJ, 2076 $\times$ 3088 pixels, 2.4-$\mu$m pixel size, 12-bit dynamic range, 58 fps). The entire assembled CM$^2$ V2 prototype weighs ~11.5 grams, of which the CMOS PCB takes most of the size (36 mm $\times$ 36 mm $\times$ 4 mm) and the weight (9 grams) whereas the custom parts weigh only 2.5 grams in-total. 

\begin{figure}[h]
\centering
\includegraphics[width=\linewidth]{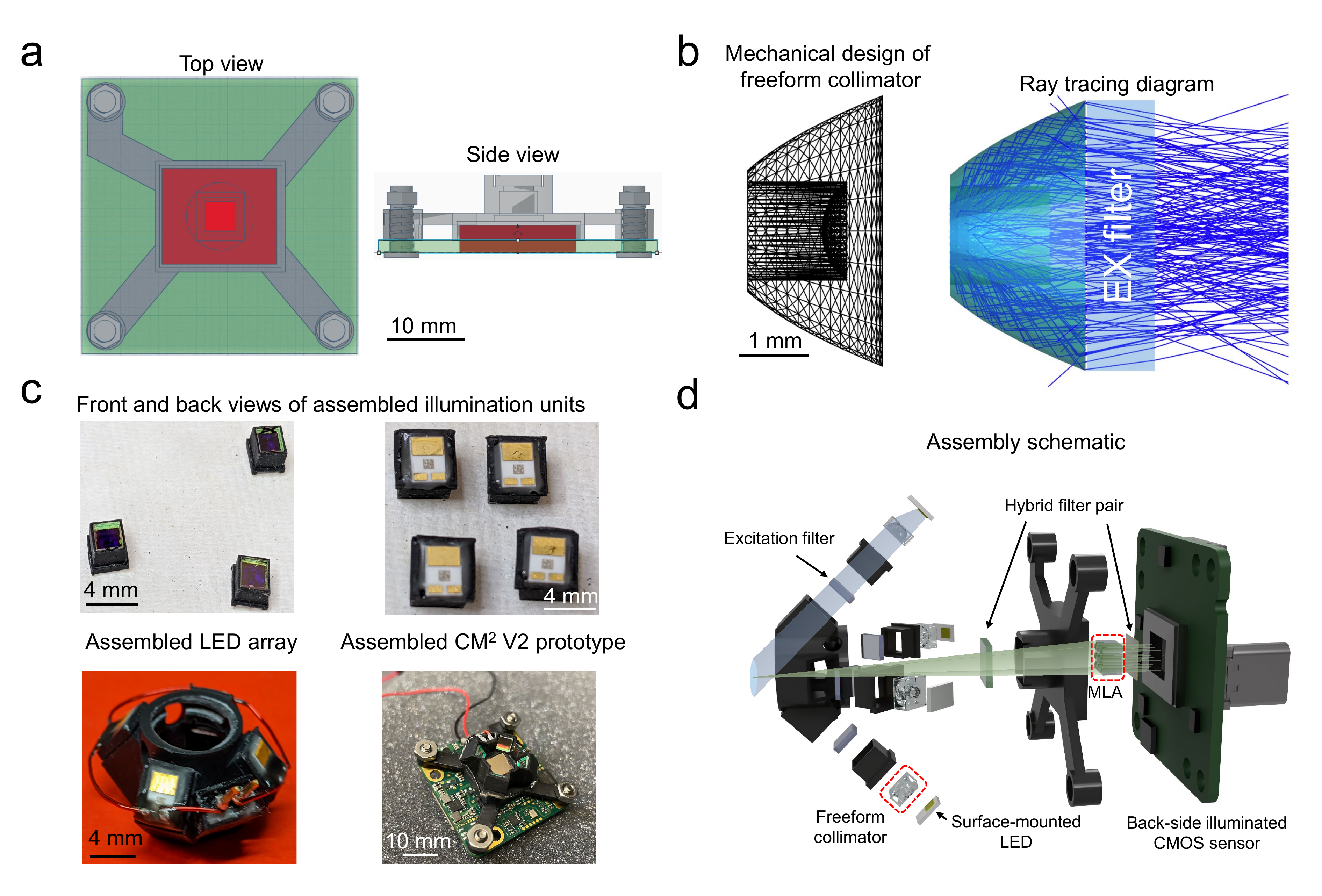}
\caption{\textbf{Additional details on the CM$^2$ V2 design and the prototype.} (a) Top and side views of the 3D printed housing (in gray) that holds the optics (not drawn) and the sensor (in red and green). (b) Mechanical design of the freeform collimator and the ray tracing diagram with the interference-excitation filter from Zemax. (c) Photos of the assembled illumination units, the 4-LED array, and the CM$^2$ V2 prototype. Each illumination unit hosts a surface mounted LED source, the 3D printed freeform illuminator, and the excitation filter. (d) The complete CM$^2$ V2 assembly schematic. }
\label{S1}
\end{figure}

 In the imaging module, we use an off-the-shelf rectangular MLA with a 100\% fill factor (Fresnel Technologies Inc., no. 630, focal length = 3.3 mm, lens pitch = 1 mm, thickness = 3.3 mm). The MLA is directly placed on top of the CMOS sensor to achieve single-shot 3D imaging with a compact form factor. The MLA is diced into a 3 $\times$ 3 array using a high-precision automatic dicing saw (Disco Dad, no.3220) to keep the sides of the MLA clear in order to reduce vignetting and edge effect.
  Fig. \ref{S1}a shows the top and side views of the 3D printed MLA housing and how it is mounted onto the CMOS PCB board (IDS Imaging, board-level IMX178). The red region indicates the sensor region of the CMOS. The green regions are the PCB that host extra electronics. 
  
 The hybrid emission filter set consists of two filters: a glass interference filter (Chroma Technology, no. 535/50, 1 mm thickness) and a thin-film absorption filter (Edmund Optics, Wratten color filter no. 12, 0.1 mm thickness). We place the interference filter in front of the MLA and the thin-film absorption filter behind the MLA. This design reduces the maximum angle of incidence onto the interference filter to achieve better filtering whereas the remaining leakage is suppressed by the absorption filter.  All the optical elements and the CMOS sensor are held by a 3D-printed housing (designed in TinkerCAD, printed on Formlabs Form 2, black resin, no. RS-F2-GPBK-04). The assembled imaging module is mounted to the CMOS PCB by four mini set screws and hex nuts (Thorlabs, no. HW-KIT3). The assembled CM$^2$ V2 prototype has a calibrated de-magnification of ~1.7$\times$ which results in an effective pixel size of 4.15 $\mu$m. The CM$^2$ V2 prototype does not require precise alignment of the optics. The field varying PSFs are experimentally calibrated and later computationally analyzed by our 3D-LSV model.

The illumination module consists of four freeform LED-illumination units held together by a 3D-printed dome-shaped base plate, which also blocks the ambient light. Each freeform LED-illumination unit contains the following parts: a surface mounted LED (Lumileds, LXML-PB01-0040), a 3D-printed freeform collimator (printed on Formlabs Form 2, clear resin, no. RS-F2-GPCL-04), a glass excitation filter (Chroma Technology, no. 470/40, 4 mm), and a 3D-printed LED housing (printed on Formlabs Form 2, black resin, no. RS-F2-GPBK-04). To achieve better surface quality, we post-process the 3D-printed collimator by the following procedure. A thin layer of clear resin (diluted 5 times with isopropyl alcohol) is applied to the outer parabolic surface of the collimator and cured under a UV lamp. The four illumination units are wired sequentially and connected to an LED driver (LED-dynamics Inc., 3021-D-E-350, 350 mA). 
In Fig. \ref{S1}b, we provide the exact mechanical design and the ray tracing diagram of the freeform illuminator. The ray tracing in Zemax has incorporated the LED source spectrum and the incidence-dependent transmittance profiles of the interference-excitation filter from the manufacturer (Chroma Technology). The ray tracing result shows the freeform illuminator efficiently refracts and reflects light from the LED source and allow them to pass through the interference coating. The small divergence of outcoming light is because the LED source has an emitting area of 1 mm$^2$.

The four LED units are placed symmetrically around the imaging path with a lateral offset of ~6.7 mm from the optical axis and tilted by ~45 degrees. The positions and orientations of the four illumination units are modeled and optimized in Zemax to achieve the maximum light delivery efficiency and overall excitation uniformity at the imaging plane. 
The assembled illuminator units (including 3D printed housing, LED source, freeform collimator, and the excitation filter) are shown in Fig. \ref{S1}c Top. The four units are installed into the LED base plate. Fig. \ref{S1}c Bottom two photos show the wired illumination array and fully assembled CM$^2$ V2 prototype, respectively. For better illustration, the CM$^2$ V2 assembly schematic is provided in Fig. \ref{S1}d.

After the CM$^2$ V2 is fully assembled, the excitation is validated on a green fluorescence calibration slide (Thorlabs, no. FSK2). The total excitation power across the designed 8-mm region is up-to 75 mW (at maximum driving current of 350 mA), measured by a power meter (Thorlabs no.PM121D). The illuminator is turned on at 300 mA continuously for 1 hour and no overheating issues were observed. 

\subsection{Spectral Properties and the Design of Hybrid Emission Filter Set}
\begin{figure}[htbp]
\centering
\includegraphics[width=.8\linewidth]{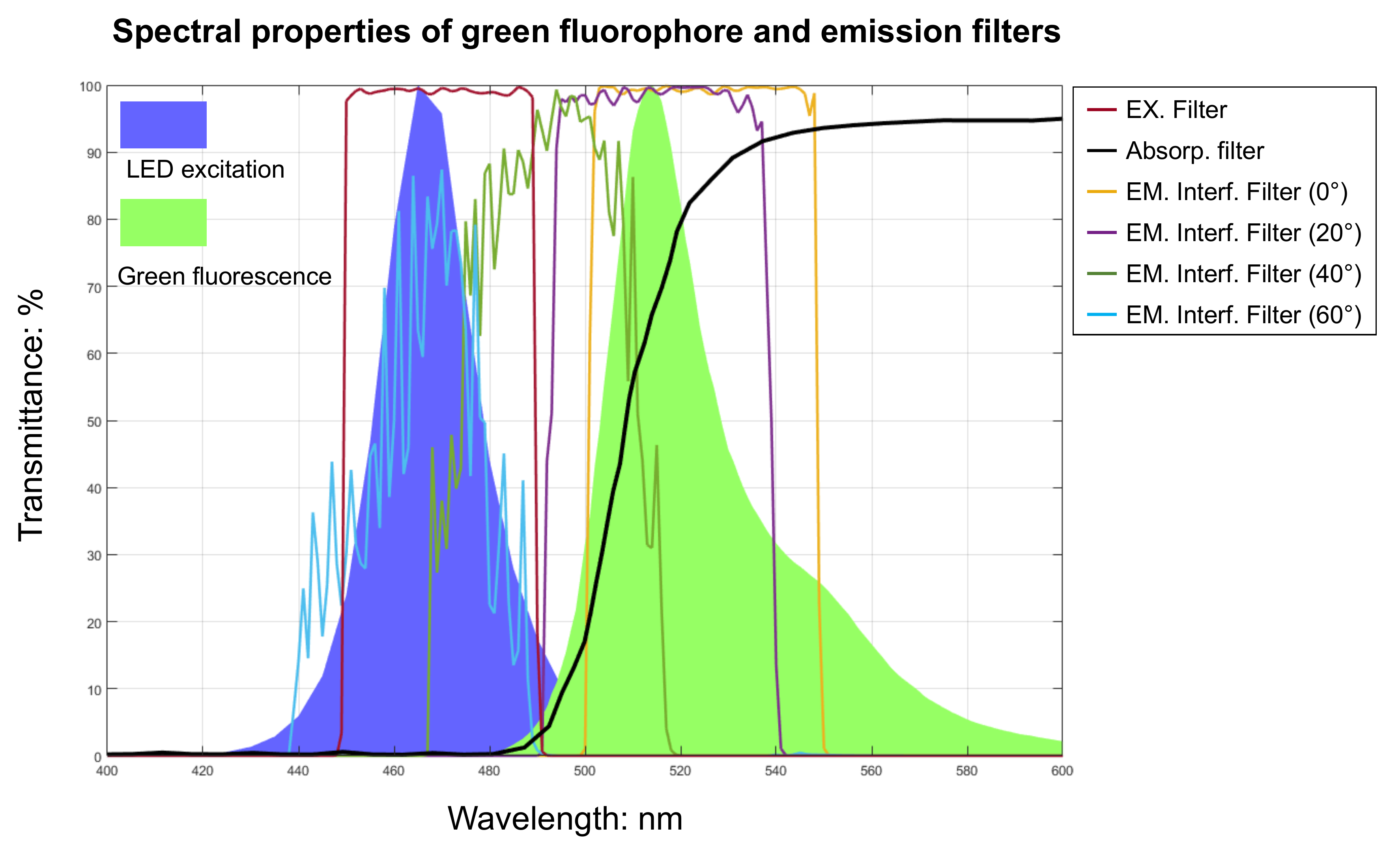}
\caption{\textbf{Spectral properties of hybrid filter design for green fluorescence.} The transmission profiles of the interference-emission filter at 0,20,40,60-degree incidence and the incidence-independent absorption filter.}
\label{S2}
\end{figure}
The non-normal incidence of uncollimated light in CM$^2$ V1 results in a substantial wavelength shift of the transmitted spectrum of the emission filter. In Fig. \ref{S2}a, we plot the spectral profiles of the LED source (blue solid region), green fluorophore emission (green solid region), excitation filter (red curve), emission filter at 0, 20, 40, 60-degree incidence (yellow, purple, green, and blue curve, respectively), and the absorption filter (black curve). The transmittance window of the interference-based emission filter shifts to shorter wavelength with a degraded profile (the oscillating curves) at non-normal incidence. It has a wide spectral overlay with the excitation spectrum, results in a background fluorescence in CM$^2$ V1 measurements. By adding an incidence-independent absorption filter, whose long-pass spectral profile is plotted in Fig. \ref{S2}a (black curve). Experimentally, we find that adding the absorption filter to the imaging path provides better suppression of background fluorescence at the cost of slightly reduced transmittance at emission wavelength. Fig. 2e in the main article has clearly shown that the hybrid emission filter provides much enhanced image contrast and reduced the background haziness by $\sim$5 times.

\subsection{CM$^2$ V2 Lateral Resolution Characterization}

\begin{figure}[!h]
\centering
\includegraphics[width=.8\linewidth]{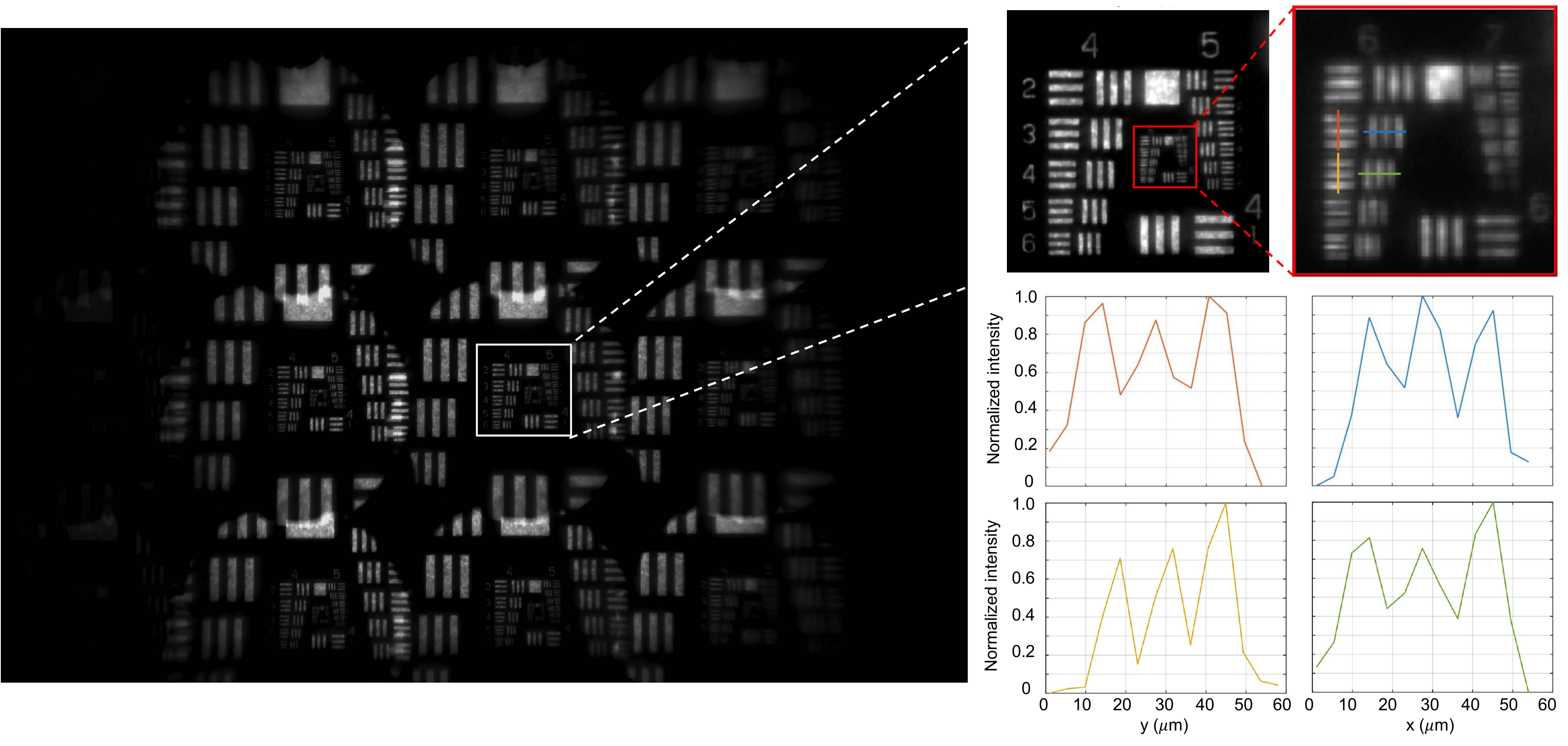}
\caption{\textbf{CM$^2$ V2 lateral resolution experimental characterization.} The experimental measurement on a 2D planar fluorescent resolution target. Zoom-in image show features at both Group 6 Element 3 (6.2-µm) and Group 6 Element 4 (5.52-µm) can be resolved. Plots show the corresponding horizontal and vertical line profiles of the selected elements on the resolution target.}
\label{S3}
\end{figure}

We experimentally characterize the lateral resolution of CM$^2$ V2 by imaging a fluorescent resolution target. Fig. \ref{S3} shows the captured raw measurement of the fluorescent resolution target. The zoom-in region shows that the features at both Group 6 Element 3 (6.2 µm) and Group 6 Element 4 (5.52 µm) can be resolved. For better illustration, we further plot the horizontal and vertical line profiles of the selected elements on the resolution target. Based on this measurement, we conclude that the lateral resolution for CM$^2$ V2 is $\sim$6 µm.

\subsection{CM$^2$ V2 Experiment setup }

\begin{figure}[!h]
\centering
\includegraphics[width=\linewidth]{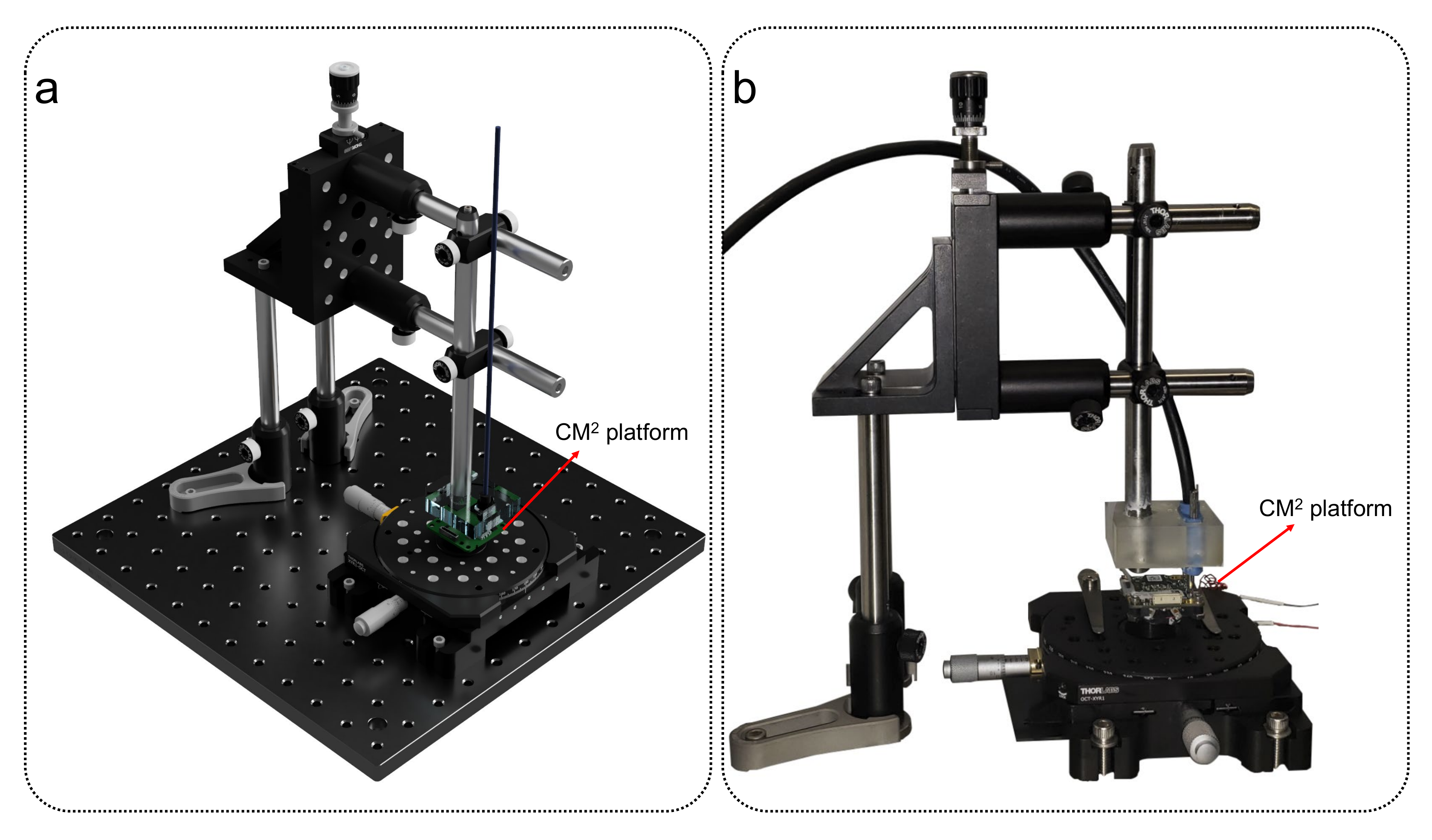}
\caption{\textbf{CM$^2$ V2 mounting stage.} (a) 3D model of the experimental setup. It consists of an XY translation stage with a rotating platform and a single-axis translation stage. A USB Type-C cable connects the sensor with the host computer for data transmission. The four LED units are powered by our custom constant current source, which can output a maximum current of 500 mA (not drawn). (b) Photograph of the mounting stage.}
\label{fig8}
\end{figure}

In order to acquire the experimental data, we mount the CM$^2$ V2 onto the setup shown in Fig.~\ref{fig8}. The position of the sample is adjusted by the XY translation stage, and the Z translation stage controls the axial focus of the CM$^2$ V2. To make sure the measurement is within the range of the calibrated PSFs, a stack of images are collected at different working distances, and the one image with the best reconstruction quality is chosen. The camera parameters are slightly sample-dependent and are generally set to be with 30 ms exposure time, 0 gain, 12- bit dynamic range, and 30 FPS to satisfy our targeted imaging requirements.

\clearpage
\section{CM$^2$ V2 Computational Pipeline}
\subsection{Sparse PSF calibration process}
The spatially varying PSFs of the assembled CM$^2$ V2 are experimentally calibrated. The calibration process collects the 3D-LSV PSFs at a set of 3D sparse locations within the targeted imaging volume. To perform the calibration, we first build a point source setup consisting of two parts: a point source and a 3-axis high-precision automatic scanning stage. The point source is built by first stacking multiple layers of thin diffusing films (Parafilm) on top of a bright green LED source (Thorlabs, M530L4), whose spectrum approximately matches with the green fluorescence. A 5-µm mounted pinhole (Thorlabs, P5D, stainless steel) is placed immediately after the diffusing films. The multi-layer diffusing film homogenizes the LED illumination before entering the pinhole. The resulting point source provides a large divergence angle (>60 degrees), i.e. a large illumination NA, which is approximated as a point source. The angular intensity distribution from this point source is characterized in Fig. \ref{S9}b. The point source is mounted on a 3-axis scanning stage (Thorlabs, MT3Z8) that is automatically controlled by a custom MATLAB script.

\begin{figure}[htbp]
\centering
\includegraphics[width=\linewidth]{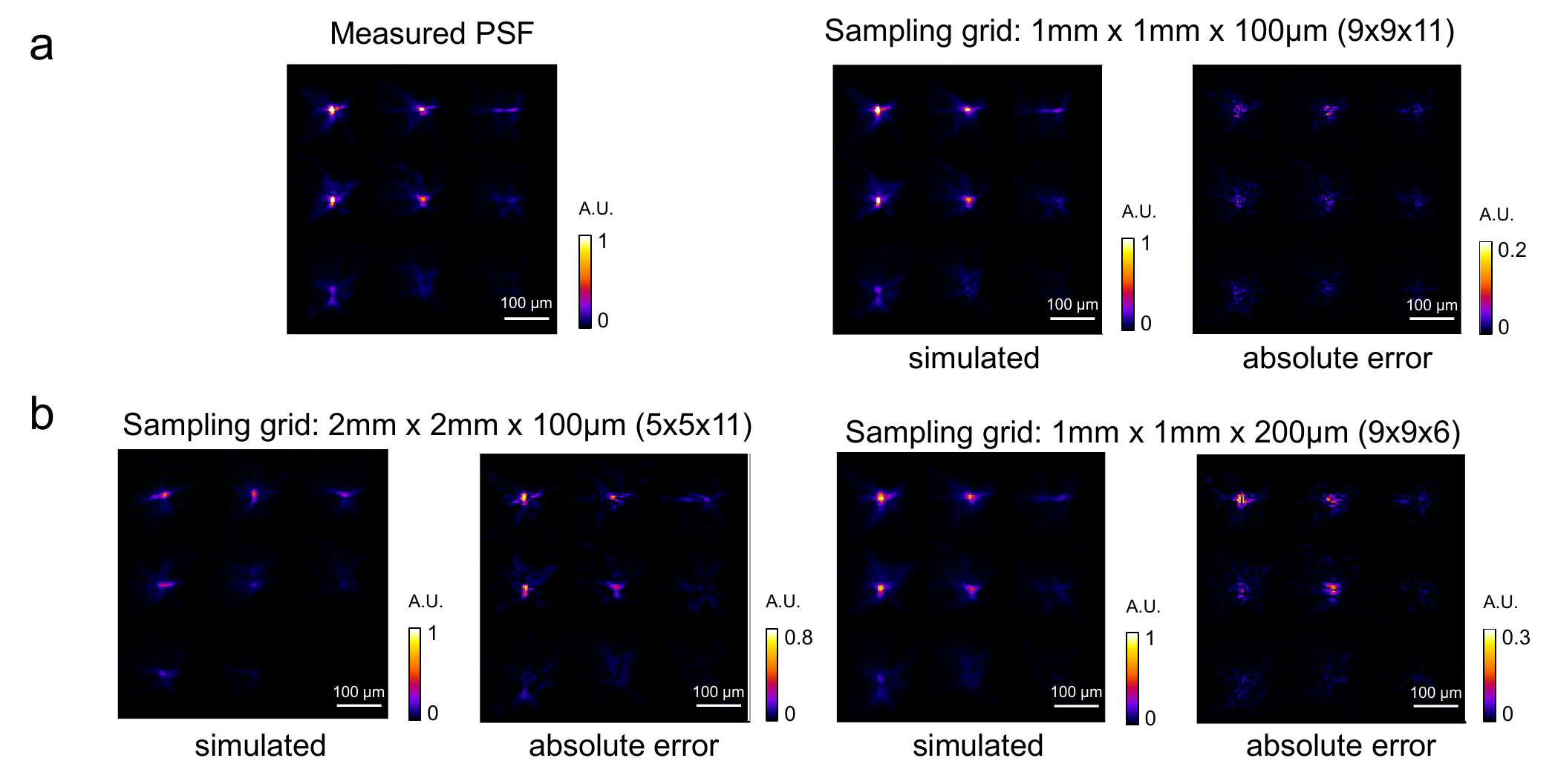}
\caption{\textbf{3D-LSV model reconstruction accuracy with different PSFs sampling grid.} 
(a) Left: Measured PSF at an unseen location. Right: Simulated PSF and the absolute error using the 3D-LSV model using the current sampling grid. (b) Simulated PSF and the absolute error with 2$\times$ down-sampled grid in the $xy$ plane (Left) and 2$\times$ down-sampled grid in the axial ($z$) direction (Right). The absolute error shows the reconstruction accuracy degrades when the sampling grid becomes sparse.}
\label{lsv_down}
\end{figure}
To calibrate the 3D LSV PSFs, we empirically choose a sparse sampling grid and scan the point source across the grid. The 3D grid spans [-4 mm to 4 mm] along both lateral dimensions and [-500 µm to 500 µm] along the axial dimension. The scanning step size is 1 mm laterally and 100 µm axially, respectively. In total, we collect PSFs on a 9 $\times$ 9 $\times$ 11 grid and 891 measurements. 

The sampling grid is essential for the 3D-LSV model to accurately characterize the system. The grid should be dense enough to capture the key variations present in the 3D PSF across the 3D volume while as sparse as possible to allow efficient physical calibration,  data storage, and computation. To demonstrate the effect of the sampling grid, we compare the reconstruction accuracy of the 3D-LSV model using the current PSF sampling grid and that from a down-sampled grid in Fig. \ref{lsv_down}. By comparing the reconstructed PSF at an ``unseen'' 3D location, the result shows that the reconstruction accuracy degrades when the sampling grid becomes sparse. Specifically, the reconstruction performance with laterally ($xy$) down-sampled PSFs suffers more severe loss of accuracy compared to that from axially ($z$) down-sampled PSFs. This is because the PSF variation is more prominent in the $xy$ plane due to the severe off-axis aberrations and PSF truncation at the peripheral FOV. 
We can further increase the 3D-LSV model accuracy at the cost of PSF calibration time and computational time and memory storage. 

After determining the PSF sampling grid, the MATLAB script controls both the stage scanning and provides the triggers for the CM$^2$ V2 image acquisition. To further account for the non-uniform angular profile of the point source, the MATLAB script adaptively adjusts the exposure times at different scanning locations. This ensures all the measured PSFs are not saturated or too dark. The exposure time is recorded to later normalize the PSF measurements before the 3D-LSV modeling. The whole calibration process takes about 2 hours to complete.

\subsection{Additional details on the 3D Linear Shift Variant (LSV) model}

\begin{figure}[h]
\centering
\includegraphics[width=\linewidth]{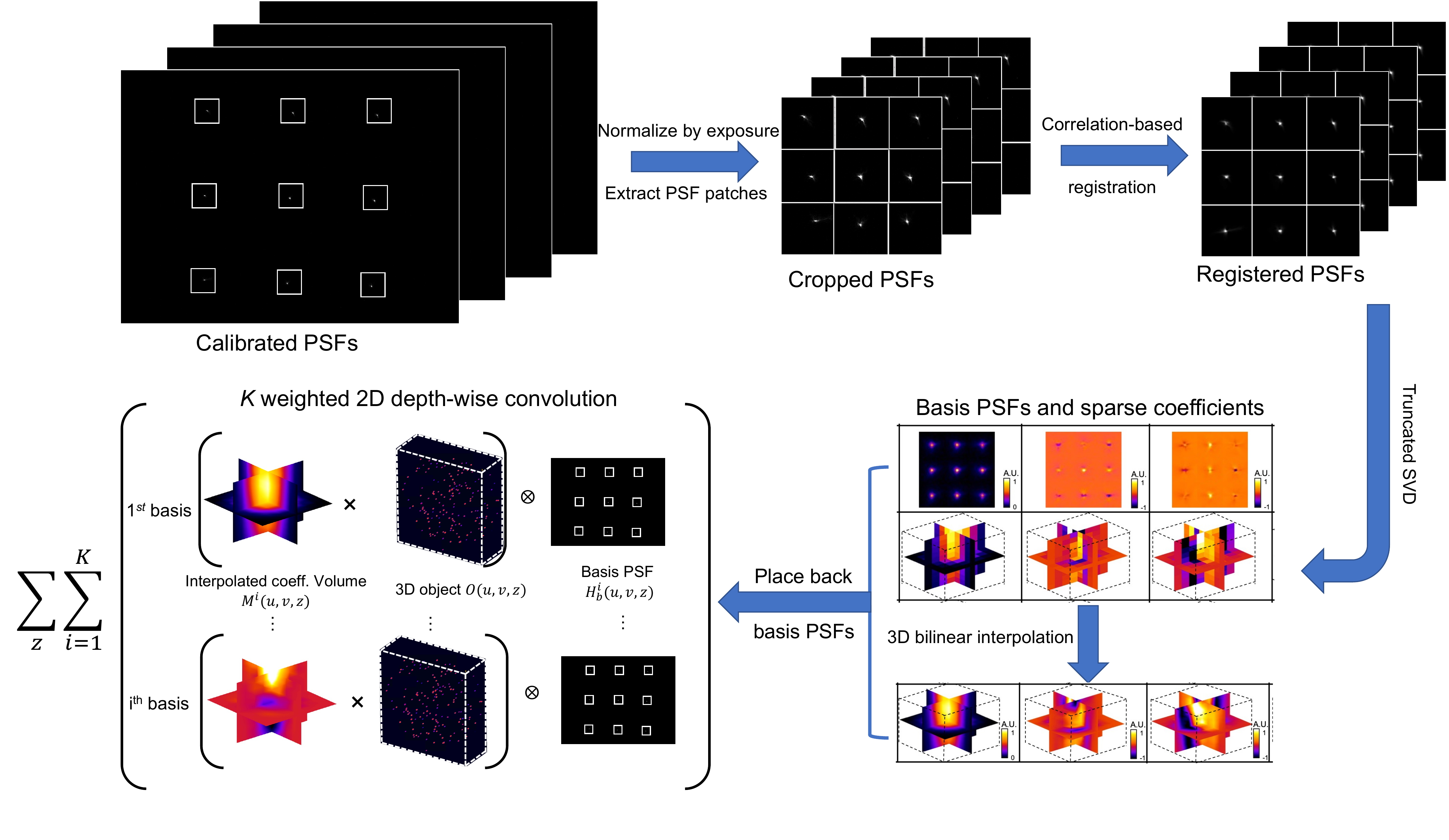}
\caption{\textbf{Procedure diagram of our 3D-LSV model.} Local small PSF patches are first cropped from the calibrated PSF measurements based on the chief ray locations. The cropped PSF patches are registered based on the estimated axial sheering and re-grouped into array PSFs. Then we perform TSVD on the registered array PSFs to obtain basis array PSFs and their coefficients at a sparse grid of calibrated locations. The coefficients are then 3D interpolated to the entire imaging volume with 3D bilinear interpolation. Next, the basis PSFs patches are placed back to their original locations. Lastly, the CM$^2$ measurement is computed by k weighted 2D convolutions between the object volume and the basis PSFs, followed by a summation along the axial dimension z.}
\label{S4}
\end{figure}

\begin{figure}[!h]
\centering
\includegraphics[width=\linewidth]{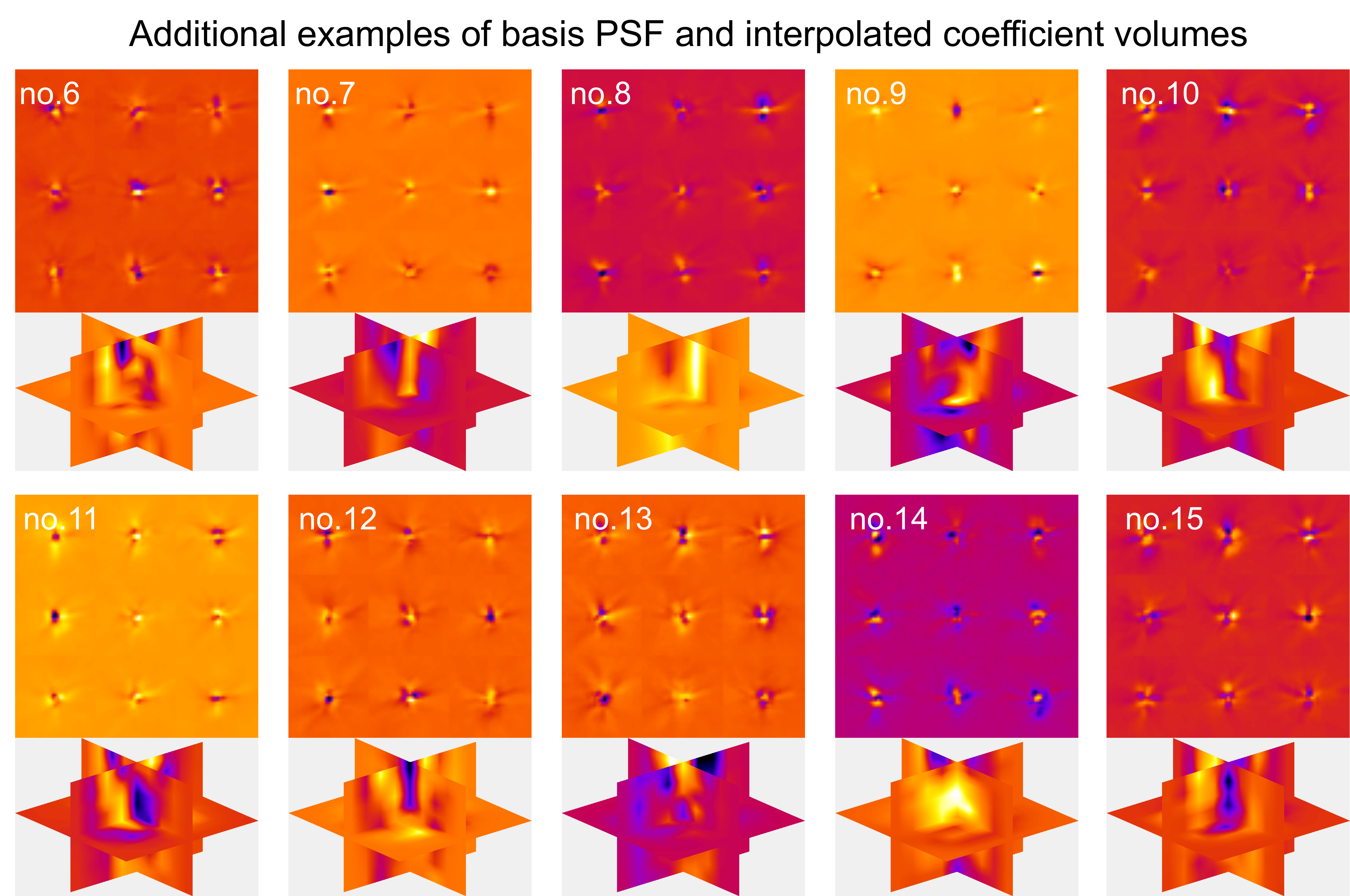}
\caption{\textbf{More examples of the decomposed basis PSFs and the interpolated coefficient volumes.}}
\label{S5}
\end{figure}

We provide additional details of the 3D-LSV model and experimentally validate the accuracy of the model. Fig. \ref{S4} shows a diagram of the 3D-LSV decomposition and interpolation process. The experimentally calibrated PSFs (2076 $\times$ 3088 pixels) are first normalized by the exposure time recorded in the calibration process. Next, small PSF patches (160 $\times$ 160 pixels) are extracted from the PSF measurements based on the chief ray locations. Due to the finite conjugate imaging geometry of the CM$^2$ system, the PSF at a different depth exhibits a different amount of lateral shift that is approximately linearly increases with the depth (i.e. axial sheering). To enable efficient PSF decomposition, we estimate the amount of axial sheering based on the on-axis PSF stack and then align the PSF patches with the in-focus PSF by numerically “undo” lateral shift. The 9 aligned PSF patches are regrouped into a 3 $\times$ 3 foci array (480 $\times$ 480 pixels). Next, the 891 calibrated array PSFs are decomposed using the singular value decomposition (SVD) and truncated to the leading 64 terms. There are two products from this TSVD process: the 64 basis PSFs H (480 $\times$ 480 pixels) and 64 coefficient volumes M (9 $\times$ 9 $\times$ 11 voxels). To match our reconstruction sampling, the coefficient volumes are further interpolated onto a dense 1920 $\times$ 1920 $\times$ 80 grid using the 3D bilinear interpolation method. To construct the full-sized (2076 $\times$ 3088 pixels) CM$^2$ measurement, the basis PSF patches are then put back to their original pixel locations. The locations are determined by the chief ray locations and the estimated axial sheering. Lastly, the measurement is computed by k weighted depth-wise 2D convolutions between the object volume and the basis PSFs, followed by a summation along the axial dimension z. In Fig. 3c of the main article, we visualize the first 5 basis PSFs and their coefficient volumes. Fig. \ref{S5} shows the next 10 basis PSFs and their interpolated coefficient volumes.

\begin{figure}[!h]
\centering
\includegraphics[width=\linewidth]{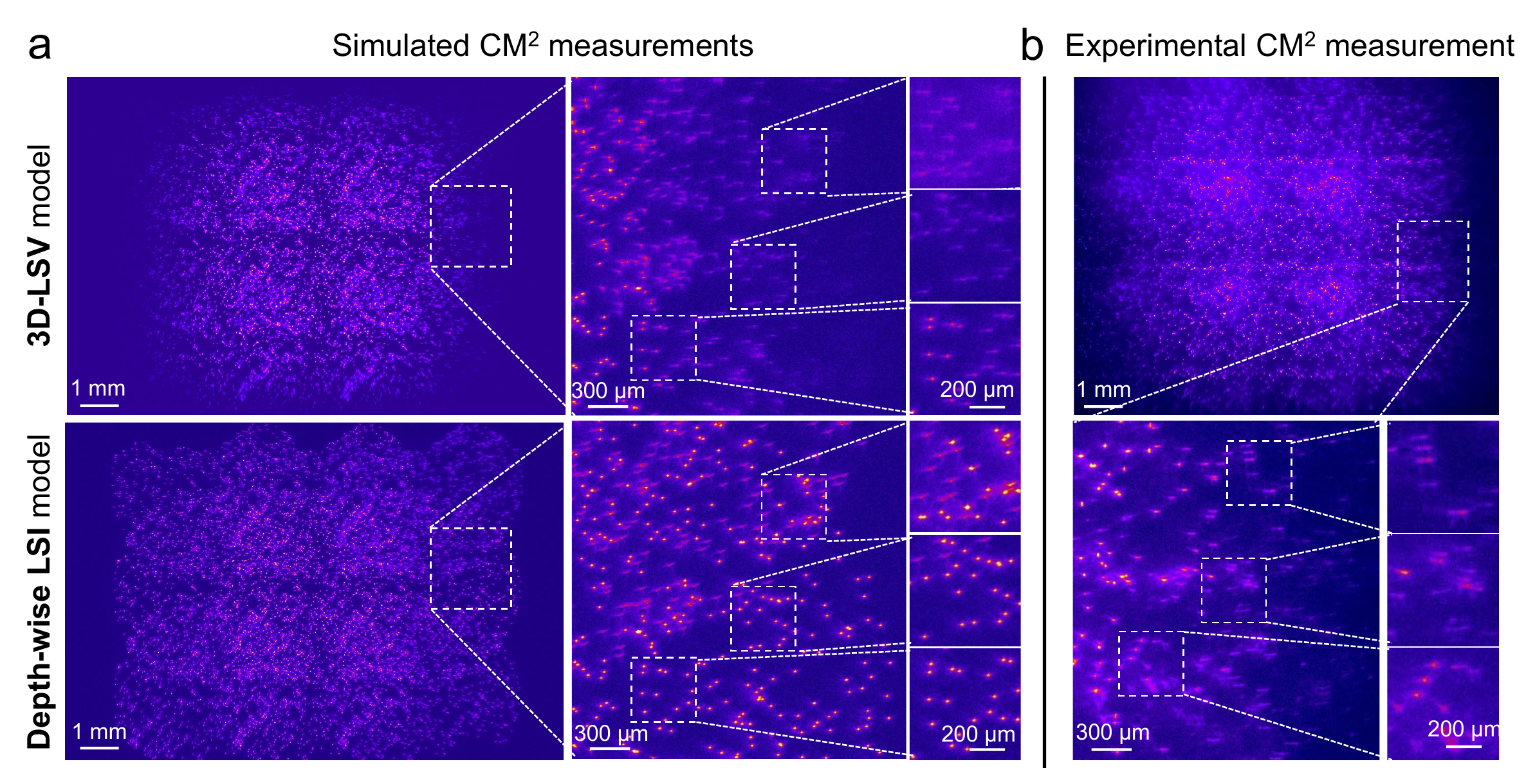}
\caption{\textbf{Experimental validation of the 3D-LSV model.}
(a) Simulated measurements using our 3D-LSV model and the depth-wise LSI model (sample: 10-µm fluorescence beads randomly placed in a 5 $\times$ 5 $\times$ 0.8 mm$^3$ volume). (b) Experimental measurement on 10-µm fluorescence beads with similar density to the simulation. This comparison shows that our 3D-LSV model can accurately capture the key aberrated features in real experiments.}
\label{S6}
\end{figure}

To visually validate the synthetic measurement generated by this 3D-LSV model, we compare a simulated image and an experimental measurement taken from a 5 $\times$ 5 $\times$ 0.8 mm$^3$ volume with 10-µm fluorescence beads with similar seeding density in Fig. \ref{S6}. The forward model used in our V1 system is based on a depth-wise linear shift invariant (LSI) approximation\cite{xueSingleshot3DWidefield2020}. To highlight the improvement of the proposed 3D-LSV model over the depth-wise LSI model, we compare the synthetic measurements from the same object using both models in Fig. \ref{S6}a. In the zoom-in panels taken from the same regions at the peripherical FOV where the aberrations are more apparent, we highlight that the 3D-LSV model can synthesize the key image features in the real experiment. This is essentially important to train CM$^2$Net based on only simulation data, yet the trained network is directly generalizable to real experimental measurements. 

\subsection{Analysis on the denoising capability of the 3D-LSV model}
\begin{figure}[htbp]
\centering
\includegraphics[width=\linewidth]{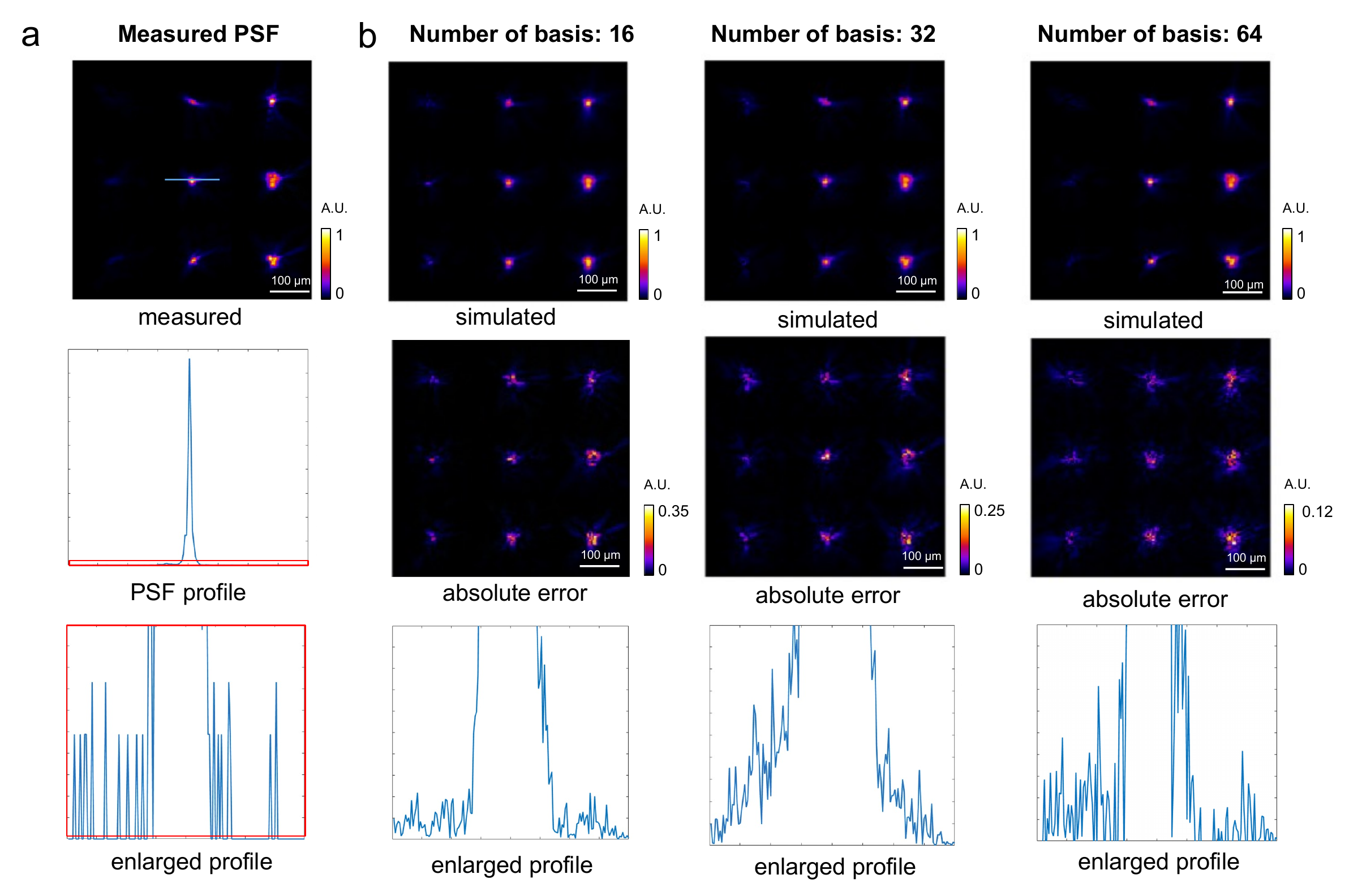}
\caption{\textbf{Denoising capability of the 3D-LSV model with different number of basis PSFs.}
(a) Top: A measured 3$\times$3 PSF array on the calibration grid. Middle: Intensity profile of the labeled PSF. Bottom: Enlarged PSF intensity profile to show the background fluctuations. (b) The simulated PSF with different number of basis PSFs, including 16, 32, and 64 (used in the main text). 
The absolute error between the simulated and measured PSF shows that the 3D-LSV model accuracy improves when the number of basis PSFs becomes larger. The enlarged profiles for all simulated PSFs show smoother background intensity profiles compared to the measured PSF, which indicates that the low-rank model can suppress the noise in the raw PSF measurement.}
\label{denoise}
\end{figure}

In this section, we analyze the denoising capability of the 3D-LSV model with different number of basis PSFs. 
We first decompose the raw calibration PSF stacks with different number of basis using the truncated Singular Value Decomposition (TSVD) and then compute the corresponding 3D-LSV model and simulate the PSF at a seen position. 
Next, we compare the absolute error between the simulated and the measured PSFs and plot the background intensity profile. 
Since when constructing our 3D-LSV model only the simulated PSF is used even at the calibration "seen" 3D locations, this procedure allows assessing the effective denoising capability of the 3D-LSV model.
Fig.~\ref{denoise} shows that all simulated PSFs have smoother background profiles, validating that the low-rank model can suppress the noise particular in the background region. 
Moreover, the denoising performance for 3D-LSV model is related to the number of basis. When the number of basis is too large, the LSV model not only well approximates the system aberrations, but also inadvertently capture the random noise present in the raw calibration PSF measurements. 
This shows a trade-off between the reconstruction accuracy and denoising capability. Since the noise level at for our raw PSF measurement is low, we choose a relatively large number of basis PSFs (K = 64) to provide a good reconstruction accuracy with a moderate denoising effect.

\subsection{Implementation details of the CM$^2$Net and its building block}
\begin{figure}[htbp]
\centering
\includegraphics[width=\linewidth]{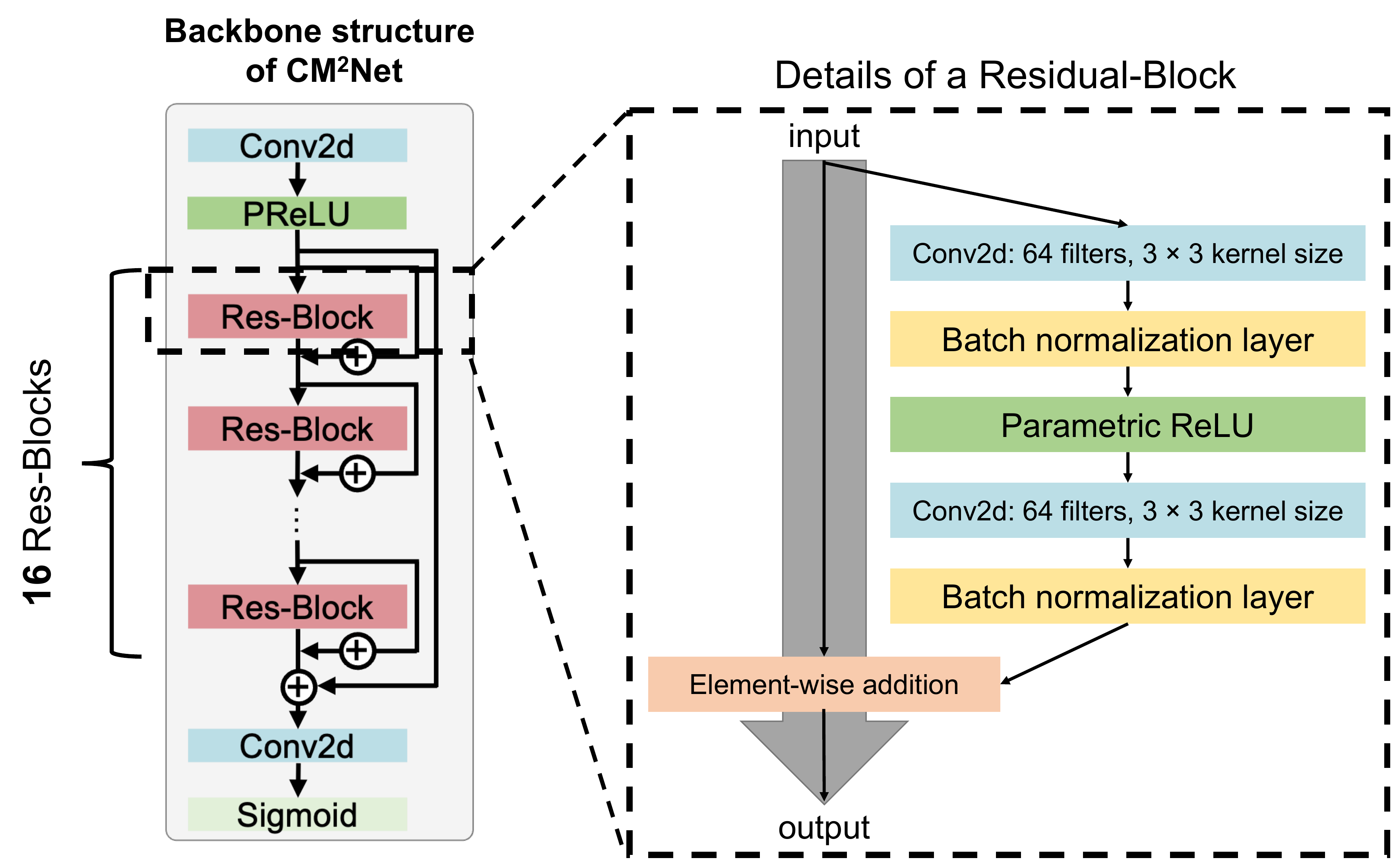}
\caption{\textbf{Implementation details of CM$^2$Net building block.}The three sub-networks in the CM$^2$Net share the same backbone structure of a deep residual network that connects 16 residual-blocks with sequential residual connections. Each residual-block consists of a 2D conv layer with batch normalization (BN), a Parametric ReLU nonlinearity, another 2D conv layer with BN, and element-wise addition.}
\label{S7}
\end{figure}

In this section, we describe the details of the network implementation and its training and testing procedure. The CM$^2$Net contains three sub-networks, including the view demixing-net, view-synthesis-net, and lightfield-refocusing enhancement-net. The three sub-networks all use the same residual network structure and only differ in the input/output dimensions. The detailed structure of a residual block is provided in Fig. \ref{S7}. The input tensor first goes through a 2D convolution layer which has 64 convolution kernels with size 3 × 3. After the 2D convolution layer, the intermediate tensor is fed into a batch normalization layer, followed by a nonlinear activation layer using Parametric ReLU (PReLU) layer, a nonlinear function with learnable slope in the negative side of the axis. The tensor then goes through another pair of 2D convolution and batch normalization layers to further increase the receptive field. Lastly, the tensor is element-wise added to the input tensor of this block, which forms a residual connection. The numbers of input and output channels, denoted by (a, b) for the demixing-net, view-synthesis-net, and enhancement-net are (9, 9), (9, 80), and (32, 80), respectively.

The CM$^2$Net takes a stack of 9 cropped views from a CM$^2$ measurement as the input. In  training phase, the input data dimension to demixing-net is 320 $\times$ 320 $\times$ 9, where the third dimension denotes the number of channels. The output size from the demixing-net remains the same. The non-learnable lightfield refocusing module applies a “shift-and-add” refocusing operation on the demixed 9 views:
\begin{equation}
RFV(x,y;\Delta z)=\sum_{u=-1}^{1}\sum_{v=-1}^{1}VS\left(u,v;x-\frac{M^2 d}{z_0}u\Delta z,y-\frac{M^2 d}{z_0}v\Delta z\right),
\label{LFR}
\end{equation}
where $RFV$ is the refocused slice with $\Delta z$ defocus distance, $(u,v)$ is the coordinates of the 3 $\times$ 3 microlens array, $VS$ is the demixed view stacks,  the amount of shift is determined by the system magnification $M$, nominal focal distance $z_0$, size of the microlens $d$, the microlens index $(u,v)$ and the defocus distance $\Delta z$. 
To implement this, we convert the amount of shift to be [-18: +17] pixels (+ means shifting towards the center view, - means the opposite), which ensure that the refocused stack consisting of 36 planes will cover the targeted 800-µm depth range.
Note that this refocusing algorithm generates artifacts around the image boundaries. Therefore, we discard the outmost 32 pixels in both lateral dimensions, resulting in a 256 $\times$ 256 $\times$ 36 refocused volume.

The enhancement-net is trained to first resample the refocused volume onto an axially finer sampling grid (256 $\times$ 256 $\times$ 80) matching the ground-truth volume, and then enhance the 3D reconstruction. The resampling is performed by a 2D convolution layer of kernel size 3 with 80 channels, which increases the depths from 36 to 80. To perform the view-synthesis, similarly only the central 256 $\times$ 256 $\times$ 9 region is extracted from the demixed views and fed into the view-synthesis-net. The output dimension from the view-synthesis-net is 256 $\times$ 256 $\times$ 80, the same as the ground-truth volume. 

In the inference phase, we directly feed the full-FOV measurement to perform a single-pass 3D reconstruction, which bypasses the stitching artifacts suffered by patch-wise reconstructions.

\subsection{Ablation study on the demixing module of the CM$^2$Net}
In this section, we show in simulation that the view-demixing task can be reliably performed and the demixing-net module substantially improves the quality in the downstream 3D reconstruction.

\begin{figure}[htbp]
\centering
\includegraphics[width=\linewidth]{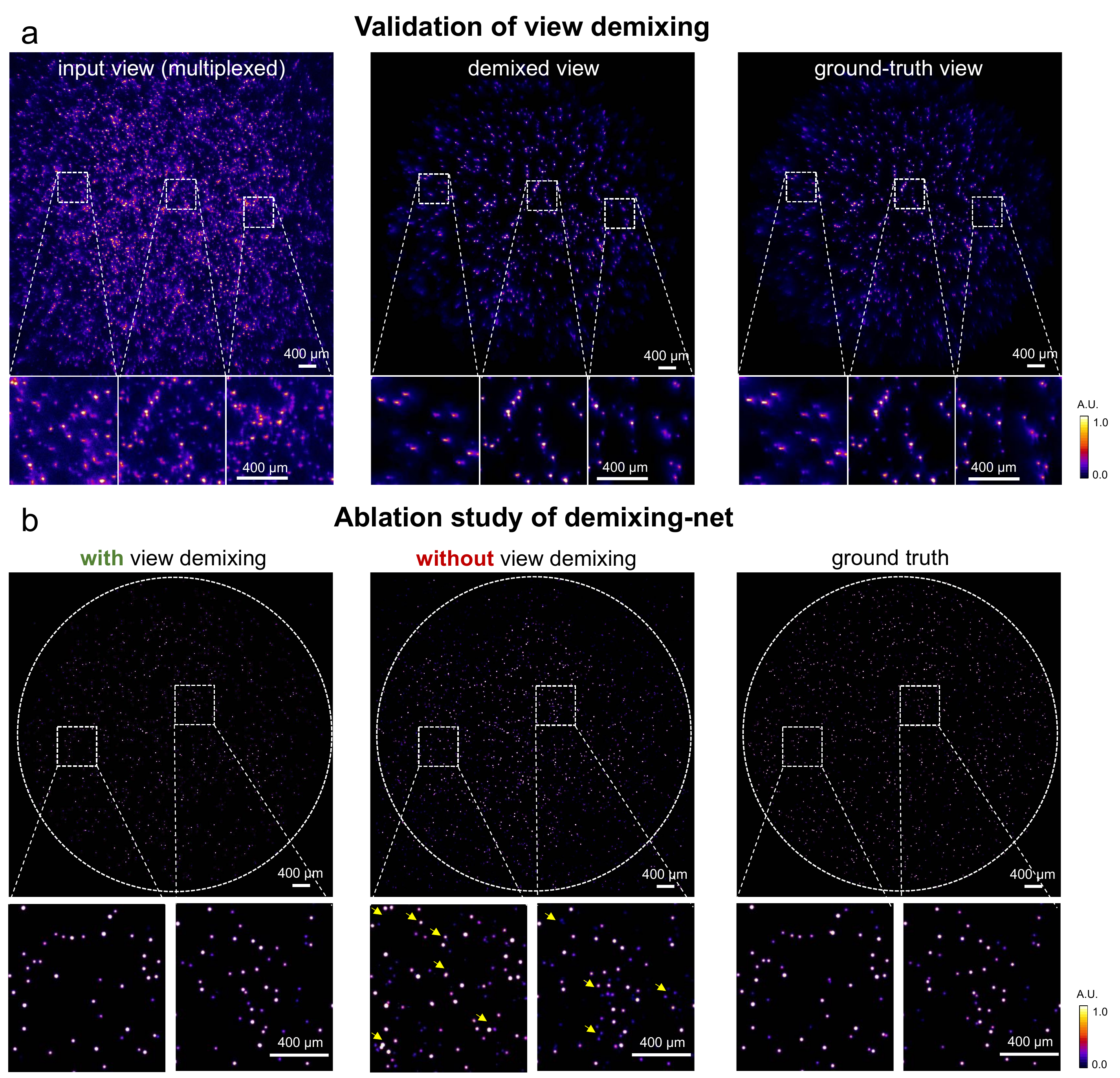}
\caption{\textbf{View-demixing network module significantly improves 3D reconstruction quality.}(a) View demixing results on a simulated CM$^2$ measurement (sample: fluorescent beads with random sizes between 10 µm and 20 µm in a cylindrical volume (7-mm diameter, 0.8 mm depth). The multiplexing artifacts in the “input view” from the raw CM$^2$ measurements are effectively demultiplexed by the demixing-net. (b) Ablation study results on the demixing-net. The XY MIPs shows that the proposed CM$^2$Net structure with the demixing-net module provides high-quality reconstruction. Removing the demixing-net module results in a large number of false positives (marked by yellow arrows) in the reconstruction.}
\label{S8}
\end{figure}

First, we demonstrate the effectiveness of view-demixing by comparing the demixing-net demixed views against the ground truth in simulation. Both the CM$^2$ measurement and the nine individual crosstalk-free views are simulated using our 3D-LSV model using the PSFs of the MLA and each microlenses, respectively. In Fig. \ref{S8}, we show results on a testing volume consisting of fluorescent beads with random sizes between 10 µm and 20 µm in a cylindrical volume (diameter $\sim$7 mm, thickness $\sim$0.8 mm). The large FOV results in strong view multiplexing in the raw CM$^2$ measurement, as evident in the example “input view” centered around the central microlens. The “demixed view” from the demixing-net closely matches with the ground-truth view from the central microlens without crosstalk artifacts, as highlighted in the three zoom-in panels taken from widely separated regions. The demixing-net can robustly perform this task since each fluorescent bead imaged by different microlenses contain distinct aberrated features.

Next, we show that the view-demixing step significantly improves the 3D reconstruction quality. Specifically, we perform the following ablation study in Fig. \ref{S8}b. The same dual-branch reconstruction network (including the lightfield-refocusing enhancement module and the view-synthesis module) is used to process either the demixing-net demixed views or the multiplexed views from the raw measurement on the same simulated data. The XY maximum intensity projections (MIPs) of the 3D reconstructions from the two networks and the ground-truth volume are shown in Fig. \ref{S8}b. As highlighted in the two zoom-in regions, the reconstruction from the network with the demixing-net closely matches with the ground truth, whereas the one without view-demixing suffers from a large number of false positives (marked by the yellow arrows).

We further quantify the reconstruction quality by recall and precision. The CM$^2$Net with the demixing-net achieves an average recall and precision of 0.7 and 0.94 respectively, while the one without the demixing-net achieves much lower recall of 0.57 and precision of 0.24. This comparison highlights that although the CM$^2$Net without the demixing-net is sensitive enough to recover 57\% of the emitters, the 3D reconstruction suffers from very low precision. The comparison on the precision values implies that the demixing-net helps to reduce the false-positive rate from 0.76 to 0.06, a ~13$\times$ improvement.

\subsection{Ablation study on the reconstruction module of the CM$^2$Net}
\begin{figure}[htbp]
\centering
\includegraphics[width=\linewidth]{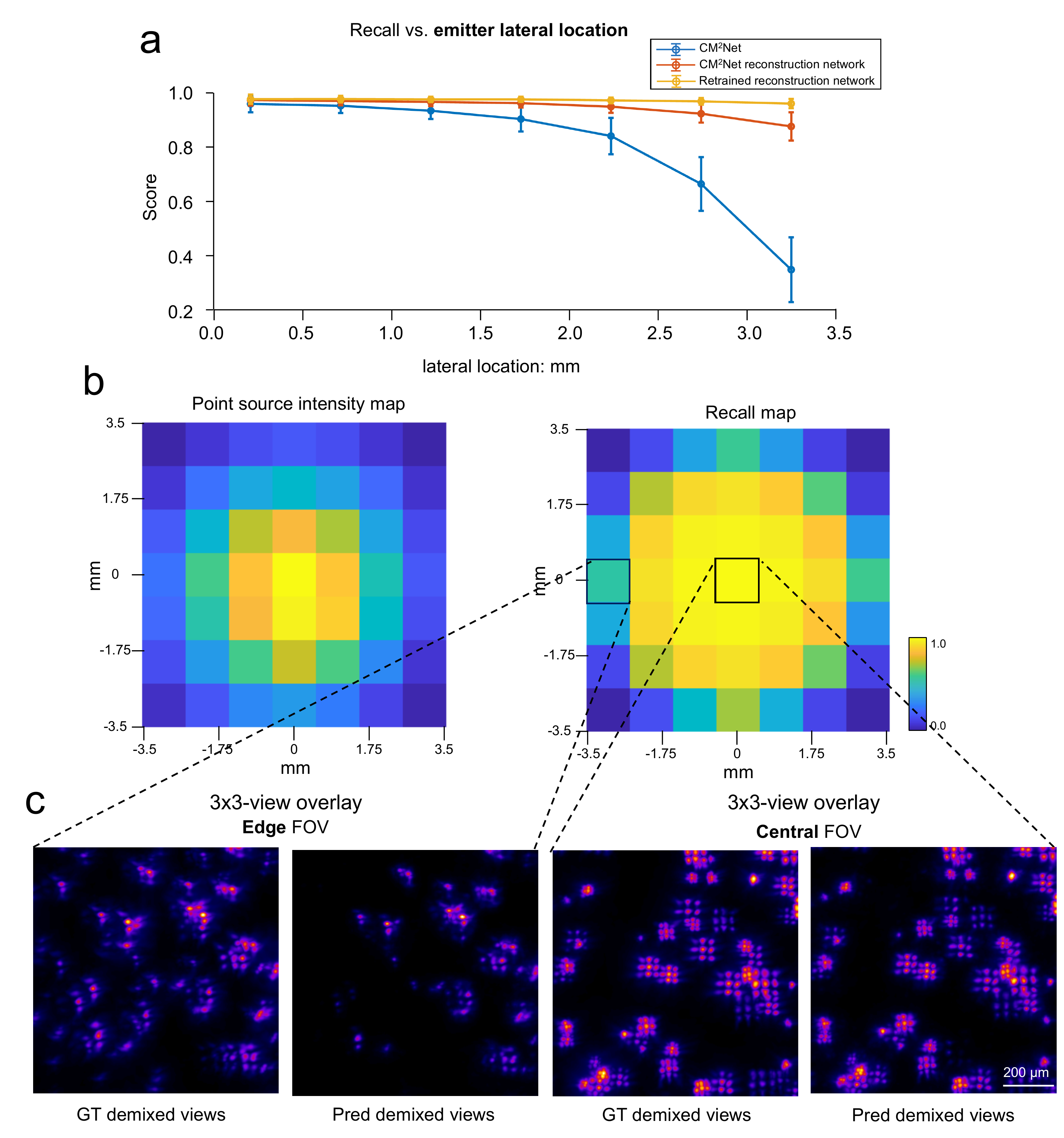}
\caption{\textbf{Ablation study on the reconstruction module of the CM$^2$Net using the ground-truth demixed views.}(a) Yellow curve: recall for a retrained reconstruction module using the ground-truth demixed views as the input. The recall is close to 1 consistently for the entire 7-mm FOV range, showing the effectiveness and robustness of our reconstruction module design. Orange curve: recall for directly feeding the ground-truth demixed views to the reconstruction module of the pre-trained CM$^2$Net. The recall remains >0.89 across the 7-mm FOV. Blue curve: recall for the trained CM$^2$Net, which is equivalent to feeding the view-demixing-net demixed views to the reconstruction module of the pre-trained CM$^2$Net. The decreased performance is attributed to the degraded view-demixing results. (b) Left: the intensity distribution of the point source used in our PSF calibration as measured by total intensity under a single microlens at different scanning positions. Right: the recall map of the CM$^2$Net is visualized by binning onto the same pixel grid as the point source intensity map. Both maps exhibit much reduced values at the outer FOV regions, which suggests that the degraded recall at the outer FOV regions may be due to the rapid intensity decay of the point source used in the model. (c) Comparison of the 3×3-view overlay view-demixing results between patches from the edge and central FOV. The central patch that achieves $\sim$1 recall has higher SNR and shows nearly perfect view-demixing result. However, the view-demixing prediction on the patch from the edge FOV suffers from low SNR and missing particles.}
\label{S9}
\end{figure}
In this section, we perform an ablation study to analyze the potential reasons for the decrease of recall near the edges of the FOV. We first separate the CM$^2$Net into demixing network and the reconstruction network (consisting of the enhancement-net and the view-synthesis-net). First, we train another network with only the reconstruction network and take the ground-truth demixed views as the input. The recall for emitters at different lateral location are then evaluated on the testing set using the same method in Section 3.2. The reconstruction results achieve recalls close to 1 consistently for the entire 7-mm FOV range (Fig. \ref{S9}a yellow curve), showing the effectiveness and robustness of our reconstruction network design. To further evaluate the trained CM$^2$Net, we quantify the recall of the reconstruction network in the pre-trained CM$^2$Net by directly inputting the ground-truth demixed views. The result (orange curve) shows a slight degradation as compared to the re-trained reconstruction network but remain >0.89 across the 7-mm FOV. Both results are much higher than that from the CM$^2$Net predictions on the raw CM$^2$ measurement (blue curve). This indicates that the CM$^2$Net’s reconstruction module provides superior performance, and the degraded performance at the outer FOV is due to imperfect view-demixing at these regions.

Upon visual inspection, we hypothesize that the view-demixing network is impacted by the low light intensity and reduced SNR of the point source near the edge of the FOV. To show this, we calculate the intensity map of the point source under the central microlens in Fig. \ref{S9}b and show that the point source’s intensity drops as much as $\sim$85\% near the edge as compared to the intensity at the central FOV. Moreover, we compare the point source intensity map and recall map on the same pixel grid. Both maps exhibit similar non-isotropic distributions and significant drops at outer FOV regions. Finally, we show comparisons between the ground-truth and the view-demixing-net predicted demixed views (1-mm$^2$ patch) for both a central and an edge image patches in Fig. \ref{S9}c. The view-demixing result for the central FOV is highly accurate, whereas the result near the edges suffer from a large amount of missing particles. We also visually observe that the SNR for the demixed views at the edge FOV are much worse than the central FOV. 

Overall, this ablation study shows that the combination of the view-synthesis-net and enhancement-net can achieve superior reconstruction performance and is robust to large SNR variations and high dynamic range of the measurements. We provide evidence that the low recall near the edge is possibly due to the unevenly distributed point source used in the PSF calibration.

\clearpage
\subsection{Ablation study on the reconstruction sub-modules of the CM$^2$Net}

In this section, we perform ablation studies on the two branches in the reconstruction module to show how the light-field refocusing (LFR) enhancement branch and the view-synthesis (VS) branch complement each other.

\begin{figure}[htbp]
\centering
\includegraphics[width=\linewidth]{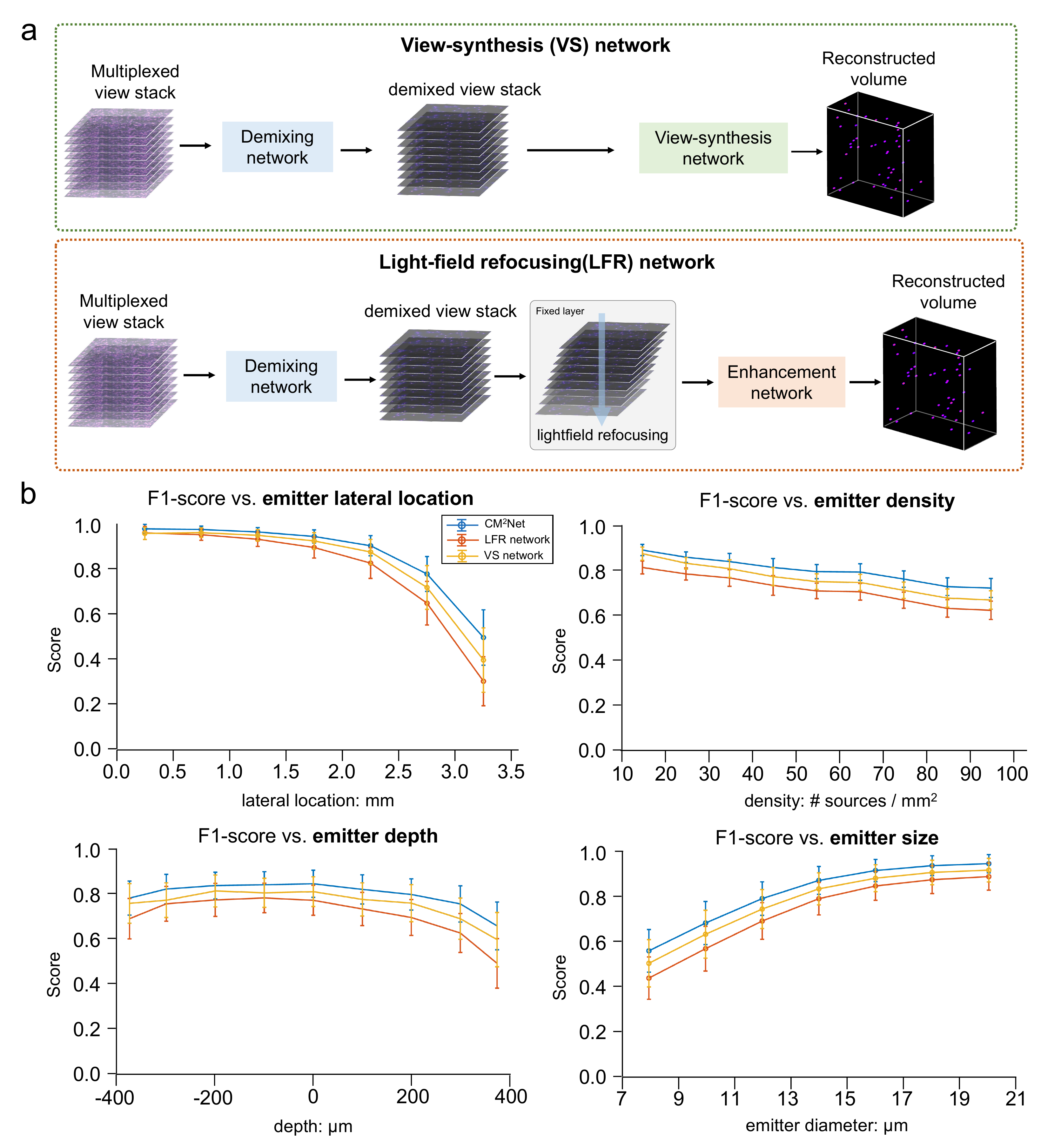}
\caption{\textbf{Ablation study on CM$^2$Net’s reconstruction sub-nets.} (a) The network structure to perform ablation study on the view-synthesis branch (VS) and the light-field refocusing enhancement branch (LFR).
(b) Quantitative analysis of evaluating the particle detection performance of the VS, LFR and CM$^2$Net. CM$^2$Net achieves the best score at all conditions by combining the complementary information from the VS and LFR branches.}
\label{ablation}
\end{figure}

First, we compare the network performance by removing either the LFR or the VS branch in the CM$^2$Net to show the effectiveness of the combined network structure. 
To do this, we build two new reconstruction networks with only the LFR or the VS branch (the network structure is shown in Fig. \ref{ablation}a). 
Both networks use the same data set and follow the same training scheme as the full CM$^2$Net. After training for 48 hours on all the networks, we test and evaluate the performance by the quantitative analysis procedure shown in Section 3.2. The results are summarized in Fig. \ref{ablation}b.
This ablation study shows that the CM$^2$Net achieves the best F1-score on all the conditions by combining both the VS and LFR branches.

\begin{figure}[!th]
\centering
\includegraphics[width=0.95\linewidth]{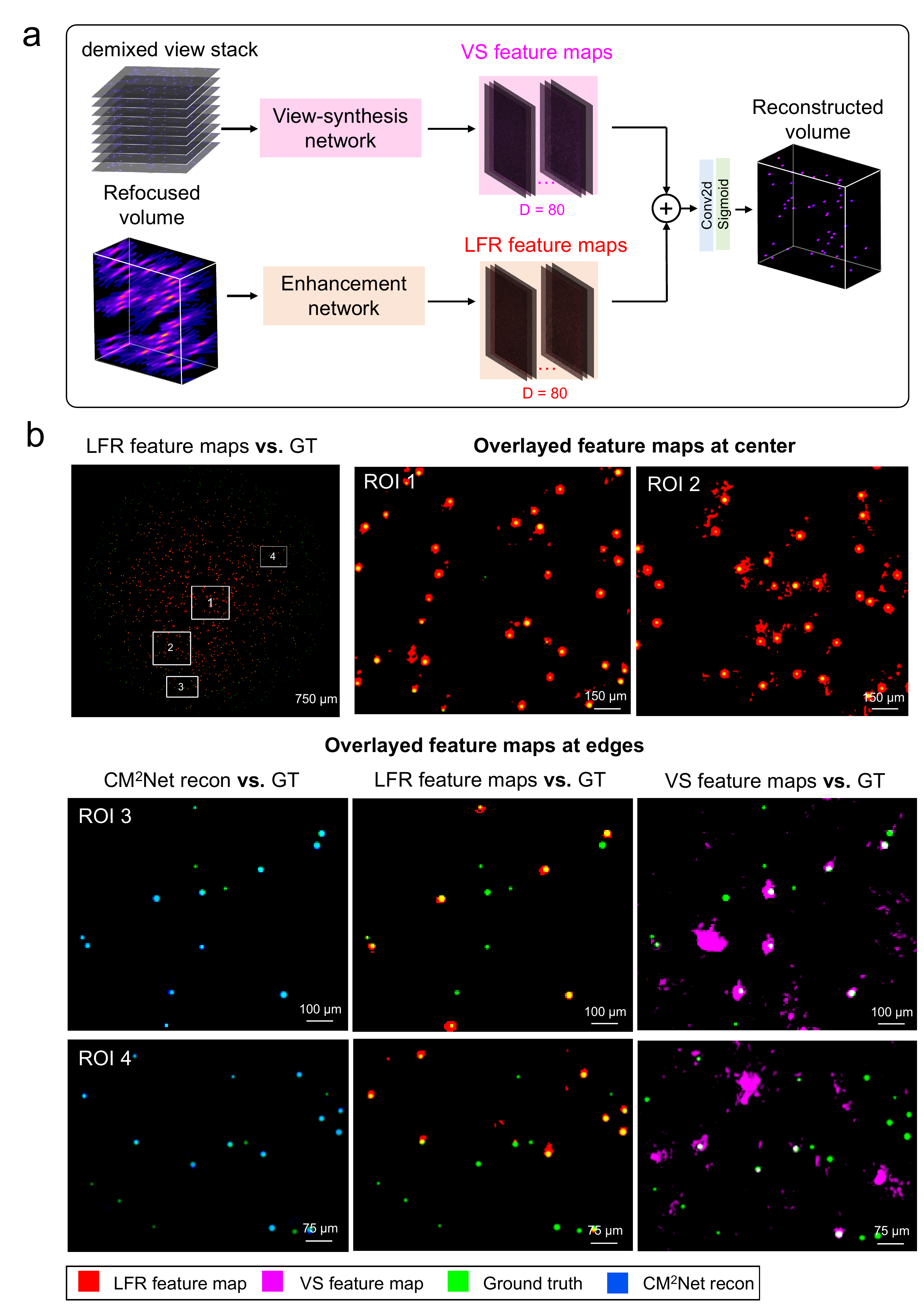}
\caption{\textbf{Feature maps analysis of the CM$^2$Net’s reconstruction sub-nets.} (a) Illustration of the extraction of feature maps from the VS and LFR branches.
(b) Top: The overlay between the MIPs of the ground truth and the feature maps from the LFR  network. The contrast for feature maps is enhanced for visualization. The ROI 1 and ROI 2 show that the LFR feature maps highly matches with the ground truth at the central FOV (yellow indicates matched particles).
Bottom: The overlay of the MIPs of the ground truth and the CM$^2$Net reconstruction, LFR feature maps MIP and VS feature maps MIP at the FOV edges. The particles detected by the CM$^2$Net (in light blue) is the combination of the particle detected by the LFR branch (in yellow) and the VS branch (in white). The CM$^2$Net learns the complementary information provided by the VS and LFR branches to achieve high detection rate across a wide FOV.}
\label{feature_maps}
\end{figure}

To further illustrate the LFR and VS networks provide complementary information. Next, we visualize the final feature maps from the two branches of the CM$^2$Net. We directly extract the feature maps from the trained CM$^2$Net (as shown in Fig. \ref{feature_maps}a) and compare the MIP of the feature maps with the MIP of the ground truth at the central and edge FOVs. 
In Fig. \ref{feature_maps}b, we show that the feature maps from the LFR branch highly matches with the ground truth at the central FOV while miss a large amount of particles at the edge.
This is because the input (lightfield refocused volume based on the demixed views) to the LFR-network suffers from artifacts at the peripheral regions from inexact view matching from the demixed views and the boundary artifacts from the ``shift-and-add'' operation. To address this challenge, CM$^2$Net uses the additional VS branch to the reconstruction module to help improve the performance at the peripheral FOV regions. 
In Fig. \ref{feature_maps}b, we compare the ground truth volume, the CM$^2$Net reconstruction, the feature maps from the LFR branch, and the feature maps from the VS branch to visualize the improvement owing to VS branch. We show that at the FOV edges, the VS branch can detect particles that are missed by the LFR branch, and thus improve the CM$^2$Net's recovery of emitters across the large FOV.

\clearpage
\section{Performance evaluation metrics}
We use particle-wise detection metrics, including recall, precision and F1-score to quantitatively evaluate the CM$^2$Net 3D reconstruction. We choose this set of metrics because they focus on quantifying the error made on the reconstructed emitters, whereas voxel-value averaged metrics, e.g. mean squared error, are biased by the background for sparse objects like ours. 

We describe the details on the computations of the three metrics using MATLAB language. Starting from the reconstructed volumes from the CM$^2$Net, we first detect all the recovered particles within the volume.  The detection is done by binarizing the recovered 3D volume with a global optimal thresholding (\verb|imbinarize|) and then extracting the locations and sizes of connected 3D components (recovered particles) from the binarized volume (\verb|bwconncomp|, \verb|regionprops3|). We then compute a distance matrix by calculating the Euclidean distance between every recovered and ground-truth particle and solve a linear assignment problem based on the distance matrix to assign the recovered particles to the corresponding ground truth (\verb|matchpairs|).

If the Euclidean distance between the recovered and the matched ground-truth particles is larger than a pre-defined distance threshold (i.e. 2$\times$ the particle size in our simulation), we count it as a False Negative (FN), meaning that the CM$^2$Net fails to correctly reconstruct the particle. If a recovered particle does not find a match in the ground truth, we count it as a False Positive (FP). A True Positive (TP) is when the distance between the recovered and the matched ground-truth particles is smaller than the distance threshold (examples are shown in Fig. \ref{S16}a). Lastly, the recall, precision, F1-score, Jaccard index (JI), and lateral and axial Root mean squared localization error (RMSE) are computed as follows:

\begin{equation}
\mathrm{Recall}= \frac{\#\mathrm{TP}}{\#\mathrm{TP} + \#\mathrm{FN}}
      = \frac{\#\mathrm{emitters\;correctly\;reconstructed}}{\mathrm{\#total\;emitters\;in\;GT\;volume}},
\label{eq3}
\end{equation}

\begin{equation}
\mathrm{Precision}= \frac{\#\mathrm{TP}}{\#\mathrm{TP} + \#\mathrm{FP}}
      = \frac{\#\mathrm{emitters\;correctly\;reconstructed}}{\#\mathrm{total\;emitters\;in\;reconstructed\;volume}},
\label{eq4}
\end{equation}

\begin{equation}
\mathrm{F1\;score}= \frac{2}{\mathrm{Recall}^{-1}+\mathrm{Precision}^{-1}}
= \frac{\#\mathrm{TP}}{\#\mathrm{TP} + 1/2(\#\mathrm{FP} + \#\mathrm{FN})},
\label{eq5}
\end{equation}

\begin{equation}
\mathrm{JI} = \frac{\#\mathrm{TP}}{\#\mathrm{TP} + \#\mathrm{FP} + \#\mathrm{FN}},
\label{JI}
\end{equation}

\begin{equation}
\mathrm{Lateral\, RMSE} =  \sqrt{\frac{1}{\#\mathrm{TP}}\sum_{i\in R\cap G}(x_i^{r} - x_i^{g})^2 + (y_i^{r} - y_i^{g})^2},
\label{LRMSE}
\end{equation}
\begin{equation}
\mathrm{Axial\, RMSE} =  \sqrt{\frac{1}{\#\mathrm{TP}}\sum_{i\in R\cap G}(z_i^{r} - z_i^{g})^2},
\label{ARMSE}
\end{equation}
where $ x_i^{r}, y_i^{r}, z_i^{r}$ is reconstructed emitter's position and $ x_i^{g}, y_i^{g}, z_i^{g}$ is the corresponding ground truth position.

\begin{figure}[htbp]
\centering
\includegraphics[width=\linewidth]{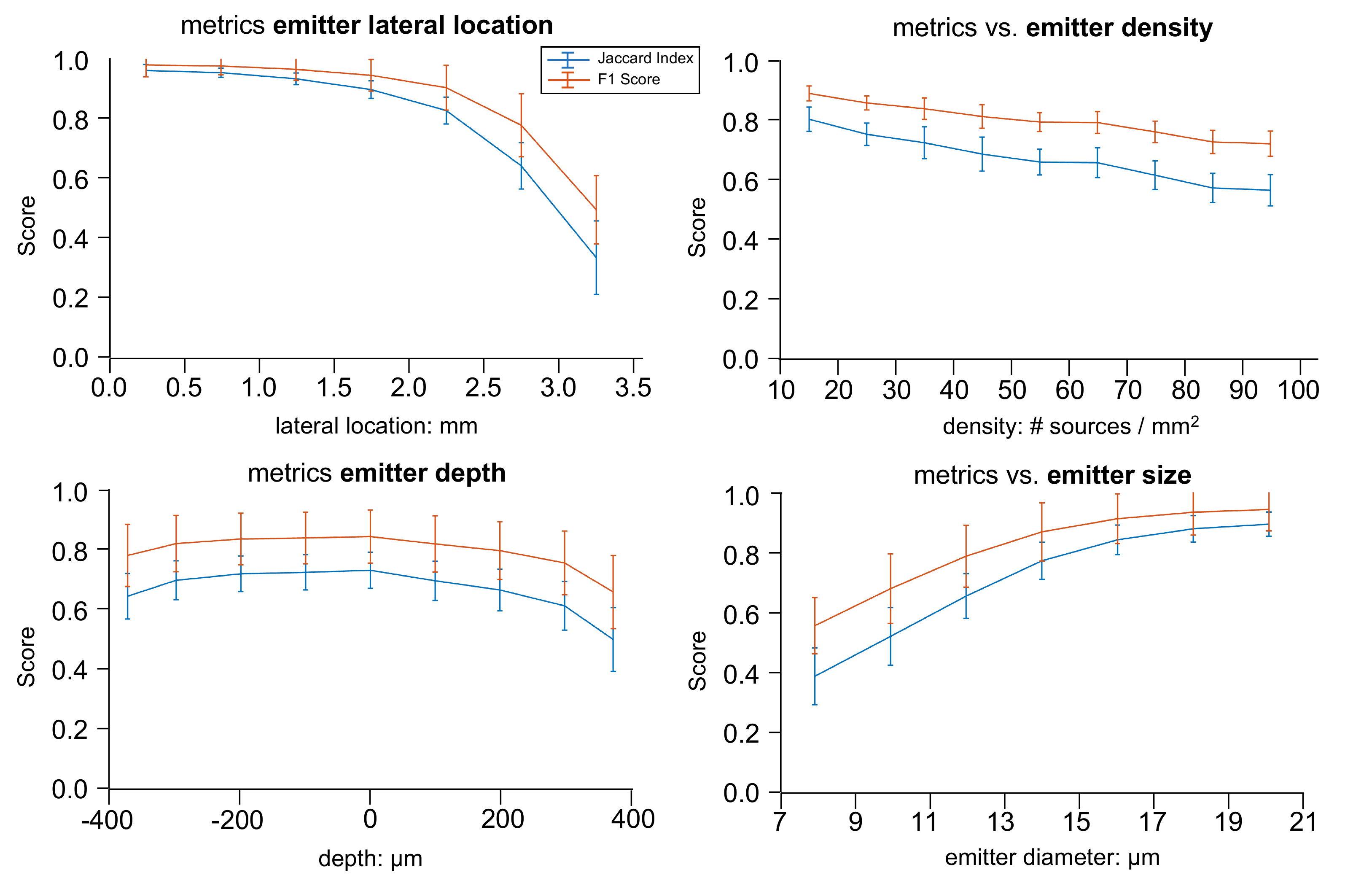}
\caption{\textbf{CM$^2$Net performance evaluated by F1-score and JI.} Quantified F1-score and JI when varying the emitter’s (a) lateral location, (b) seeding density, (c) depth, and (d) size. F1-score and JI show the same trends since they both combine recall and precision in a similar way.}
\label{f1_ji}
\end{figure}

\begin{figure}[htbp]
\centering
\includegraphics[width=\linewidth]{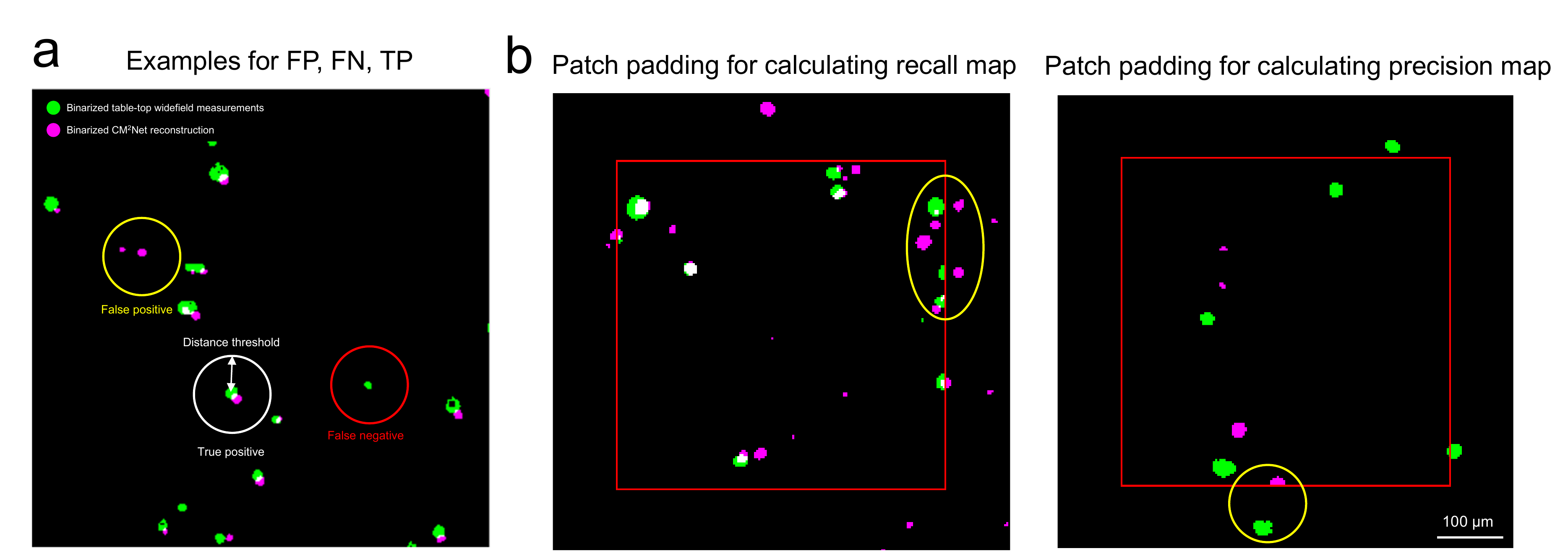}
\caption{\textbf{Details for the procedure of CM$^2$Net quantification analysis.}(a) An example overlay image patch between the registered and binarized widefield measurement and the CM$^2$Net reconstruction. The circle indicates the distance threshold for performing linear assignment. The yellow, red, and white circles show examples of false positive, false negative, and true positive, respectively. (b) Example patch padding procedure for computing the recall (left) and precision (right) map, respectively. The red box shows the effective 500 $\times$ 500-µm$^2$ non-overlap FOV used to quantify the metrics. To maintain the distance threshold at the FOV edge, we choose a larger CM$^2$Net reconstruction patch (700 $\times$ 700-µm$^2$) when calculating the recall map, and a larger widefield measurement patch (700 $\times$ 700-µm$^2$) when calculating the precision map. The yellow circle shows the particles around the edges, which requires the patch padding to find a proper match within the defined distance threshold.}
\label{S16}
\end{figure}

Recall and precision quantify the detection performance in two complementary aspects. Recall quantifies the CM$^2$Net’s sensitivity by the fraction of emitters it correctly reconstructed out of the total in the ground-truth volume. Precision measures the fraction of emitters it correctly reconstructed out of the total reconstructed emitters. Both F1-score and JI are metrics combining recall and precision, which are highly correlated. Lateral and axial RMSE measure the localization accuracy. The global threshold for binarizing the recovered volume is set by maximizing the F1-score on the evaluation set~\cite{bao2021segmentation}.
The recall, precision, lateral and axial RMSE under different conditions are reported in the main text (Section 3.2 and Figure 6). The F1-score and JI under the same set of conditions are shown in Fig. \ref{f1_ji}.

The metric maps reported in Sections 3.2 and 3.3 are computed by the metric value on non-overlapping patches (250 µm $\times$ 250 µm in Section 3.2 and 500 µm $\times$ 500 µm in Section 3.3). When quantifying the metrics on experimental data (Sections 3.3 and 3.4), we first co-register the CM$^2$Net reconstruction and the widefield measurement by manually selecting a few matched feature points in the central FOV. We then extract particles by binarizing both the widefield measurement and the CM$^2$Net reconstruction. Next, we follow the same evaluation pipeline as above. Since the amount of image distortion suffered by the widefield and CM$^2$ measurements are different, good registration can only be achieved around the central region, while particle pairs at the peripheral region are expected to have a larger amount of separations. Therefore, the distance threshold we choose for the experimental data is $\sim$5$\times$ the particle diameter, which is larger than the one used for the simulation (2$\times$ the particle diameter).

To maintain the same distance threshold for particles at the edge of each patch, we perform particle-pairing on slightly expanded patches (see in Fig. \ref{S16}b). Specifically, when computing the recall map on a 500 $\times$ 500-µm$^2$ patch, we select a larger CM$^2$Net reconstruction patch (700 $\times$ 700-µm$^2$) and pad 0 to the corresponding widefield measurement patch, since recall requires detecting all the true positives. We visually show that without considering the distance threshold at the edge, the recall value will decrease since some matched pairs (yellow circle) between the reconstructed and the matching ground-truth are separated by the edge of the patch (red box) while within the distance threshold. When calculating the precision map, we select a larger widefield measurement patch (700 $\times$ 700-µm$^2$) and pad 0 to the corresponding CM$^2$Net reconstruction patch in order to identify all the reconstructed particles corresponding to the matching widefield measurement. We visually show that without considering the distance threshold at the edge, some matched pairs (yellow circle) are separated by the edge of the patch (red box) while within the distance threshold, which contaminates the precision quantification. 

The binarized MIP image pairs are shown in Figs. \ref{S13}b and \ref{S15}b for the 10-µm and mixed-size bead phantom, respectively. Moreover, since the particles in the experiment are not uniformly distributed, we observe \verb|NaN| values (both the widefield measurement and the reconstruction in this image patch are empty) when perform quantification for the experimental data. For fair evaluation, we exclude the \verb|NaN| values when calculating the mean and standard deviation. When making the metric maps, we linearly interpolate the pixels having \verb|NaN| values, and applies a binary mask to indicate the sample region.

In Fig. 6, the line plots show the quantitative metrics under different conditions. Each point and the associated error bar represent the mean and standard deviation, respectively. The testing data set contains in-total 180 random volumes and $\sim$5$\times$10$^5$ emitters. Each point in Fig. 6b is computed on $\sim$20 volumes for each range of emitter density. Each point in Fig. 6a, c and d is computed on emitters for each range (labeled on the horizontal axis) of lateral location, depth, and emitter size.

\clearpage
\section{Experimental Preparation and Result Analysis}
\subsection{Preparation of phantom objects and table-top widefield measurements}
Fluorescent 3D objects are prepared according to the following protocol. Green fluorescent particles (Thermo Fisher Scientific, Fluoro-Max Dry Fluorescent Particles, 10-µm and 15-µm) are added to $\sim$1 mL of clear resin (Formlabs, no. RS-F2-GPCL-04). The mixture is then diluted and moved onto a standard 3 $\times$ 1 inch microscope slide (Thermo Fisher Scientific, no. 125493) and curved under a UV lamp for $\sim$30 minutes. We use a micropipette (Thermo Fisher Scientific, Adjustable Volume Pipette, 10-100 µL, no. FBE00100) to control the transferred volume of mixture ($\sim$40 µL) so that the object is $\sim$800 µm thick and 7-mm wide.

To verify the 3D reconstructions from CM$^2$Net, we experimentally collect axial focal stacks (z-stack) on a commercial table-top epi-fluorescence microscope (Nikon TE2000-U) with GFP filter sets (Thorlabs, no. MDF-GFP) and a scientific CMOS camera (PCO Imaging, Pco.edge 5.5). To acquire the full-FOV widefield measurements, we use a low-magnification objective lens (Nikon, CFI Plan Apo Lambda 2$\times$, 0.1 NA) with 50-µm axial step size across the 800-µm depth range. To acquire a high-resolution axial scan at zoom-in regions, we use a high-magnification objective lens (Nikon, CFI Plan Achromat 20$\times$, 0.4 NA) with 25-µm axial step size across the 800-µm depth range.

\subsection{Preprocessing steps on experimental data}
\begin{figure}[htbp]
\centering
\includegraphics[width=.8\linewidth]{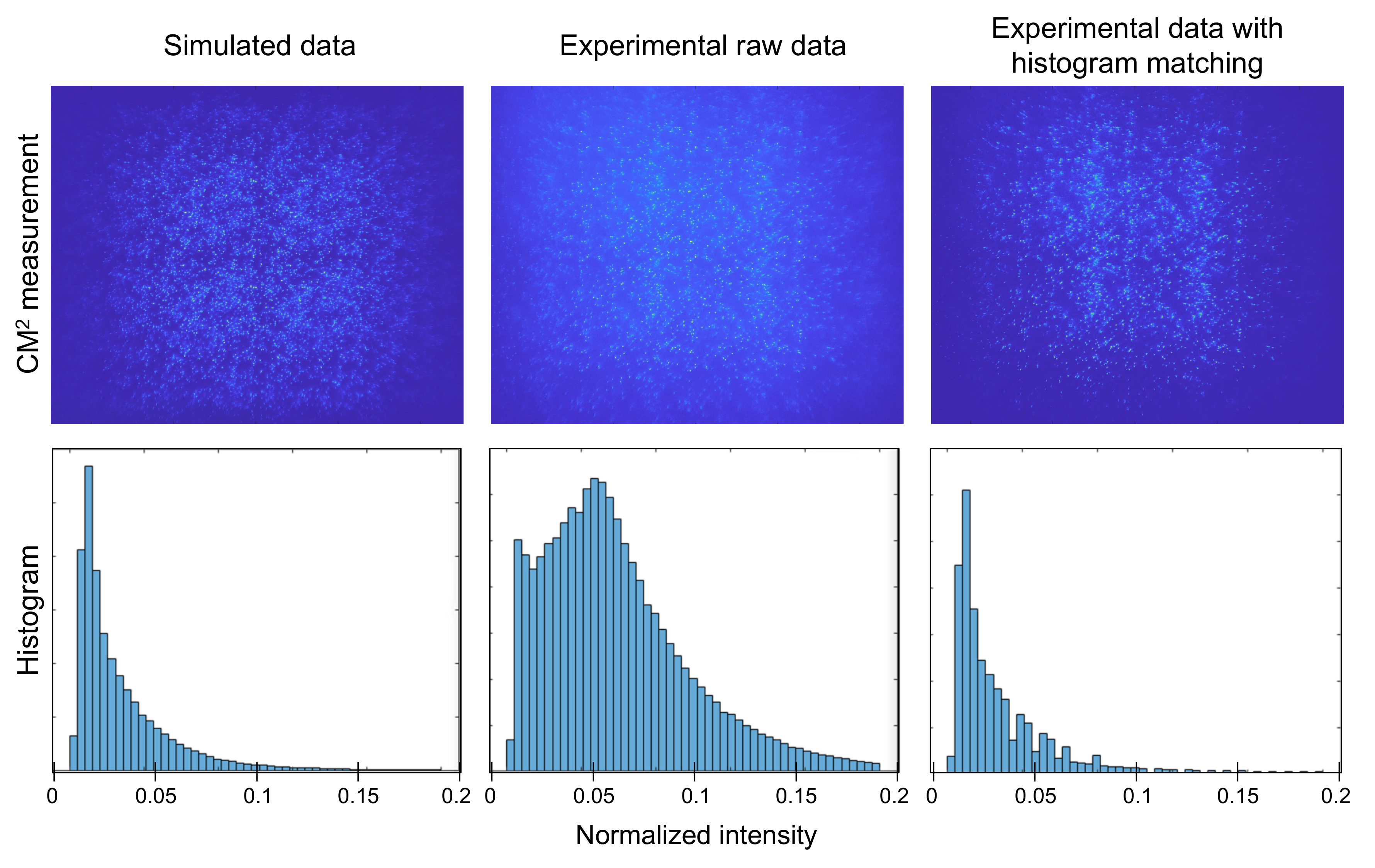}
\caption{\textbf{Preprocessing on experimental CM$^2$ measurements with histogram matching.}Preprocessed experimental measurements have a suppressed background and a better match with the intensity distribution of the simulated training data.}
\label{S10}
\end{figure}

In this section, we provide additional details on the preprocessing of experimental CM$^2$ measurements. The goal is to match the intensity distribution between the simulated and the experimental data so that our simulator-trained CM$^2$net can generalize well to real measurements. The main discrepancy between the measured and simulated images is the low-frequency fluorescent background due to the reflection from the microscope slide. To remove the background and match the intensity statistics, we preprocess the experimental measurements with histogram matching. The reference histogram is estimated from the entire training dataset. The histogram matching builds a monotonic mapping from the experimental data to the simulated data. A comparison between the histograms of the simulated data, the experimental data with and without histogram matching are provided in Fig. \ref{S10}, where one can clearly see the background is much suppressed after the preprocessing. This can be further verified in their histograms. The distribution of preprocessed experimental data matches simulated data much better than raw experimental measurements. We find that other commonly used background removal methods (such as thresholding, morphological opening, etc.) do not generalize well on real experimental measurements. Another preprocess step is to adjust the centers of the cropped 9 views to compensate for the small displacement of the chief ray positions between the simulation and experiments. The adjustment is done by manually aligning a few on-axis particles in each view.

\subsection{Recall, precision, and F1-score maps in phantom experiments}
In this section, we report the details of recall, precision, and F1-score evaluations on the two experiments in the main article. The recall and precision maps are computed on non-overlapping patches (500 µm $\times$ 500 µm) across the entire FOV.

\begin{figure}[htbp]
\centering
\includegraphics[width=.85\linewidth]{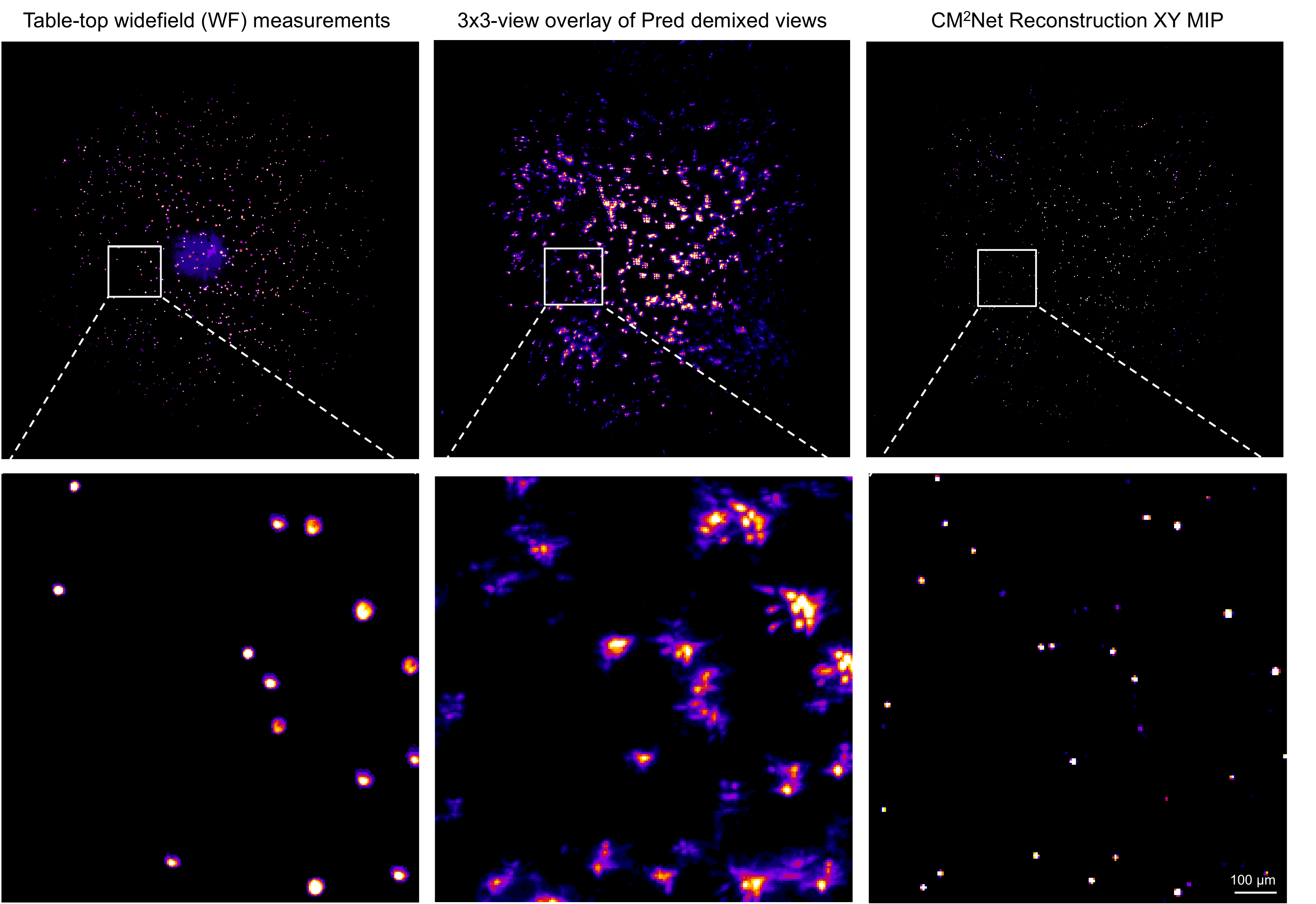}
\caption{\textbf{Comparison of the experimental widefield measurement, 3×3 overlay of predicted view-demixing result, and CM$^2$Net reconstruction MIP.}The false positives in the CM$^2$Net reconstruction MIP can find correspondence in the 3×3 overlay predicted demixed views, indicating that the performance decrease for the experimental data originates from the view-demixing net.}
\label{S11}
\end{figure}

\begin{figure}[htbp]
\centering
\includegraphics[width=.9\linewidth]{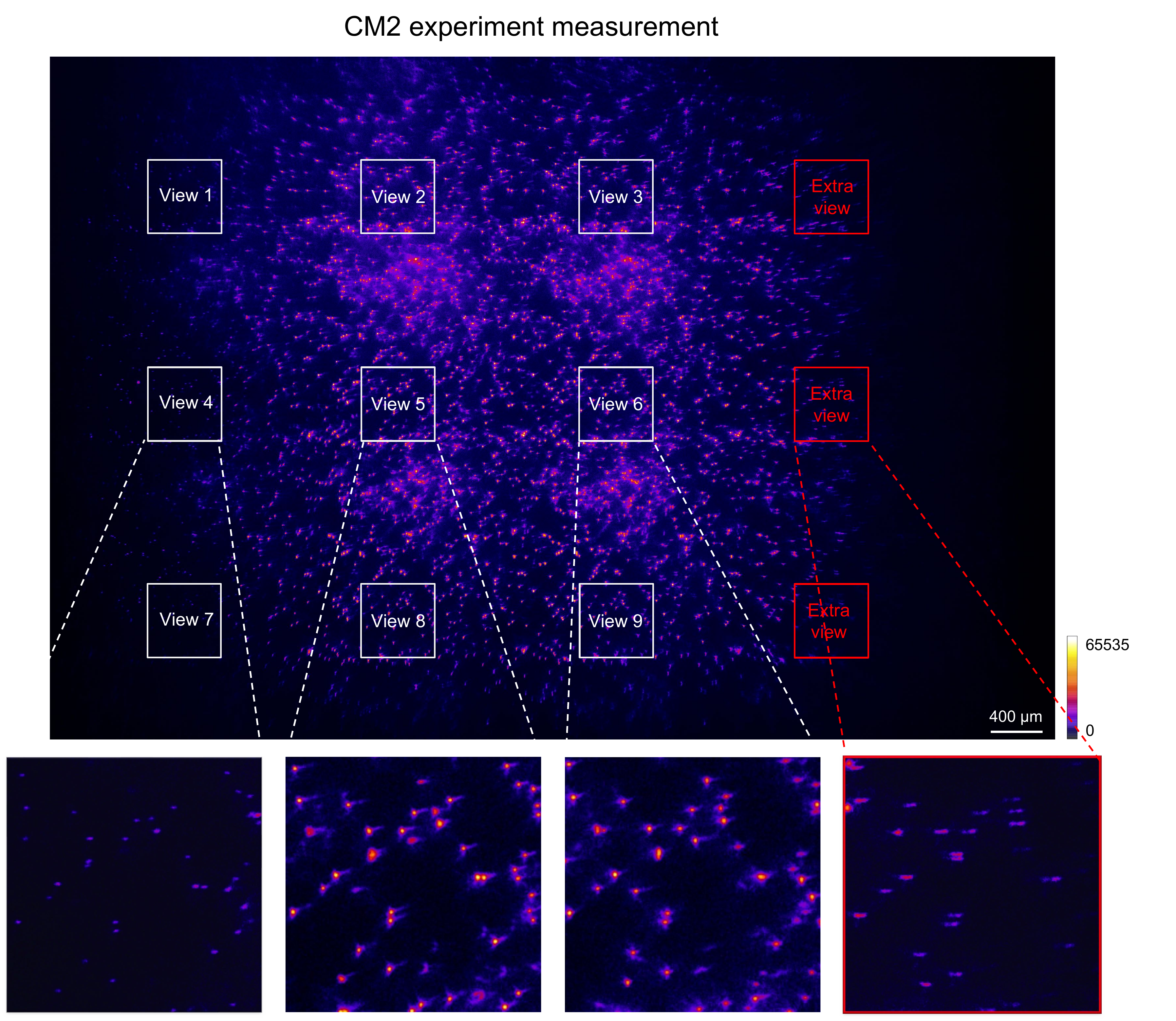}
\caption{\textbf{Extra views in the 10-µm bead experimental measurement.}The partial microlenses adjacent to the 3 $\times$ 3 microlens array generate extra views in the experimental measurement. The extra views contaminate the 9 central views, which results in increased false positives and mis-detections in the experiments.}
\label{S12}
\end{figure}

\begin{figure}[htbp]
\centering
\includegraphics[width=\linewidth]{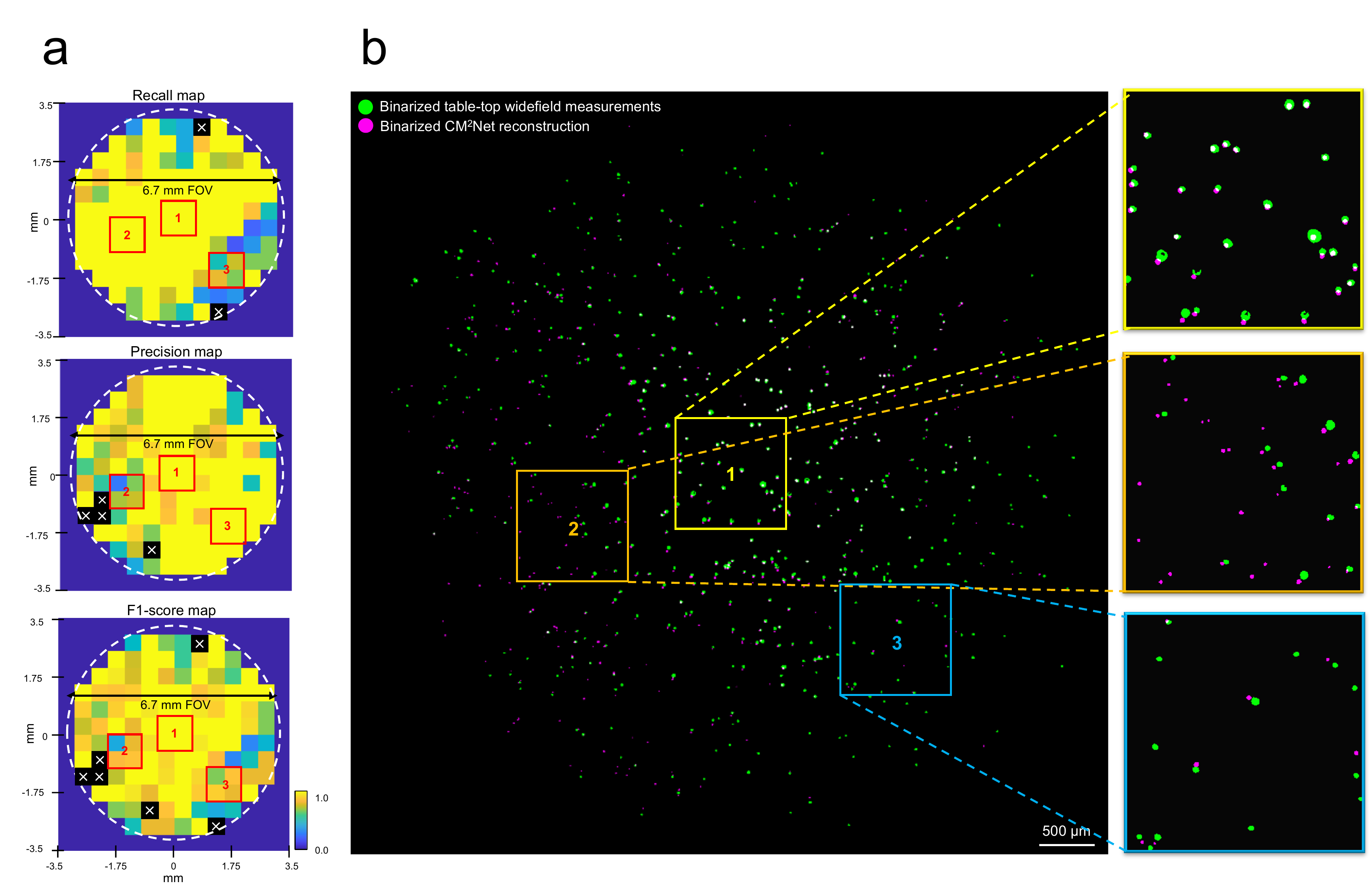}
\caption{\textbf{Quantitative evaluation on 10 µm bead experimental data.}(a) The recall, precision, and F1-score maps for the measurement in Fig. 9 of the main text (sample: 10-µm fluorescent beads in a cylindrical volume with 6.7-mm diameter and 0.5-mm depth). The ‘x’ label indicates the metric value is zero. The dashed circles show the expected boundary of the phantom object. (b) The overlay between the registered and binarized widefield measurement and the CM$^2$Net reconstruction. Patch 1 (size: 1-mm$^2$) is from the central FOV that achieves the highest F1 = 1.0. In region 1, the binarized map for widefield measurement match perfectly with the CM$^2$Net reconstruction. Patch 2 is from the edge FOV that has a low precision but with perfect recall (i.e. all ground-truth particles are matched but with unmatched falsely reconstructed particles). Patch 3 is from the edge FOV and has a low recall but with perfect precision (all detected particles are matched but with unmatched ground-truth particles). Note that since the image wrapping is unavoidable in the real experimental data and the registration is based only on the central particle pairs, the particle pairs at the peripheral FOV are expected to exhibit slight separations.}
\label{S13}
\end{figure}

For the 10 µm beads phantom in Fig. 7, the recall and precision maps are shown in Fig. \ref{S13}a. The CM$^2$Net reconstruction achieves a recall $\sim$0.78 and precision $\sim$0.80. In comparison, the recall and precision in simulation at the corresponding density range is $\sim$0.83 and $\sim$0.97, respectively (Fig. 6b). This shows that the simulator-trained CM$^2$Net degrades slightly on experiments with $\sim$5\% higher mis-detection rate and $\sim$7\% higher false-positive rate at this imaging condition. To further analyze the potential reason that causes the performance decrease in the experiment, we compare the widefield measurement, the 3×3 overlay of the predicted view-demixing results, and the CM$^2$Net reconstruction MIP. We find that the decreased performance originates from the view-demixing results (see Fig. \ref{S11}).  We attribute the reduced performance to the undesired extra views in the experimental measurements due to an extra column of partial microlenses adjacent to the main 3 $\times$ 3 microlens array (see Fig. \ref{S12}).

\begin{figure}[!h]
\centering
\includegraphics[width=\linewidth]{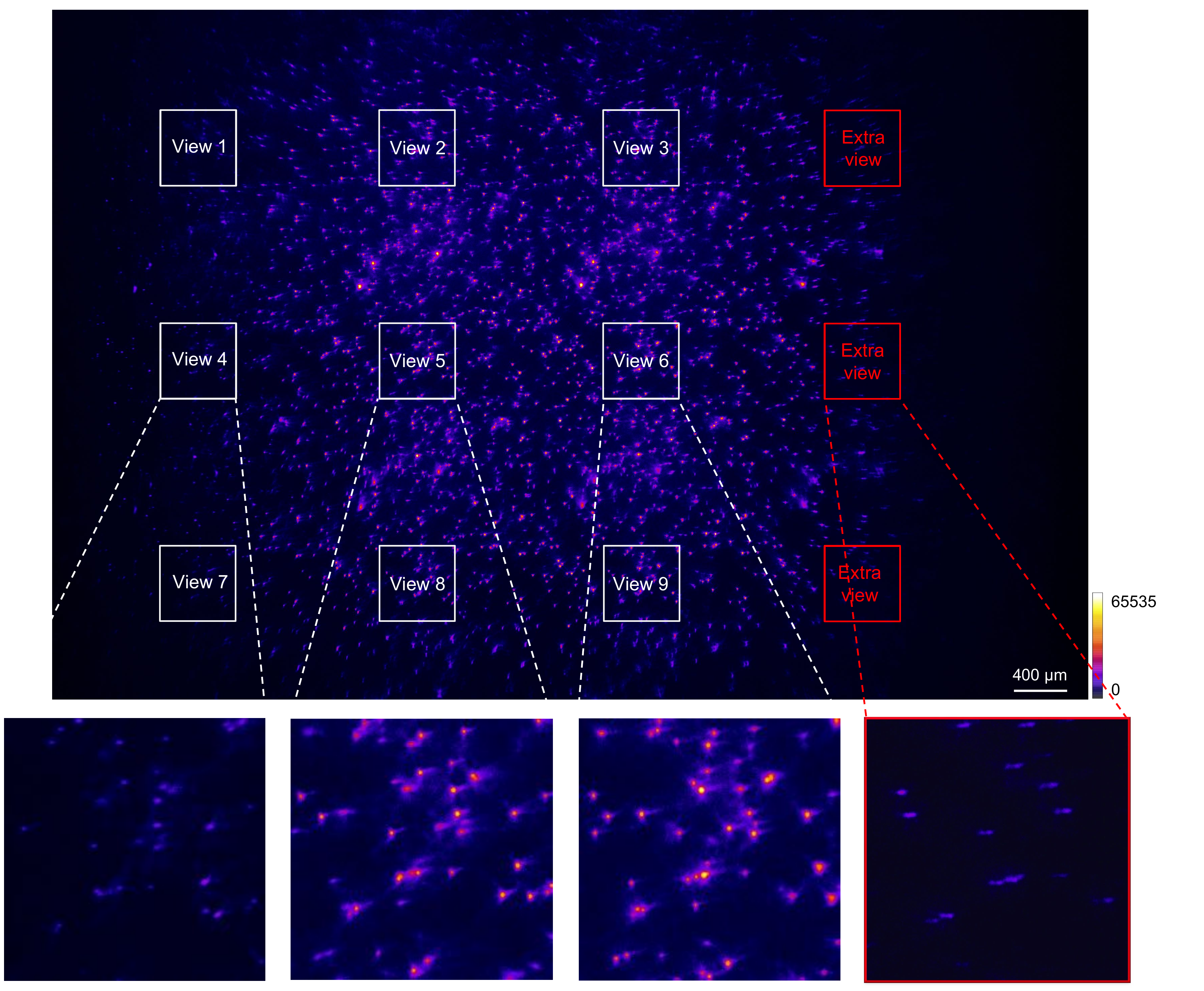}
\caption{\textbf{Extra views in the mixed-size bead experimental measurement.}Compared to the mono-10µm bead experiment, the reduced FOV (6.7mm to 6.5mm) reduces the level of contamination from the extra views, which results in decreased false positives and higher precision in the reconstruction.}
\label{S14}
\end{figure}

\begin{figure}[!h]
\centering
\includegraphics[width=\linewidth]{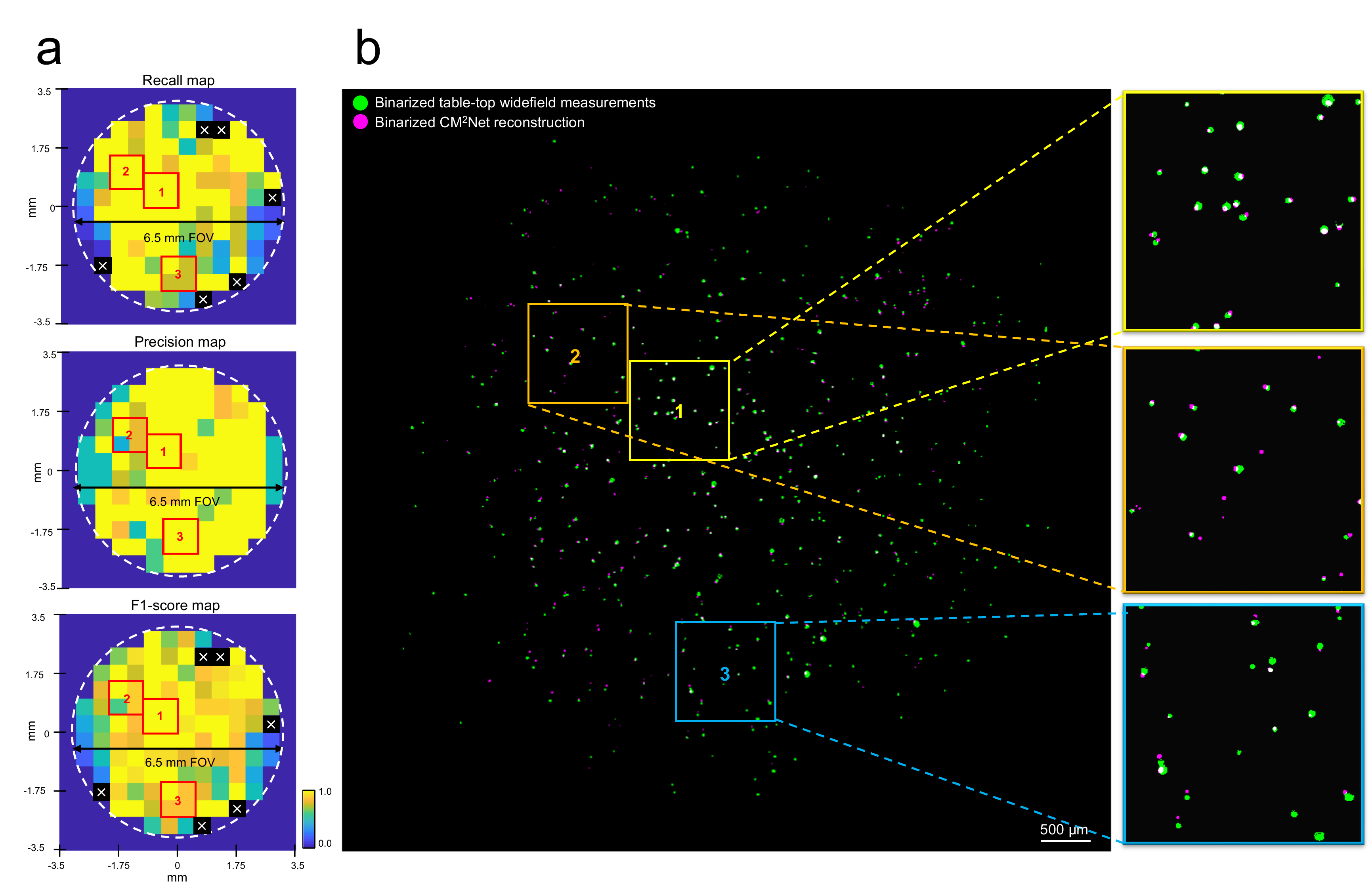}
\caption{\textbf{Quantitative evaluation on mixed-size bead experimental data.}(a) The recall, precision, and F1-score maps for the object in Fig. 10 of the main text (sample: 10-µm and 15-µm mixed fluorescent beads in a cylindrical volume with 6.5-mm diameter and 0.8-mm depth). The x labels indicate the metric value is zero and the dashed circles show the expect 6.5-mm diameter region of the phantom object. (b) The overlay of the registered and binarized widefield measurement and CM$^2$Net reconstruction. Patch 1, 2, and 3 (size: 1-mm$^2$) show example regions of having high F1-score, low recall, and low precision, respectively.}
\label{S15}
\end{figure}

\clearpage
For the mixed-size beads phantom in Fig. 8, the recall and precision maps are shown in Fig. \ref{S15}a.  The CM$^2$Net achieves averaged recall $\sim$0.73 and precision $\sim$0.84 across the entire 6.5-mm FOV. As compared to the mono-10µm bead experiment, we hypothesize that the decreased recall is attributed to the greater intensity and SNR variations caused by bead size variation in the measurement. The increased precision is due to the reduced FOV and less contaminations from the extra views (see Fig. \ref{S14}).

For better visualization, we further show the overlays of the registered and paired binarized maps between the widefield measurement and the CM$^2$Net reconstruction. We zoom-in on three patches (size: 1 mm$^2$) in the metric maps and the corresponding overlay map. Patch 1 from the central FOV achieves the highest F1 score 1.0. The patch with index 2 and 3 is extracted from the edge FOV that either has a low precision with perfect recall (all ground truth particles are matched but with redundant detection) or a low recall with perfect precision (all detected particles are matched but failed to detect some ground truth particles). Patch 2 is from the edge FOV that has a low precision but with perfect recall (i.e. all ground-truth particles are matched but with unmatched falsely reconstructed particles). Patch 3 is from the edge FOV and has a low recall but with perfect precision (all detected particles are matched but with unmatched ground-truth particles). Note that since the image wrapping is unavoidable in the real experimental data and the registration is based only on the central particle pairs, the particle pairs at the peripheral FOV are expected to exhibit slight separations. For direct visualization, we enlarge the corresponding overlap maps in Figs. \ref{S13}b and \ref{S15}b. We show in patch 1, the particles detected from the widefield measurement match perfectly with the particles reconstructed by the CM$^2$Net. In patches 2 and 3, we show unpaired particle detected from either the CM$^2$Net reconstruction (low precision) or the widefield measurement (low recall), which are consistent with our metric map, which validates the accuracy of our evaluation procedure.


\clearpage
\section{Numerical study on complex objects}

\subsection{Simulation on cortex-wide 3D neuronal imaging}

In this section, we perform pilot numerical study on imaging fluorescently labeled neurons across the entire mouse cortex to show the potential capability of applying CM$^2$ V2 and the trained CM$^2$Net for large-scale neural imaging. 
One challenge for in-vivo neural imaging is that the neurons are distributed in the highly curved cortex, which requires the imaging device to be robust to the complex 3D surface geometry. 
In the main text, we performed thorough analysis on the cylindrical volume embedded with spherical fluorescent emitters. 
Here, to demonstrate the computational pipeline can be applied to image neurons with irregular shapes, we test CM$^2$Net (trained using the dataset in the main text) and directly apply to simulated CM$^2$ measurements on the open-source dataset on VIP neurons labeled mouse brain that were imaged by the mesoscale selective plane-illumination microscope (mesoSPIM)~\cite{voigt2019mesospim}. 

\begin{figure}[htbp]
\centering
\includegraphics[width=\linewidth]{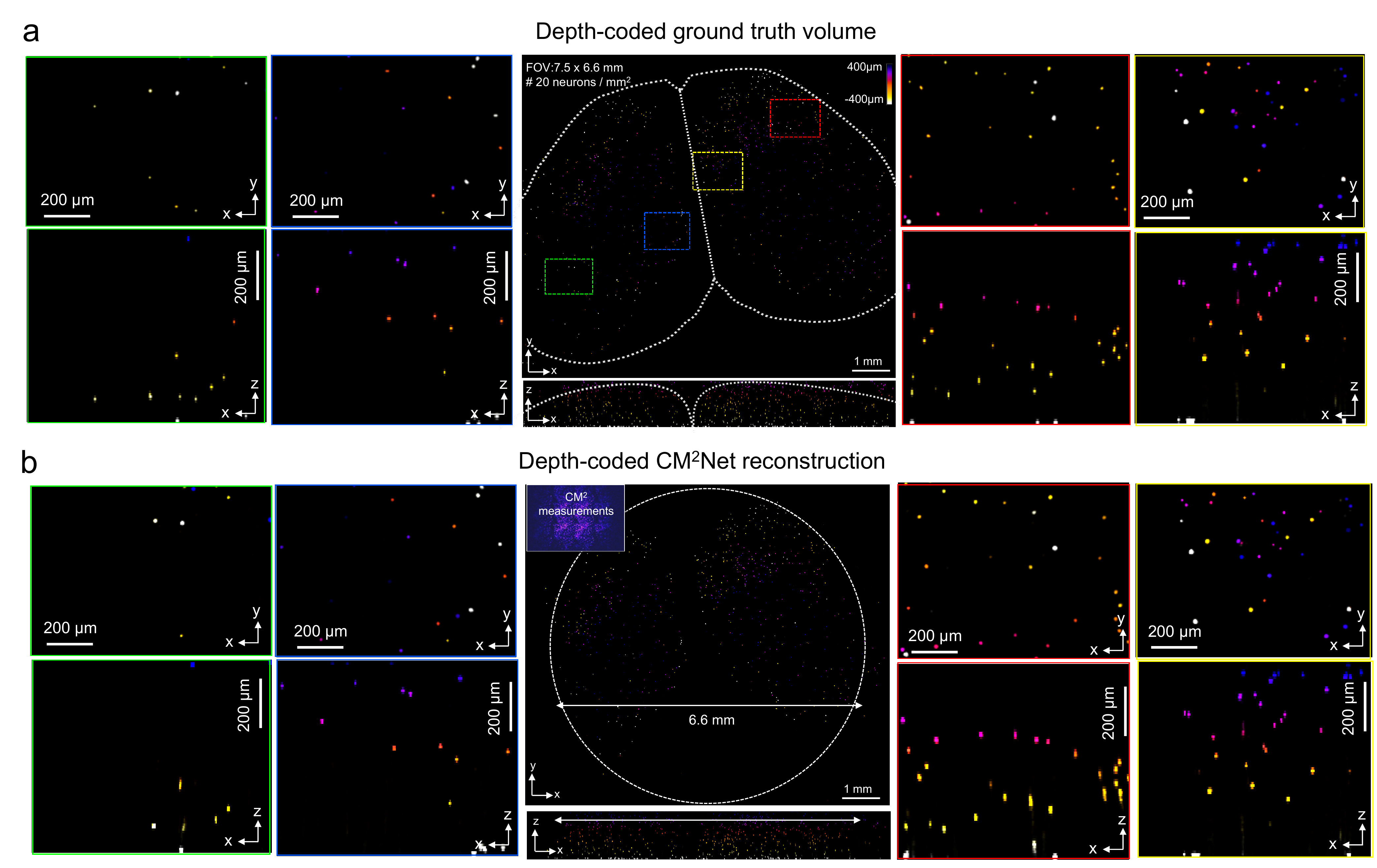}
\caption{\textbf{Numerical study on sparsely labeled neurons in mouse cortex.} (a) Visualization of the depth-encoded MIP of the ground truth volume. The volume size is  7.5 $\times$ 6.6 $\times$ 0.8 mm$^2$ and the neuron density is around 20 neurons / mm$^2$. The curved irregular contour indicates the left and right cerebral cortex region. The four patches cross the  central and edge FOVs are enlarged for comparison. (b) Visualization of depth-encoded MIP CM$^2$Net reconstruction. The synthesized CM$^2$ measurement is shown as the inset. At this low neuron density, the CM$^2$Net full-FOV reconstruction is in good agreement with the ground truth.}
\label{neuron1}
\end{figure}

\begin{figure}[htbp]
\centering
\includegraphics[width=\linewidth]{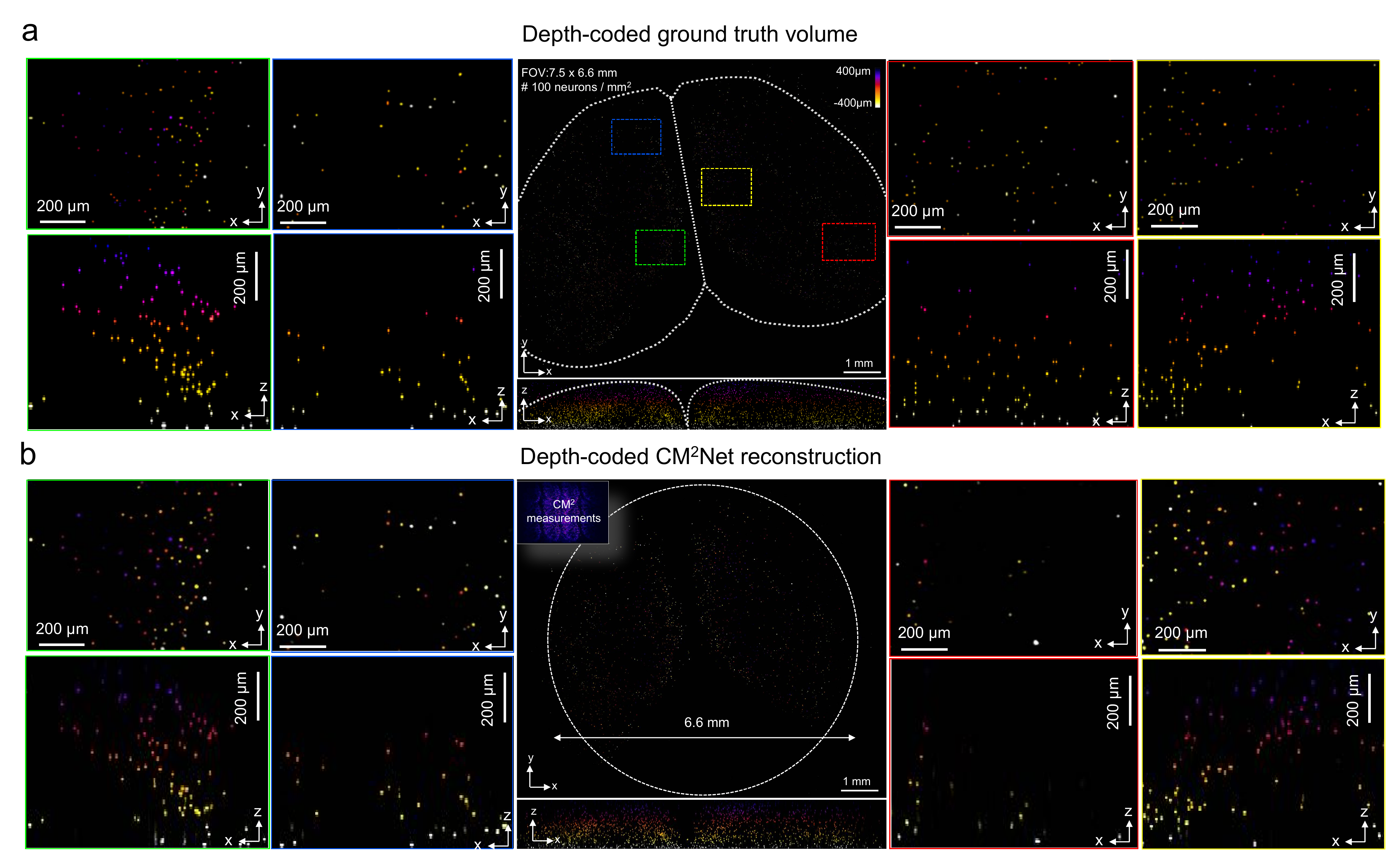}
\caption{\textbf{Numerical study on densely labeled neurons in mouse cortex.}(a) Visualization of the ground truth of the imaging volume as the color-encoded MIP. The volume size is  7.5 $\times$ 6.6 $\times$ 0.8 mm$^3$ and the neuron density is around 100 neurons / mm$^2$. The curved irregular contour labels the left and right cerebral cortex region. The four patches at the FOV central and edges are enlarged for comparison. (b) Visualization of  the CM$^2$Net reconstruction with depth encoded MIP. The synthesized CM$^2$ measurement is shown as the inset. At this high neuron density, the CM$^2$Net provides high-quality reconstruction in the central FOV (around 6.6 mm, shown in blue, yellow and green patches). The performance largely degrades at the edges (shown in the red patch), which follows the observation in Section 3.2.}
\label{neuron3}
\end{figure}

To generate the testing data, we first segment the cerebral cortical region (7.5 $\times$ 6.6 $\times$ 0.8 mm$^3$) within a 800 µm depth range from the whole mouse brain volume. We then scale the segmented volume to match with the CM$^2$ sampling grid. 
Next, we apply a threshold to remove the fluorescence background and extract the locations and sizes of the connected 3D components (neurons) from the brain volume in MATLAB (\verb|bwconncomp|, \verb|regionprops3|). The original dataset aggregates the neuronal activity across 7-8 min in a single 3D volume, resulting a high neuron density of $\sim$530/mm$^2$. We randomly select the neurons from the brain volume to down-sample the number of neurons in each measurement, which simulates the sparse neuronal activity at each time point expected in practice. 
We simulate two imaging volumes with two neuron densities, including 20/mm$^2$ and 100/mm$^2$. Next, we apply the 3D-LSV model on these two  volumes to synthesize the  CM$^2$ measurements (as shown in the inset in Figs.~\ref{neuron1}b and ~\ref{neuron3}b, respectively). Finally, the CM$^2$ measurement is input to the CM$^2$Net to perform 3D reconstruction. 

The imaging volume and CM$^2$Net reconstruction for sparsely labeled neurons is shown in Fig.~\ref{neuron1}. First, to visualize the curved brain geometry, we register the ground truth volume with the cerebral cortex contour extracted from Allen Mouse Brain Atlas~\cite{lein2007genome}(shown by the white dotted line). Next, we assess the full-FOV reconstruction by comparing the depth-encoded MIP of the ground truth volume and the CM$^2$Net 3D reconstruction. We further compare four zoom-in regions from the center (blue and yellow patches) and edge (green and red patches) FOVs. By visual inspection, CM$^2$Net successfully reconstruct the irregular 3D brain geometry and robustly reconstruct the neurons with different intensities and sizes with isotropic high resolution. The reconstruction FOV is limited to around 6.6 mm (labeled by the white circle). There are several miss-detections when exceeds the effective FOV (shown in green patch). These results agree well with the FOV analysis in Section 3.2. 

The imaging volume and the CM$^2$Net reconstruction for a densely labeled neuronal volume is shown in Fig. \ref{neuron3}. The left and right cerebral cortex contour is labeled in the ground truth volume (white dotted line). First, we validate the full-FOV reconstruction by comparing the XY and XZ MIPs of the depth-coded ground truth volume and the CM$^2$Net 3D reconstruction. To further assess the reconstruction at greater resolution, four regions from the center (green and yellow patches) and corner (blue and red) FOVs are enlarged for comparison. By visual inspection, the reconstruction matches well to the ground truth at the central 6.6$\times$6.6 mm$^2$ FOV (labeled with white circle). The performance largely degrades when exceeds the effective FOV (example shown in red patch) and it is more severe compared to the low density case, which follows the quantitative analysis in Section 3.2.

\subsection{Simulation on 3D imaging of mouse brain vessels}

\begin{figure}[htbp]
\centering
\includegraphics[width=\linewidth]{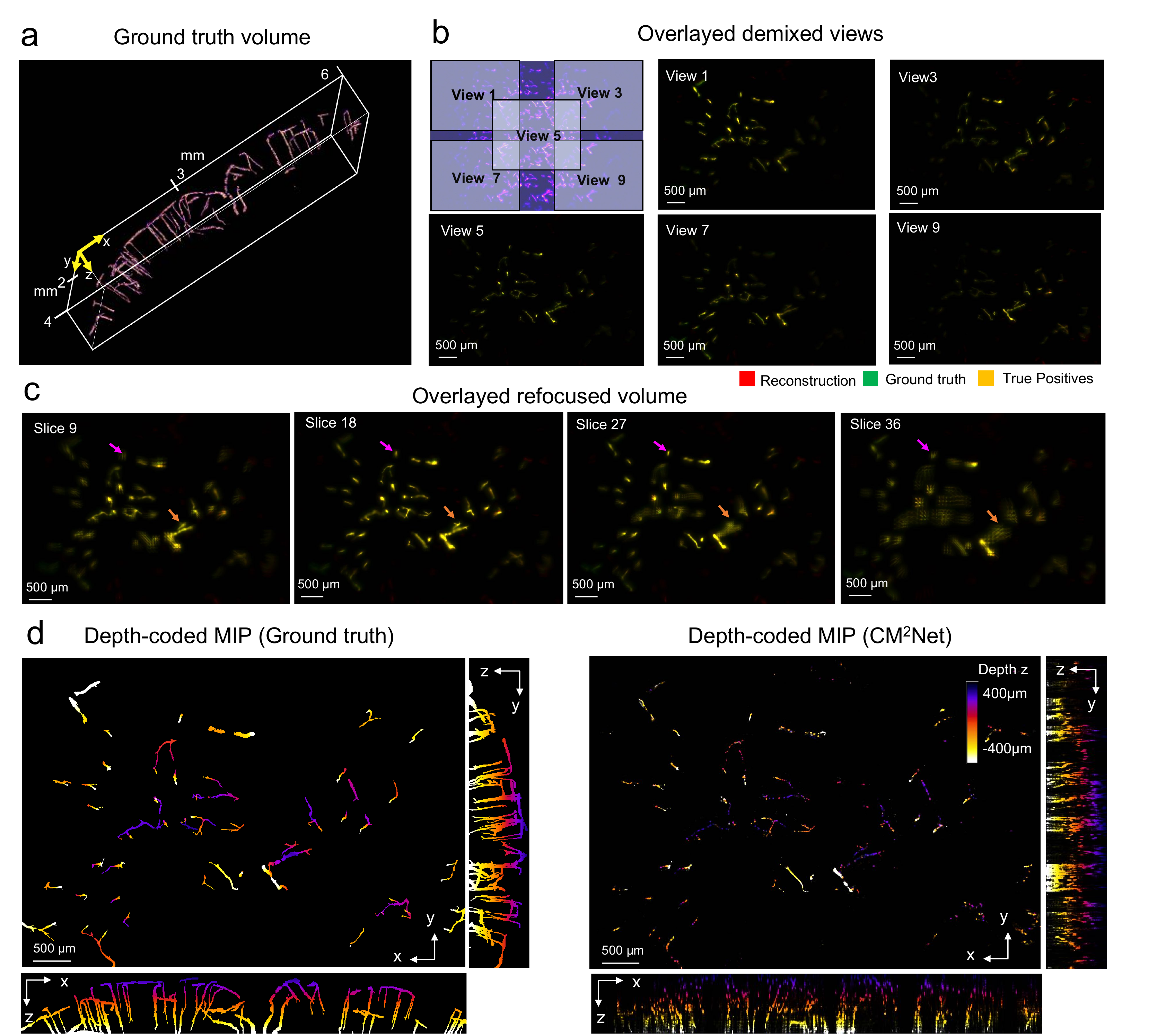}
\caption{\textbf{Numerical study on a mouse vessel network.} (a) Visualization of the ground truth volume. The volume size is 6 $\times$ 4 $\times$ 0.8 mm$^3$. The small vessels are axially overlapped and distributed in the curved mouse brain. (b) Overlap between the ground truth (green) and network prediction (red) of the demixed views.
The first figure is the synthesized CM$^2$ measurement. The labeled patches are the corresponding input for extracting the demixed views. The result shows that the network view-demixing result is matched well with the ground truth (yellow). (c) Overlap between the ground truth (green) and network prediction (red) of lightfield refocused volume (generated by the shift-and-add operation from the network demixed view stack). The slices shown in the figure are uniformly distributed across the entire refocused depth range. The orange and magenta arrows indicated the small vessels focused at the 18th and 27th slices. The result shows that the refocused volume provides accurate estimation of the 3D information. Results from (b) and (c) show that the particle-trained view-demixing network can generalize well to the continuous dense object.
(d) Depth-encoded MIPs of the ground-truth volume and CM$^2$Net reconstruction. The CM$^2$Net can accurately recover the volumetric distribution of the vessels but with discontinuity artifacts due to the sparsity enforced by the particle-trained CM$^2$Net.}
\label{vas}
\end{figure}

In this section, we perform a numerical study on a mouse vasculature to demonstrate the ability of  CM$^2$ to image more complex structures. 
A major concern for deep learning based reconstruction is the lack of generalization ability. 
Although we trained the CM$^2$Net on only sparse particles to mimic the neuron-like structures, here we study how the CM$^2$Net can be applied to more complex and non-particle objects. 
We test the CM$^2$Net on the open-source whole mouse brain vessels segmentation data set~\cite{todorov2020machine} to show the particle trained CM$^2$Net can still recover the volumetric mouse vasculature.

The preprocessing steps are similar to the previous section. First, We segment the cerebral cortical region within the 800 µm depth (volume size: 6$\times$ 4.2 $\times$ 0.8 mm$^3$) and extract the vessels from the brain volume. 
Next, we keep only small vessels within the volume with size ranging from 800-4000 pixels to form the imaging volume. Next, we scale the volume to match the CM$^2$ sampling grid and simulate the CM$^2$ measurement by our 3D-LSV model (first figure in Fig.~\ref{vas}b). Finally, the measurement is input to CM$^2$Net for 3D reconstruction.

The ground-truth volume and the CM$^2$Net view-demixing and reconstruction results are shown in Fig.~\ref{vas}. First, we show the complexity of the imaging volume in Fig.~\ref{vas}a, where axial overlapped small vessels are distributed in the  curved mouse brain. 
Next, we assess the view-demixing results by visualizing the overlay between the ground-truth demixed views (green) and the network demixed results (red). 
Figure \ref{vas}b shows that the network demixed views are matched well with the ground truth (overlay shown in yellow), demonstrating that the view-demixing network is robust to demultiplex the overlapped views for dense continuous objects, even though it is trained entirely on the sparse discrete particle data set. 
The successful generalization of the CM$^2$Net view-demixing module lays the foundation for robust reconstruction of the dense object. 
On the one hand, the demixed views are the images captured from different microlenses, which captures the information within different FOV regions. 
With an accurately demixed views stack, the VS branch can learn high quality information across the full FOV. 
On the other hand, the lightfield refocused volume generated from an accurate view stacks can provide focus information for the LFR enhancement branch, which helps the enhancement branch to accurately recover the object's depth information. To demonstrate this, we show two small vessels digitally refocused at different depth in Fig. \ref{vas}c.
Figure \ref{vas}c shows the highly matched overlay between the ground truth and the predicted lightfield refocusing volumes at different depth planes. 
Finally, we validate CM$^2$Net final Reconstruction  by comparing the XY, XZ and YZ MIPs of the depth-encoded ground-truth and CM$^2$Net reconstruction in Fig.~\ref{vas}d. 
By visual inspection, the CM$^2$Net reconstruction  matches well with the ground truth in all projections, showing that the CM$^2$Net can preform high-quality volumetric reconstruction for such dense object. 
However, we also observe discontinuity artifacts in the reconstruction result. 
This is attributed by the fact that the CM$^2$Net was only trained on the sparse particle objects, which implicitly enforces a sparsity prior on the reconstruction volume. 
We expect one can remove the discontinuity artifacts by fine tuning the CM$^2$Net on a more complex training dataset.

\clearpage

\bibliography{si_refs}